%% file: FINAL_main.tex
\documentclass[10pt,a4paper]{article}

\usepackage[a4paper,margin=1in]{geometry}
\usepackage[T1]{fontenc}
\usepackage{lmodern}
\usepackage{microtype}
\usepackage{booktabs}
\usepackage{longtable}
\usepackage{pdflscape}
\usepackage{placeins}
\usepackage{multirow}
\usepackage{tabularx}
\usepackage{threeparttable}
\usepackage{siunitx}
\usepackage{caption}
\usepackage{amsmath,amssymb,amsfonts}
\usepackage{bm}
\DeclareMathOperator{\logit}{logit}
\usepackage{mathrsfs}

\usepackage{booktabs}
\usepackage{array}
\usepackage{threeparttable}
\usepackage{tabularx}

\usepackage{graphicx}
\usepackage[export]{adjustbox}
\graphicspath{{./images/}}
\usepackage{booktabs}
\usepackage{subcaption}

\usepackage{xcolor}
\usepackage{tikz}
\usetikzlibrary{arrows.meta,positioning,shapes.geometric}
\newcommand{\sourcevar}[1]{\par{\footnotesize\texttt{#1}}}

\usepackage[backend=bibtex,style=authoryear,natbib=true,maxcitenames=2,maxbibnames=99]{biblatex}
\usepackage{authblk}

\usepackage[hidelinks]{hyperref}

\newcommand{\keywords}[1]{\vspace{0.5em}\noindent\textit{Keywords:} #1}

\title{\textbf{A general class of shared gamma frailty models for risk prediction, with application to breast cancer screening adherence}}

\author[1]{Marco Bonetti}
\author[2]{Edoardo Ratti}
\author[3]{Solveig Hofvind}

\affil[1]{\small Bocconi University, Milan, Italy}
\affil[2]{
University of Milano--Bicocca, Monza, Italy}
\affil[3]{BreastScreen Norway, Norwegian Institute of Public Health, Oslo, Norway}

\date{}

\begin{document}

\maketitle

\begin{abstract}
We introduce a new class of mixed models for longitudinal or clustered binary data. The formulation is inspired by the shared frailty multivariate survival model for clustered data, and it allows for closed forms for all the relevant quantities, which allows for an efficient and fast computational solution of the associated inferential and prediction problems. We focus on an easily interpretable specification of the model, discuss the associated individual risk prediction, and compare our results to the traditional GLMM model for binary data with random intercept. We implement the methods on a large motivating dataset on the individual longitudinal adherence to the breast cancer screening program of the Oslo region.\end{abstract}


\section{Introduction}
Clustered binary outcomes arise when individuals are nested within centers,
members belong to the same household, or repeated observations are collected
from the same individual. Responses within a cluster may remain associated
after accounting for measured covariates. Accounting for this dependence is
important both for inference on covariate effects and for predicting individual
outcomes. Two widely used approaches to modeling correlated discrete responses are
marginal models and random-effects models
\citep{MolenberghsVerbeke2005}. Marginal models specify the population mean
response separately from the association structure. Under this approach, Generalized Estimating
Equations (GEE) had provided an influential framework for estimating such models without
specifying a full joint distribution \citep{LiangZeger1986}.

Random-effects models introduce latent cluster-specific quantities that
aim to capture residual heterogeneity and induce dependence among responses. Within this approach, Generalized Linear Mixed Models (GLMMs) extend generalized linear models to
clustered data, retaining the familiar interpretation of regression
coefficients while allowing flexible dependence through the random-effects
structure \citep{BreslowClayton1993}. Predicting these latent effects supports
model diagnostics and comparisons between clusters
\citep{SkrondalRabeHesketh2009}. The latent effects may also be of direct interest,
as in small-area estimation for the prediction of area-specific risks
\citep{JiangLahiri2001,Jiang2003,RaoMolina2015} and in identifying clusters with
unusually high or low outcome risks
\citet{NeuhausMcCullochBoylan2025}.

Despite these advantages, GLMMs has several potential drawbacks. Parameter estimation can be computationally
demanding. In logistic-normal GLMMs, the likelihood requires integration over
the random effects, generally without a closed-form solution
\citep{Demidenko2013,JiangNguyen2021}. Common strategies include Laplace
approximations, marginal and penalized quasi-likelihood
\citep{BreslowClayton1993}, adaptive Gauss--Hermite quadrature
\citep{LiuPierce1994,PinheiroBates1995}, and Monte Carlo likelihood methods
\citep{McCulloch1997,BoothHobert1999}. These strategies require to balance computational
cost against approximation or simulation error. Quasi-likelihood methods can
be inaccurate for binary responses with small clusters or substantial
heterogeneity \citep{RodriguezGoldman1995}, while quadrature becomes expensive
as the dimension of the random effects increases. Alternatives to design models with tractable conditional and marginal
relationships were investigated such as marginally specified logistic-normal models
\citep{Heagerty1999}, bridge distributions \citep{WangLouis2003}, and
specialized link--distribution combinations
\citep{KazemiHassanzadeh2021}.
Explicit marginal means, however, do not by themselves imply a closed-form joint marginal likelihood.

Random-effect prediction presents similar challenges. For known model
parameters, the posterior mean of the random effect given the observed cluster data minimizes
expected squared-error loss. Replacing the parameters by their maximum likelihood estimates yields the
empirical Bayes posterior mean, also called an empirical best predictor
\citep{Robinson1991,SkrondalRabeHesketh2004,SkrondalRabeHesketh2009,JiangNguyen2021}.
For non-normal outcomes, its computation typically involves a ratio of integrals, evaluated using
quadrature or Laplace-based approximations such as those of \citet{TierneyKadane1986}.
\textcite{SkrondalRabeHesketh2009} discuss posterior prediction in multilevel generalized linear models, including the use of adaptive quadrature to evaluate posterior moments. Other proposals target prediction in the specific GLMMs logistic settings. \textcite{TamuraGiampaoli2010} investigate procedures based on empirical best prediction and \textcite{TamuraGiampaoli2013} develop a computationally simpler prediction method based on regressing fitted random effects on aggregated group covariates. In the context of prediction, \textcite{SunEtAl2018} propose classified mixed logistic prediction that links new observations from a validation data to clusters available in a training data through a classification step.

A less developed literature obtains analytically tractable likelihoods and random effect predictions by
combining binary response models with positive distributed random effects.
\citet{Conaway1990} used a complementary log-log link and log-gamma random
effects to integrate the likelihood analytically.
\citet{CoullHousemanBetensky2006} extended this work to multivariate
gamma random effects, allowing more flexible dependence while retaining
computational tractability while \citet{ZhangChan2014} adoped correlated
exponential latent effects. \citet{RiceJohnsonStrawderman2019} model repeated HIV-testing indicators
using a shared frailty framework. For this application, they specify the
frailty distribution within the power-variance-function family, which includes
the possibility of a point mass at zero to represent potential never-testers.

Following these methodological developments, we propose a class of models for
clustered binary responses based on a shared gamma frailty and a flexible,
covariate-dependent reference probability. The term \textit{shared gamma frailty}
is inherited from the survival-analysis literature. In our framework, indeed,
the random effect does not enter the linear predictor additively. Instead, it
acts multiplicatively on a non-negative function of the covariates. The proposed class has several advantages. First, likelihood inference,
marginal probabilities, and posterior predictive quantities are available in
closed form, avoiding numerical integration or
Monte Carlo methods and leading to a straightforward implementation. Second,
the joint distribution of binary responses within a cluster has a
computationally simple expression. Third, the reference probability, or more
generally the covariate-dependent function defining the model, can be chosen
flexibly according to the scientific setting and the predictive objective.
Finally, specifying a logistic reference probability preserves the usual
interpretation of regression coefficients and facilitates a direct comparison
with the logistic random intercept GLMM, one of the most
widely used approaches for clustered binary data. The proposed model can
therefore be considered as a computationally tractable alternative to the
logistic random intercept GLMM.

The remainder of the paper is organized as follows. Section~\ref{sec:meth} introduces the
general model class and develops its likelihood, marginal probabilities, and
posterior predictive quantities, with particular attention to the logistic
reference probability. Section~\ref{sec:simulation-study} investigates the performance of the proposed
model through extensive simulation studies. Section~\ref{sec:comp_BLMM_GLMM}
compares it with the logistic random intercept GLMM. Section~\ref{sec:CaseStudy} presents an application to the
longitudinal adherence data of women in the catchment region of the Oslo breast cancer screening program in Norway. Section~\ref{sec:Discussion} discusses the results and outlines directions
for further research.

\newpage
\section{Methods}\label{sec:meth}
\subsection{A general class with shared gamma frailties}\label{sec:class}

Consider independent clusters $i=1,\ldots,n$, containing binary responses $Y_{ij}$, $j=1,\ldots,n_i$, and fixed covariate vectors $\bm x_{ij}$. Write $\bm Y_i=(Y_{i1},\ldots,Y_{in_i})'$ and let $\bm X_i$ contain the corresponding design vectors. Conditional on a positive cluster-specific frailty, assume
\begin{equation}\label{eq:kernel}
\begin{aligned}
  &Y_{ij}\mid\gamma_i,\bm X_i
 \overset{\mathrm{ind}}{\sim}\operatorname{Bernoulli}\{\pi_C(\gamma_i;\bm x_{ij})\},\\[0.4cm]
 &E(Y_{ij} \mid \gamma_i, \bm x_{ij})=\pi_C(\gamma_i;\bm x_{ij})=\exp\{-\gamma_i u(\bm x_{ij};\bm\beta)\}. 
\end{aligned}
\end{equation}

Here $u(\bm x;\bm\beta)$ is measurable, finite, strictly positive on the covariate and parameter domains, and independent of $\theta$.  We suppress parameter arguments in probabilities when unambiguous and write $u_{ij}=u(\bm x_{ij};\bm\beta)$. Independently across clusters, and independently of the covariates, we assume
\begin{equation}\label{eq:gamma}
 \gamma_i\sim\operatorname{Gamma}(\theta,\theta),\qquad
 E(\gamma_i)=1,\quad\operatorname{Var}(\gamma_i)=\theta^{-1},\quad\theta>0.
\end{equation}
This formulation implies that a larger frailty decreases the success probability.
A useful specification of \eqref{eq:kernel} starts with a reference function $B:\mathbb R\to(0,1)$. Denoting the linear predictor as $\eta_{ij}=\bm x_{ij}'\bm\beta$, the reference function $B$ can be defined such that 
\begin{equation}\label{eq:reference}
 u_{ij}(\eta_{ij})=-\log B(\eta_{ij}),\qquad
 \pi_C(\gamma_i;\bm x_{ij})=B(\eta_{ij})^{\gamma_i}.
\end{equation}
Thus $B(\eta_{ij}) = \pi_C(1;\bm x_{ij})$ is the probability at unit frailty, i.e. at a mean latent risk. This convenient formulation allows one to choose an appropriate - and interpretable - baseline risk function on which the individual latent risk then acts as a modifier. A strictly increasing $B$ gives success probabilities increasing in $\eta$. A cumulative density function (CDF) on $\mathbb R$ is an appropriate choice, but being a CDF is not required for the calculations. Table~\ref{tab:links} lists possible examples. On the other hand, differentiability is needed for standard likelihood inference.  Note that the shared-frailty restriction of \citet{CoullHousemanBetensky2006}, which builds on \citet{Conaway1990}, is a member of this class with $B(\eta)=\exp(-e^\eta)$.

\begin{table}[!htbp]
\centering\small
\caption{Reference probabilities and the corresponding positive functions. $\Phi$ denotes the standard normal CDF.}\label{tab:links}
\begin{tabularx}{\textwidth}{@{}X l l c@{}}
\toprule
Specification & $B(\eta)$ & $u(\eta)=-\log B(\eta)$ \\
\midrule
Logistic & $(1+e^{-\eta})^{-1}$ & $\log(1+e^{-\eta})$ \\
Probit & $\Phi(\eta)$ & $-\log\Phi(\eta)$ \\
Complementary log-log & $1-\exp(-e^\eta)$ & $-\log\{1-\exp(-e^\eta)\}$ \\
Increasing log-log & $\exp(-e^{-\eta})$ & $e^{-\eta}$\\
Decreasing log-log, (Coull et al. 2006, for a shared scalar frailty) & $\exp(-e^\eta)$ & $e^\eta$\\
\bottomrule
\end{tabularx}
\end{table}

Linearity in $\gamma$ in the exponent allows to recover marginal probability of the outcome, i.e., averaged over the population frailty distribution while holding the covariates fixed. Recall that the gamma Laplace transform is
\begin{equation}\label{eq:lt}
 L_\theta(s)=E(e^{-s\Gamma})=(1+s/\theta)^{-\theta},\qquad s\geq0.
\end{equation}
Thus, the marginal probabilty of the outcome is
\begin{equation}\label{eq:pm}
\begin{aligned}
  \pi_M(\bm x_{ij})&=E_\Gamma\{\pi_C(\Gamma;\bm x_{ij})\}
 =L_\theta\{u(\bm x_{ij};\bm\beta)\}
 =\left(\frac{\theta}{\theta+u(\bm x_{ij}, \bm\beta)} \right)^{\theta}.  
\end{aligned}   
\end{equation}

For finite $\theta$, strict convexity gives $\pi_M(\bm x_{ij})>\pi_C(1;\bm x_{ij})$; for fixed $u$, the two converge as $\theta\to\infty$.

We can also characterize the correlation between two observations induced by the model defined in \eqref{eq:kernel}. For two distinct observations $Y_{ij}, Y_{ik}$ in the same cluster, with covariates $\bm x_{ij}$ and $\bm x_{ik}$, the correlations can be recovered from \eqref{eq:corr1}. 
\begin{equation}\label{eq:corr1}
 \operatorname{Corr}(Y_{ij},Y_{ik} \mid \bm x_{ij}, \bm x_{ik})
 =\frac{L_\theta(u_{ij}+u_{ik})-\pi_M(\bm x_{ij})\pi_M(\bm x_{ik})}
 {\{\pi_M(\bm x_{ij})(1-\pi_M(\bm x_{ij}))\pi_M(\bm x_{ik})(1-\pi_M(\bm x_{ik}))\}^{1/2}}.
\end{equation}
The covariance is positive for finite $\theta$ and positive $u_{ij},u_{ik}$. For equal covariates, that is $\bm x_{ij} = \bm x_{ik} = \bm x$, the covariate equals $\operatorname{Var}_\Gamma\{\pi_C(\Gamma;\bm x)\}
 =L_\theta(2u)-L_\theta(u)^2$ and the correlation is then
\begin{equation}\label{eq:corr2}
 \operatorname{Corr}(Y_{ij},Y_{ik} \mid \bm x)
 =\frac{L_\theta(2u)-L_\theta(u)^2}{\pi_M(\bm x_{ij})\{1-\pi_M(\bm x_{ij})\}}.
\end{equation}
These expressions and their limits are derived in Appendix~\ref{app:corr}.

\subsubsection{Coefficient interpretation }\label{sec:ident}

For differentiable $u(\bm x_{ij};\bm\beta)$, the effect of a single covariate $x_k$ on the marginal probability can be expressed as 
\begin{equation*}\label{eq:geffect}
 \frac{\partial\pi_M(\bm x_{ij})}{\partial x_k}
 =-\frac{\theta}{\theta+u(\bm x_{ij};\bm\beta)}
       \pi_M(\bm x_{ij})\frac{\partial u(\bm x_{ij};\bm\beta)}{\partial x_k}.
\end{equation*}
The derivative is local and is taken with all other covariates fixed. From \eqref{eq:geffect} it is apparent that covariate effect is generally not straightforward to interpreter and depends of the choice of $u(\bm x_{ij};\bm\beta)$. Nevertheless, the logistic reference function $B$ is particularly useful because it gives the familiar log-odds interpretation of regression coefficients. In addition, such specification leads itself to a natural comparison to widely used binary normal random intercept model with link logit from the class of generalized linear mixed models (GLMM).  In the rest of the article we thus focus on this specialization. In Appendix~\ref{app:lik}, we derive closed-form expressions for the main quantities of
interest under the logistic reference probability and identify the steps that
extend directly to other members of the shared gamma model class such as those in Table~\ref{tab:links}. The closed-form likelihood follows the marginal-probability
inversion of \textcite{Conaway1990}.

\subsection{The Binary Lehmann Mixed Model (BLMM)}\label{sec:blmm}

Choosing $B(\eta_{ij})=\operatorname{expit}(\eta_{ij})$ in~\eqref{eq:reference} gives $u(\bm x_{ij};\bm\beta)=\log(1+e^{-\bm x_{ij}'\bm\beta})$. Thus

\begin{equation}\label{eq:pc}
 \pi_C(\gamma_i;\bm x_{ij})=\left\{\frac{e^{\bm{\beta}'\bm{x}_{ij}}}
{1+e^{\bm{\beta}'\bm{x}_{ij}}}\right\}^{\gamma_i}.
\end{equation}
We call this the Binary Lehmann Mixed Model (BLMM) because it applies a Lehmann power transformation used in porportional hazard and frailty models for survival data to the logistic reference probability \citep{Lehmann1953,VaupelEtAl1979,BalanPutter2020}. Under the LBMM, a larger frailty therefore corresponds to a smaller probability of success.
From \eqref{eq:lt}, the marginal probability is
\begin{equation}\label{eq:bpm}
 \pi_M(\bm x_{ij})=\left\{\frac{\theta}{\theta-\log\pi_C(1;\bm{x}_{ij})}\right\}^{\theta}.
\end{equation}

\begin{figure}[!htbp]
\centering\includegraphics[width=0.72\textwidth]{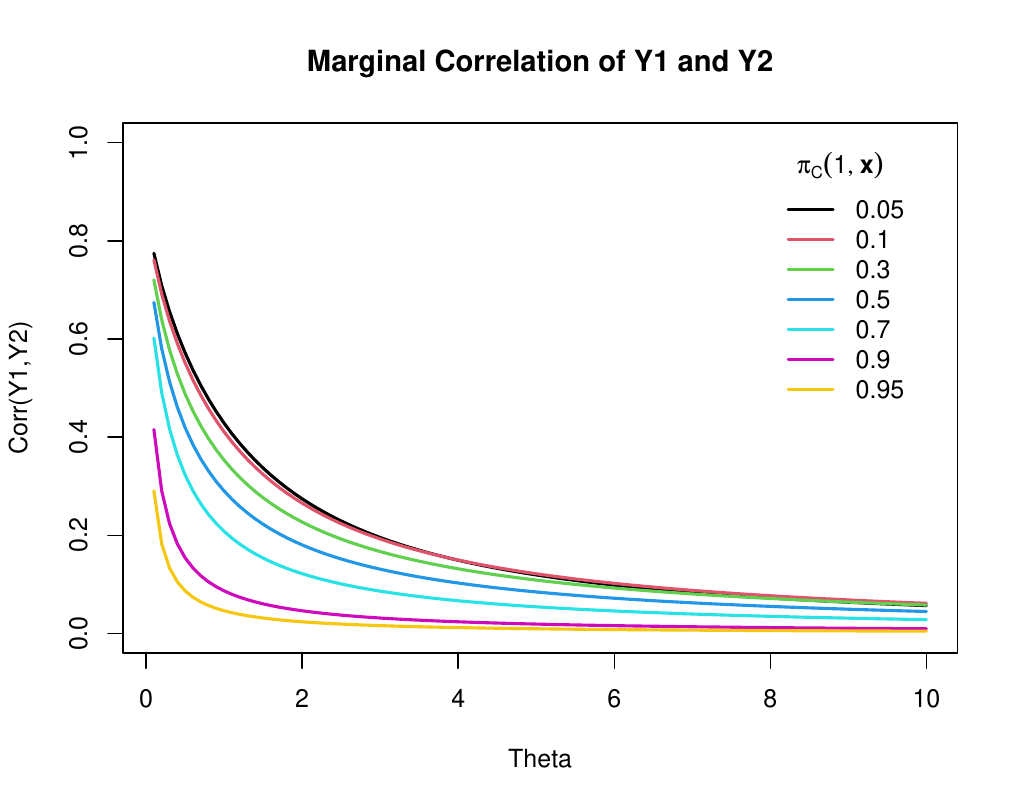}
\caption{BLMM marginal correlation between two same-cluster outcomes with equal covariates, as a function of $\theta$. Curves correspond to the reference probabilities $\pi_C(1;\bm x)$ shown in the legend.}\label{fig:corr}\label{fig:corrY1Y2}
\end{figure}

For $0<\pi_C(1;\bm{x}_{ij})<1$, the map $\gamma\mapsto\{\pi_C(1;\bm{x}_{ij})\}^{\gamma}$ is strictly convex. Since the gamma distribution is nondegenerate for finite $\theta>0$ and has mean one, Jensen's inequality gives
\[
\pi_M(\bm{x}_{ij})>\{\pi_C(1;\bm{x}_{ij})\}^{E(\gamma_i)}=\pi_C(1;\bm{x}_{ij}).
\]
Moreover,
\[
\lim_{\theta\to\infty}\left\{1-\frac{\log\pi_C(1;\bm{x}_{ij})}{\theta}\right\}^{-\theta}
=\exp\{\log\pi_C(1;\bm{x}_{ij})\}=\pi_C(1;\bm{x}_{ij}).
\]
The correlation structure induced by the BLMM is recovered from \eqref{eq:corr1} and \eqref{eq:corr2}, that is
\begin{equation*}
\operatorname{Corr}(Y_{ij},Y_{ik}) = 
\begin{cases} 
\frac{
\left\{\dfrac{\theta}{\theta-\log\pi_C(1;\bm{x}_1)-\log\pi_C(1;\bm{x}_2)}\right\}^{\theta}
-\pi_M(\bm{x}_1)\pi_M(\bm{x}_2)
}{\sqrt{\pi_M(\bm{x}_1)\{1-\pi_M(\bm{x}_1)\}
\pi_M(\bm{x}_2)\{1-\pi_M(\bm{x}_2)\}}}, & \text{if } \bm x_{ij} \neq \bm x_{ik} \\[0.9cm] 
\frac{\left\{\dfrac{\theta}{\theta-2\log\pi_C(1;\bm{x})}\right\}^{\theta}
-\left\{\dfrac{\theta}{\theta-\log\pi_C(1;\bm{x})}\right\}^{2\theta}}
{\pi_M(\bm{x})\{1-\pi_M(\bm{x})\}}, & \text{if } \bm x_{ij} = \bm x_{ik} 
\end{cases}
\end{equation*}
The nonnegativity of the variance follows from
\[
\{\theta-\log\pi_C(1;\bm{x})\}^2-\theta\{\theta-2\log\pi_C(1;\bm{x})\}
=\{\log\pi_C(1;\bm{x})\}^2\geq0.
\]
Figure~\ref{fig:corr} illustrates the induced correlation for equal covariates.

Since $\logit\{\pi_C(1;\bm x_{ij})\}=\bm x_{ij}'\bm\beta$, $\beta_k$ is the change in log-odds for a unit increase in a covariate entering linearly without interactions, with other covariates fixed and $\gamma=1$. Its corresponding odds ratio is $e^{\beta_k}$ and is interpreted as the odds ratio at a mean frailty. The meaninffullness of these intepretation depends on how the heterogeneous is the observed sample. hetergeness can be asseed by looking at the estimate of $\theta$, where larger values correpsonds to lower heterogeneity.

Looking at the marginal effects, the local effects on the marginal probability of a covariate $x_k$ entering linearly without interactions is
\begin{equation*}
\begin{aligned}
      \frac{\partial\pi_M(\bm x)}{\partial x_k}
&=\frac{\theta}{\theta+u(\bm x_{ij};\bm\beta)}
       \pi_M(\bm x_{ij})(1-\operatorname{expit}(\bm x'_{ij}\bm \beta))\beta_k\\[0.2cm]
&=s(\bm x_{ij};\theta)\beta_k,  
\end{aligned} 
\end{equation*}\label{eq:bprob}
where $s(\bm x_{ij};\theta)$ is a positive, profile-specific scaling factor that translates the unit-frailty log-odds coefficient \(\beta_k\) into the local effect of \(x_k\) on the marginal success probability.
The induced marginal logit is 
\begin{equation*}\label{eq:blogit}
   \logit\{\pi_M(\bm x_{ij})\}=-\log\left\{\left(1+\frac{\log(1+e^{-\bm x_{ij}'\bm\beta})}{\theta}\right)^\theta-1\right\}.
\end{equation*}

Therefore, the marginal odds ratio for two covariate profiles with probabilities $\pi_M(\bm x_{1}),\pi_M(\bm x_{0})$ is
\begin{equation*}\label{eq:orm}
 \mathrm{OR}_M(\bm x_1,\bm x_0) = \exp\{ \logit\{\pi_M(\bm x_{1})\} -\logit\{\pi_M(\bm x_{0})\} \}
 =\frac{r_0^\theta-1}{r_1^\theta-1},
\end{equation*}
where we denoted $r_h=1+\log(1+e^{-\bm x'_{ij}\bm \beta})/\theta$, $h=0,1$ for ease of notation. 

\subsubsection{Parameter estimation}\label{sec:est}

Let $\bm{\gamma}=(\gamma_1,\ldots,\gamma_n)'$ denote the frailty vector. For cluster $i$, let $\bm{X}_i=(\bm{x}_{i1},\ldots,\bm{x}_{in_i})'$ be the $n_i\times p$ design matrix. The complete response vector and design matrix are $\bm{Y}=(\bm{Y}_1',\ldots,\bm{Y}_n')'$ and $\bm{X}=(\bm{X}_1',\ldots,\bm{X}_n')'$, respectively. Their observed values are denoted by $\bm{y}$ and $\bm{X}$.

For each cluster, define the index sets
\[
\mathcal{J}_{i1}=\{j\in\{1,\ldots,n_i\}:y_{ij}=1\},\qquad
\mathcal{J}_{i0}=\{j\in\{1,\ldots,n_i\}:y_{ij}=0\},
\]
and let $n_{i0}=|\mathcal{J}_{i0}|$. For every subset $A\subseteq\mathcal{J}_{i0}$, set
\begin{equation*}\label{eq:rate}
a_i(A)
=\theta-\sum_{j\in\mathcal{J}_{i1}}\log\pi_C(1;\bm x_{ij})
-\sum_{j\in A}\log\pi_C(1;\bm x_{ij}).
\end{equation*}
Here, $|A|$ is the number of elements in $A$, and an empty sum equals zero. Dependence of $a_i(A)$ on $\bm{\beta}$ and $\theta$ is implicit. Since $0<\pi_C(1;\bm x_{ij})<1$ and $\theta>0$, every $a_i(A)$ is positive. Integrating out the independent frailties gives the marginal likelihood
\begin{equation}\label{eq:lik}
\begin{aligned}
L(\bm{\beta},\theta)
&=f(\bm{y}\mid\bm{X};\bm{\beta},\theta)
=\prod_{i=1}^n D_i(\bm{\beta},\theta),\\
D_i(\bm{\beta},\theta)
&=\sum_{A\subseteq\mathcal{J}_{i0}}
(-1)^{|A|}\left\{\frac{\theta}{a_i(A)}\right\}^{\theta}.
\end{aligned}
\end{equation}
The sum runs over all $2^{n_{i0}}$ subsets, including the empty set and $\mathcal{J}_{i0}$ itself. The terms for subsets of size zero, one, two, and so on correspond to the alternating expansion of the Bernoulli failure factors. The contribution $D_i(\bm{\beta},\theta)=f(\bm{y}_i\mid\bm{X}_i;\bm{\beta},\theta)$ depends on the observed data of cluster $i$, which are suppressed in its arguments. Appendix~\ref{app:lik} gives the full expansion and integration using the same notation.

Maximizing~\eqref{eq:lik} yields the maximum likelihood estimates $\widehat{\bm{\beta}}$ and $\widehat{\theta}$. Optimization is performed with $\theta=\exp(\psi)$ to enforce positivity. Standard errors are obtained from the observed information and transformed to the original parameter scale by the delta method; Appendix~\ref{app:se} gives the details.

The fully specified marginal likelihood allows likelihood-based model comparison and assessment, including likelihood-ratio or deviance comparisons between nested models. In addition, probability-based metrics provide complementary, interpretable comparisons between fitted probabilities and observed outcomes. On the fitting data they describe apparent fit, while on appropriately external data they validate prediction performances. Appropriate metrics and strategies for model assessment and validation in the context of cluster data are discussed by \textcite{VanOirbeekLesaffre2012}  \textcite{BouwmeesterEtAl2013,WynantsEtAl2018} \textcite{WynantsEtAl2015} \textcite{BarrenadaEtAl2026,MunozEtAl2025}. The metrics considered in this work for evaluating the BLMM are reported in Table~\ref{tab:perf}.

\subsubsection{Prediction}\label{sec:pred}

The conditional and marginal probabilities in~\eqref{eq:pc} and~\eqref{eq:pm} address different prediction targets. Conditional prediction incorporates the frailty of a particular cluster, inferred from its observed responses. Marginal prediction averages over the fitted population frailty distribution. We first obtain posterior summaries of the cluster-specific frailties, then use them to construct conditional predictions and their intervals. We also describe the corresponding marginal predictions and confidence intervals.

\paragraph{Predicting cluster-specific frailties}\label{sec:frail}

At fixed values of $\bm{\beta}$ and $\theta$, the posterior distribution of $\gamma_i$ depends on the data through $(\bm{y}_i,\bm{X}_i)$. Its normalizing constant is the same cluster likelihood contribution $D_i$ appearing in~\eqref{eq:lik}:
\begin{equation*}\label{eq:post}
f(\gamma_i\mid\bm{y}_i,\bm{X}_i;\bm{\beta},\theta)
=\frac{f_{\Gamma}(\gamma_i;\theta)}{D_i(\bm{\beta},\theta)}
\prod_{j=1}^{n_i}
\{\pi_{ij}(\bm{\beta})^{\gamma_i}\}^{y_{ij}}
\{1-\pi_{ij}(\bm{\beta})^{\gamma_i}\}^{1-y_{ij}}.
\end{equation*}
Posterior means, medians, and equal-tailed intervals follow from this distribution. Substitution of the maximum likelihood estimates gives their empirical Bayes versions. The algebra is provided in Appendix~\ref{app:post}.

Using the subsets and denominators defined in Section~\ref{sec:est}, let
\begin{equation*}\label{eq:ki}
K_i(\bm{\beta},\theta)
=\sum_{A\subseteq\mathcal{J}_{i0}}
(-1)^{|A|}\left\{\frac{\theta}{a_i(A)}\right\}^{\theta+1}.
\end{equation*}
The posterior mean and its empirical Bayes predictor are respectively
\begin{equation*}\label{eq:mean}
E(\gamma_i\mid\bm{y}_i,\bm{X}_i;\bm{\beta},\theta)
=\frac{K_i(\bm{\beta},\theta)}{D_i(\bm{\beta},\theta)}, \qquad
\widehat{\gamma}_i
=\frac{K_i(\widehat{\bm{\beta}},\widehat{\theta})}
{D_i(\widehat{\bm{\beta}},\widehat{\theta})}.
\end{equation*}

In addition, let $F_{\Gamma}(t;a,b)$ denote the CDF of a gamma random variable with shape $a$ and rate $b$. The posterior CDF is
\begin{equation*}\label{eq:cdf}
\begin{aligned}
F_i(t;\bm{\beta},\theta)
&:=P(\gamma_i\leq t\mid\bm{y}_i,\bm{X}_i;\bm{\beta},\theta)\\
&=\frac{1}{D_i(\bm{\beta},\theta)}
\sum_{A\subseteq\mathcal{J}_{i0}}(-1)^{|A|}
\left\{\frac{\theta}{a_i(A)}\right\}^{\theta}
F_{\Gamma}\{t;\theta,a_i(A)\},\qquad t\geq0.
\end{aligned}
\end{equation*}
Define $\widehat{F}_i(t)=F_i(t;\widehat{\bm{\beta}},\widehat{\theta})$ and
$\widehat{F}_i^{-1}(u)=\inf\{t\geq0:\widehat{F}_i(t)\geq u\}$ for $0<u<1$.
The posterior median predictor and an equal-tailed interval with posterior probability $1-\alpha$, at the fitted parameter values, are
\begin{equation}\label{eq:quant}
\bar{\gamma}_i=\widehat{F}_i^{-1}(0.5),
\qquad
[L_i,U_i]=\left[\widehat{F}_i^{-1}(\alpha/2),
\widehat{F}_i^{-1}(1-\alpha/2)\right].
\end{equation}
The CDF is available in closed form, while its quantiles are obtained by numerical inversion. These intervals condition on the estimated model parameters.

\paragraph{Predictive probabilities}\label{sec:prob}

Conditional and marginal predictions use the probabilities defined in Section~\ref{sec:blmm} \citep{SkrondalRabeHesketh2009}. The fitted conditional probability is obtained by substituting $\widehat{\bm{\beta}}$ into~\eqref{eq:pc} and evaluating the frailty at $\widehat{\gamma}_i$ and $\bar{\gamma}_i$. Evaluating it at $\gamma_i=1$ instead gives the prediction for a cluster with mean frailty. The fitted marginal probability $\widehat{\pi}_M(\bm{x}_{ij})$ is obtained from~\eqref{eq:pm} by substituting $\widehat{\bm{\beta}}$ and $\widehat{\theta}$.

An approximate $(1-\alpha)$ confidence interval for $\pi_M(\bm{x}_{ij})$ is obtained by applying the delta method on the logit scale and transforming back:
\[
\left[\frac{\exp(\xi_L)}{1+\exp(\xi_L)},
\frac{\exp(\xi_U)}{1+\exp(\xi_U)}\right],
\]
where $\xi_L$ and $\xi_U$ are the Wald limits for $\operatorname{logit}\{\pi_M(\bm{x}_{ij})\}$, defined in Appendix~\ref{app:ci}. This construction ensures that the interval will be contained in $(0,1)$.

At the fitted parameter values, an equal-tailed interval $\left[\widehat{\pi}_C(U_i;\bm{x}_{ij}),
\widehat{\pi}_C(L_i;\bm{x}_{ij})\right]
$ for the cluster-specific conditional probability follows from transforming the frailty interval in~\eqref{eq:quant}. Note that the order of the endpoints is reversed because $\pi_C(\gamma_i;\bm{x}_{ij})$ decreases in $\gamma_i$.

\section{Simulation study}
\label{sec:simulation-study}
To evaluate the performances of the proposed BLMM, we conducted a simulation study inspired by the longitudinal attendance data of women in the catchment region of the Oslo breast cancer screening program in Norway (see, e.g., \cite{BjornsonHolenEtAl2022}) analyzed in Section \ref{sec:CaseStudy}. The design, execution, analysis, and reporting of the simulation study were structured according to the ADEMP framework for simulation studies (\cite{morris2019simulation}).

\subsection{Design}
\paragraph{Aims.}\label{subsec:simulation-aims}
The primary aim of the simulation study was to assess the validity of the proposed BLMM both in terms of inference and predictions performance.

\paragraph{Data-generating mechanisms.}
\label{subsec:data-generating-mechanisms}
We considered a fully factorial set of simulation scenarios designed to evaluate how inferential and predictive performance varied across a range of conditions. Three factors were varied: the size of the training dataset,
$n_{\mathrm{train}} \in \{500,\allowbreak 1000,\allowbreak 5000,\allowbreak 10.000\}$, the number of measurement occasions per subject, $n_j \in \{3,6,10\}$, which was kept equal for all subjects within each scenario, and the frailty parameter, $\theta \in \{0.5,1,2,5\}$. This resulted in a total of $4 \times 3 \times 4 = 48$ simulation scenario. The size of the externally generated validation dataset was kept fixed at $n_{\mathrm{val}} = 5.000$ in all scenarios, so that variation in predictive performance could be attributed primarily to the size of the training dataset, the number of repeated occasions, and the degree of between-subject heterogeneity. Note that, under this experimental design, the size of the training and validation dataset coincides with the number of clusters available in each datasets. 
For each simulation repetition, both the training and validation datasets were generated anew, including the covariates. This choice was made to avoid conditioning the evaluation on a fixed set of covariate values and to obtain a conservative assessment of the methods under repeated sampling from the full data-generating mechanism.

For subject $i$ at occasion $j$, the conditional probability of the binary outcome was defined as $\pi_C(\gamma_i,\bm{x}_{ij}) = \left[1/\{1+\exp(-\bm{\beta}'\bm{x}_{ij})\}\right]^{\gamma_i}$, where $\bm{x}_{ij}$ denotes the vector of covariates and $\gamma_i$ is a subject-specific frailty term. The frailties were independently generated from a Gamma distribution, $\gamma_i \sim \Gamma(\theta,\theta)$, parameterized so that $\mathbb{E}(\gamma_i)=1$ and $\mathrm{Var}(\gamma_i)=1/\theta$. Conditional on $\gamma_i$ and $\bm{x}_{ij}$, the binary outcome was generated as $Y_{ij} \mid \gamma_i,\bm{x}_{ij} \sim \mathrm{Bernoulli}\left(\pi_C(\gamma_i,\bm{x}_{ij})\right)$.

Three subject-specific covariates were included in the data-generating mechanism. First, we generated a  binary covariate, $C_i \sim \mathrm{Bernoulli}(0.5)$, which was constant across measurement occasions for the same subject. Second, we generated a continuous baseline age, $A_{i0} \sim \mathcal{N}(\mu_{\mathrm{age}}, \sigma_{\mathrm{age}}^2)$, with $\mu_{\mathrm{age}} = 52$ and $\sigma_{\mathrm{age}}^2 = 2$. We  allowed age to increase deterministically over occasions according to $X_{ij}^{(\mathrm{age})} = A_{i0} + \Delta_{\mathrm{age}}(j-1)$, where $\Delta_{\mathrm{age}}$ was set equal to 2 and it denotes the fixed increase in age between consecutive occasions. Third, we generated a continuous time-varying covariate with an autoregressive within-subject correlation structure. For each subject $i$, the vector $\bm{Z}_i = (Z_{i1},\ldots,Z_{in_j})'$ was sampled from a multivariate normal distribution, $\bm{Z}_i \sim \mathcal{N}_{n_j}(\bm{0}, \bm{\Sigma})$, with covariance matrix $\Sigma_{ab} = \sigma_Z^2 \rho^{|a-b|}$, where $a,b=1,\ldots,n_j$, $\rho = 0.5$ and $\sigma_Z^2 = 1$. 
The regression coefficient vector $\bm{\beta} = (\beta_0,\beta_C,\beta_{\mathrm{age}},\beta_Z)'$ was fixed across all simulation scenarios. Values of betas where set to -1.00 -0.80  0.03 -0.60 respectively. The marginal prevalence of the outcome induced with these values for each simulation scenarios is reported in Appendix, Sec. \ref{app:simuresults}.

\paragraph{Estimands.}
\label{subsec:estimands-targets}
To evaluate the inferential performance of the BLMM, the estimand was defined as the regression coefficients vector $\bm{\beta}$ and the coefficient $\theta$. To evaluate the predictive performance of the BLMM, the target quantities were defined as the performance measures described below and reported in Table \ref{tab:perf}.

\paragraph{Methods compared.}
\label{subsec:methods-compared}
For the assessment of inferential performance, no competing methods were compared. 
For the assessment of predictive performance, we compared alternative ways of deriving individual frailties and conditional predicted probabilities from the fitted BLMM. In particular, we considered predictions based on the posterior mean frailty $\widehat{\gamma}_i$, the median frailty $\bar{\gamma_i}$ the corresponding predicted conditional probabilities $\widehat{\pi}_c(\widehat{\gamma}_i, \bm x_{ij} )$, $\widehat{\pi}_c(\bar{\gamma_i}, \bm x_{ij} )$. These approaches were compared in terms of the predictive performance measures reported in Table \ref{tab:perf}.
We also explore a rank-based calibration approach to derive individual frailties. This approach maps the fitted posterior means to population gamma quantiles:
\begin{equation*}\label{eq:rec}
 \widetilde{\gamma}_i
 =F_{\Gamma}^{-1}\left\{
 \frac{\operatorname{Rank}(\widehat{\gamma_i})-1/2}{n};
 \widehat\theta,\widehat\theta\right\}.
\end{equation*}
Then, the corresponding probability  $\widehat{\pi}_c(\widetilde{\gamma}_i, \bm x_{ij} )$ is evaluated.  

\paragraph{Performance measures}\label{subsec:performance-measures}

In evaluating the inferential performance of the BLMM, we considered the relative bias and the empirical coverage of the 95\% confidence intervals for the regression coefficient vector $\bm{\beta}$ and for the parameter $\theta$, estimated from the training dataset.

In evaluating the predictive performance of the BLMM, we considered several measures computed on the validation dataset. For the subject-specific frailties, we evaluated the root mean squared error, the mean absolute prediction error, and the empirical coverage and average width of the 95\% prediction intervals. Coverage was defined as
$n_{\mathrm{val}}^{-1}\sum_{i=1}^{n_{\mathrm{val}}}\mathbb{I}(L_i \leq \gamma_i \leq U_i)$. Note that, for the rank-based calibration approach, the intervals for the re-calibrated frailty are obtained by translating the original interval, i.e., $\left[L_i+\widetilde{\gamma}_i-\widehat{\gamma},
U_i+\widetilde{\gamma}_i-\widehat{\gamma}_i\right]$.
The interval's width thus remains unchanged.

For the predicted probabilities, we evaluated the mean absolute prediction error, computed both for the conditional predicted probabilities and for the marginal predicted probabilities. We also assessed the empirical coverage and average width of the corresponding 95\% prediction intervals.

Calibration and overall probability error were assessed using the Brier score, the observed-to-expected (O/E) ratio, and the integrated calibration index (ICI) (\cite{AustinSteyerberg2019}). Their empirical expressions are collected in Table~\ref{tab:perf}. For conditional predictions, the posterior mean, the reconciled predictor, or the posterior median frailty can be used in $\pi_C$, as described in Section~\ref{sec:frail}. In the table, the conditional expressions use the posterior mean.

\begin{table}[t]
\centering
\caption{Theoretical and empirical probability-based performance measures.}
\label{tab:perf}
\begin{minipage}{\textwidth}
\centering
\renewcommand{\arraystretch}{2.2}
\resizebox{\textwidth}{!}{%
\begin{tabular}{@{}lll@{}}
\hline
Measure & Theoretical expression & Empirical expression \\
\hline
Brier score &
$\displaystyle \mathbb{E}_{Y,\bm X}\!\left[(Y-\pi(\bm X))^2\right]$ &
$\displaystyle \frac{1}{\sum_i n_i}\sum_{i=1}^{n}\sum_{j=1}^{n_i}\left(y_{ij}-\widehat{\pi}(\bm x_{ij})\right)^2$ \\[0.6em]
O/E ratio &
$\displaystyle \frac{P(Y=1)}{\mathbb{E}[\pi(\bm X)]}$ &
$\displaystyle \frac{\sum_{i=1}^{n}\sum_{j=1}^{n_i}y_{ij}}{\sum_{i=1}^{n}\sum_{j=1}^{n_i}\widehat{\pi}(\bm x_{ij})}$ \\[0.6em]
ICI$^{\dagger}$ &
$\displaystyle \int_0^1 \left|\pi(\bm X)-\mathbb{E}[Y\mid\pi(\bm X)]\right|\phi(\pi(\bm X))\,d\pi(\bm X)$ &
$\displaystyle \frac{1}{\sum_i n_i}\sum_{i=1}^{n}\sum_{j=1}^{n_i}\left|\widehat{\pi}(\bm x_{ij})-\widehat{\mathbb{E}}[Y\mid\widehat{\pi}(\bm x_{ij})]\right|$ \\[0.6em]
c-index &
$P\!\left(\pi(\bm X_a)>\pi(\bm X_b)\mid Y_a=1,Y_b=0\right)$ &
$\displaystyle \frac{1}{|\mathcal P_{10}|}\sum_{(a,b)\in\mathcal P_{10}}\mathbb{I}\!\left\{\widehat{\pi}(\bm x_a)>\widehat{\pi}(\bm x_b)\right\}$ \\
\hline
\end{tabular}%
}
\par\smallskip
\raggedright\footnotesize\emph{Note.}
Replacing $\pi(\bm x)$ with $\pi_M(\bm x)$ or
$\pi_C(\gamma_i;\bm x)$ yields the marginal or conditional
version of each measure, respectively. For conditional empirical
measures, the frailty can be replaced by the posterior mean
$\widehat{\gamma}_i$, the reconciled predictor
$\widetilde{\gamma}_i$, or the posterior median $\bar{\gamma}_i$.
For the c-index, $\mathcal P_{10}$ is the set of ordered
pairs $(a,b)$, with $y_a=1$ and $y_b=0$.
Choosing overall, between-cluster, or within-cluster pairs
changes the corresponding index in addition to the choice of
predicted probability. See Appendices~\ref{app:cb}
and~\ref{app:cw} for the BLMM comparisons.

\par\smallskip
\textsuperscript{$\dagger$}\,
For the ICI, $\widehat{\mathbb{E}}[Y\mid\widehat{\pi}(\bm x_{ij})]$
denotes an estimated calibration function, obtained for example
using a locally weighted smoother or a logistic regression model
with restricted cubic splines. See \citet{AustinSteyerberg2019} for more details.
\end{minipage}
\end{table}

Discrimination was evaluated using the c-index. For clustered binary data, a pair can be drawn either from the same cluster (within-cluster) or from different clusters (between-cluster); the overall index includes both. Using conditional or marginal predicted probabilities therefore yields six indices: $c_{OC}$, $c_{BC}$, and $c_{WC}$, and $c_{OM}$, $c_{BM}$, and $c_{WM}$, respectively \citep{VanOirbeekLesaffre2012}. Unlike the measures in Table~\ref{tab:perf}, their calculation depends additionally on the comparison of pairs. For the BLMM, marginal between-cluster comparisons can be made using $\bm\beta'\bm x$, and the within-cluster indices satisfy $c_{WC}=c_{WM}$. Conditional between-cluster comparisons must retain the conditional probabilities. The corresponding arguments are in Appendices~\ref{app:cb} and~\ref{app:cw}. The empirical forms of the indices follow \citet{VanOirbeekLesaffre2012}.

We conducted 1000 simulations under each scenario, and empirical estimates of all performance measures were reported.
\paragraph{Software.}
We have implemented all algorithms in R (\cite{R2026}) Version 4.5.3, with bits of C++ code to increase speed (\cite{EddelbuettelFrancois2011}) and the assistance of CODEX (\cite{Codex2026}) for the more challenging tasks such as the intensive parallel, multi-core simulation study performed on a High Performance Computer cluster provided by our institution.

\subsection{Results}
\label{subsec:simulation-interpretation}
\paragraph{Inferential performance.}
Overall, the relative bias of the regression coefficients decreased as the training sample size increased (Figure~\ref{fig:beta_relative_bias}). The largest relative bias was observed for the intercept ($\beta_1$), especially in the smallest sample-size scenarios, whereas the relative bias for $\beta_2$ and $\beta_4$ remained comparatively small across scenarios. For $\beta_3$, the bias was more visible than for $\beta_2$ and $\beta_4$, but still decreased with increasing sample size. Coverage was generally close to the nominal 95\% level across the simulation scenarios (Figure~\ref{fig:beta_coverage}).
The estimates of $\theta$ were generally centered around the true parameter value (Figure~\ref{fig:theta_boxplot}, Appendix). The dispersion of $\widehat{\theta}$ was larger in smaller training samples and became narrower for larger $n_{\mathrm{train}}$ and larger numbers of repeated measurements per subject. The empirical coverage was close to, and in several scenarios slightly above, the nominal 95\% level (Figure~\ref{fig:theta_coverage}, Appendix).

\begin{figure}[!htbp]
    \centering
    \makebox[\textwidth][c]{%
    \includegraphics[width=.9\textwidth]{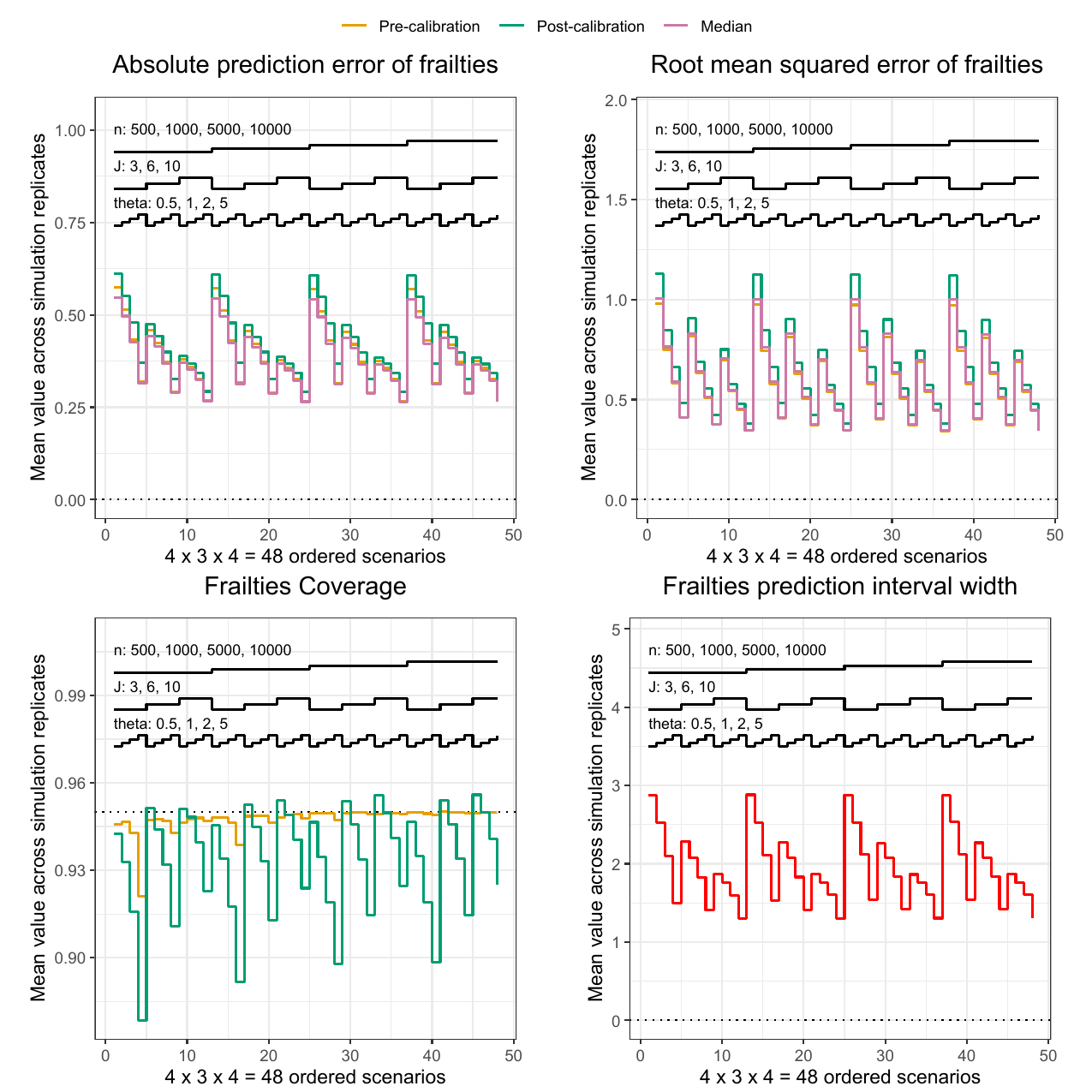}
    }
    \caption{Predictive performance for the subject-specific frailties: mean absolute prediction error, root mean squared error, empirical coverage, and prediction interval width. We denoted the posterior mean frailty and the rank-based calibration frailty as pre-calibration and  as post-calibration respectively. Prediction interval widths are reported only for the pre-calibration frailty intervals, since the post-calibration procedure only induces a location shift and therefore leaves the interval width unchanged. Note that $n$ corresponds to the number of cluster of the training dataset, i.e. the number of subjects.}
    \label{fig:frailty_prediction}
\end{figure}

\paragraph{Prediction of subject-specific frailties.} The predictive performance for the subject-specific frailties is summarized in Figure~\ref{fig:frailty_prediction}. The mean absolute prediction error and the root mean squared error decreased with the true value of $\theta$, reflecting the increasing difficulty of predicting the individual frailties when the frailty distribution induces greater variability across subjects and decreases as the number of occasions increase. Across scenarios, the posterior mean (pre-calibration), the rank-based calibration (post-calibration) and the posterior median showed broadly similar patterns, with better performance for the posterior mean and median. The coverage of the frailty prediction intervals was generally close to the nominal level for the pre-calibration frailties while for the post-calibration was more sensible to the variability of of the frailties and the number of occasions. The average width of the prediction intervals increased in the scenarios with larger frailty variability and lower number of occasions.

\paragraph{Conditional predicted probabilities.}
The predictive performance of the conditional predicted probabilities is reported in Figure~\ref{fig:conditional_probability_prediction}, Appendix. The mean absolute prediction error was generally small and decreased as the number of occasions increased. The pre-calibration, post-calibration, and median-based predictors showed similar trends, with slightly better performance for the posterior median. The coverage of the conditional predicted-probability intervals reflected the coverage of the corresponding frailty intervals, since, for fixed covariates, the individual conditional predicted probabilities are monotone transformations of the subject-specific frailties.

\begin{figure}[!htbp]
    \centering
    \makebox[\textwidth][c]{%
    \includegraphics[width=0.9\textwidth]{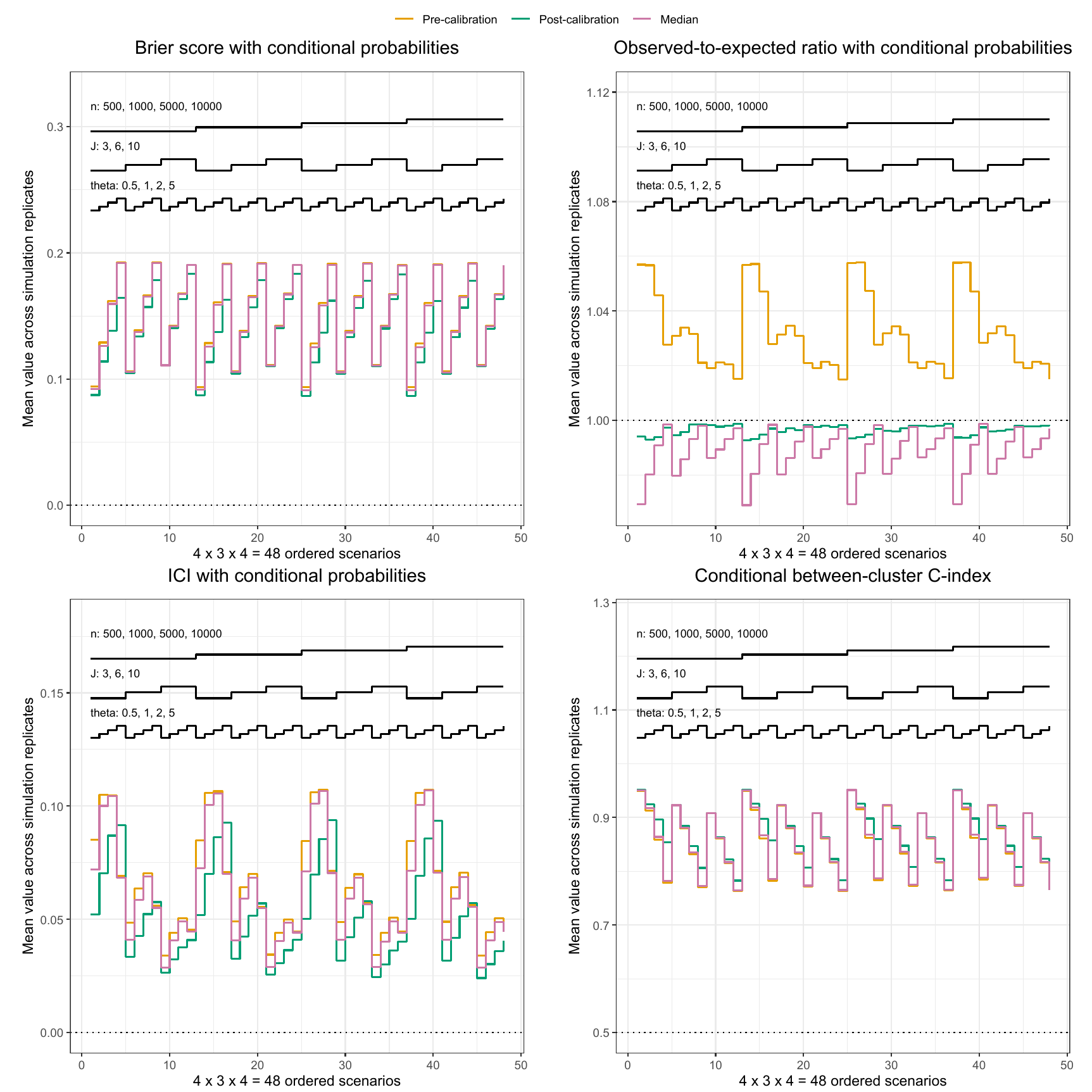}
   }

    \caption{Calibration and discrimination measures for the conditional predicted probabilities: Brier score, observed-to-expected ratio, ICI and conditional between-cluster C-index. We denoted the posterior mean frailty and the rank-based calibration frailty as pre-calibration and  as post-calibration respectively. Note that $n$ corresponds to the number of cluster of the training dataset, i.e. the number of subjects.}
    \label{fig:conditional_calibration_discrimination}
\end{figure}

Discrimination and calibration measures based on the conditional probabilities are shown in Figure~\ref{fig:conditional_calibration_discrimination}. Discrimination increased mainly with increasing variability of the individual frailties. The conditional overall C-index was essentially equivalent to the conditional between-cluster C-index. A justification is provided in Appendix, Section~\ref{overall_between_cindex}.

Regarding calibration, both the O/E ratio and the ICI were mainly influenced by the number of repeated occasions per subject, with better calibration as the number of occasions increased. Miscalibration was more evident in scenarios with stronger frailty heterogeneity and fewer occasions. The pre-calibration approach tended to underestimate the outcome probability, whereas the post-calibration and median-based approaches showed slight overestimation. Overall, post-calibration provided the best calibration, with O/E ratios closer to one and the lowest ICI values across most scenarios. The Brier score improved as frailty heterogeneity increased. Differences between the three conditional prediction approaches were small, although post-calibration tended to perform slightly better.

\paragraph{Marginal predicted probabilities.}
The predictive performance of the marginal predicted probabilities is shown in Figure~\ref{fig:marginal_probability_prediction}, Appendix. The mean absolute prediction error was small across all scenarios. The empirical coverage of the marginal predicted-probability intervals was close to the nominal 95\% level and was slightly conservative in several settings. Their average width was substantially smaller than that of the conditional predicted-probability intervals and decrease with increasing training sample size.

Discrimination and calibration measures based on marginal predicted probabilities are reported in Figure~\ref{fig:marginal_calibration_discrimination}, Appendix. The marginal Brier score was stable across scenarios and improved as the number of occasions increased. The marginal O/E ratio remained close to one, indicating good overall calibration. Similarly, the marginal ICI was low across scenarios and decreased with increasing training sample size. The marginal overall and between-cluster c-index were essentially identical for the same reason discussed above for the conditional c-index; see Appendix, Section~\ref{overall_between_cindex}. Marginal between-cluster discrimination was moderate and improved when the variability between individual frailties decreased and the number of occasions increased. The within-cluster C-index showed a similar pattern, increasing with a larger number of occasions but decreasing with lower frailty heterogeneity.

Note that our simulations do not include a comparison with the widely used GLMM model for binary outcomes. Indeed, it would not have been fair to perform such a comparison between the two models with data being generated from either one fo them. Nevertheless, such a comparison is clearly of interest, and in the next section we study the formal relationship between the two model specifications. Additional elements will be found in the analysis of the screening data in Section \ref{sec:CaseStudy}.

\FloatBarrier

\section{Comparison with the normal random-intercept GLMM}\label{sec:comp}\label{sec:comp_BLMM_GLMM}

The logistic random-intercept GLMM assumes conditional independence of the binary outcomes and
\begin{equation*}\label{eq:glmm}
 \pi_C(\alpha_i;\bm x_{ij})
 =\operatorname{expit}(\bm x_{ij}^{\mathsf T}\bm b+\alpha_i),
 \qquad\alpha_i\sim N(0,\sigma^2).
\end{equation*}
We distinguish its coefficients $\bm b$ from the BLMM coefficients $\bm\beta$. A coefficient $b_k$ is a conditional log odds ratio at every fixed random intercept, whereas $\beta_k$ has that interpretation at $\gamma=1$. Larger $\alpha_i$ increases success probabilities; larger $\gamma_i$ decreases them. It is clear from the two model specifications that in both cases, the elements of the $\bm \beta$ parameter represent the effect of a unit change in the corresponding covariate on the log-odds $\log \left[ \pi_C(1;\bm x)/(1-\pi_C(1;\bm x)) \right]$ of the outcome $Y=1$, for an observation / subject with average frailty or random effect, keeping all other covariates fixed. Clearly, given the different impact and distribution of the random effect vs. frailty random variables, one does {\em not} expect the two models to be equivalent. We studied which distributional form of $\Gamma$ in the BLMM would produce a model which would be identical to the GLMM model with $N(0,\sigma^2)$ random effect, and how far the Gamma$(\theta ,\theta$ specification is from that distribution. To avoid confusion, in this section {\em only} we refer to the random effect in the GLMM model by $\alpha$ and the corresponding terminology $\pi_C(\alpha_i;{\bm x_{ij}})$, while maintaining the notation $\gamma$ for the frailty term in the BLMM model. This will also help identify the two models more easily.

Let us call $r({\bm x})= {\bm \beta}'{\bm x}$. Setting the two conditional probabilities equal to one another yields
\begin{eqnarray*}
\pi_C(\alpha;{\bm x}) = \pi_C(\gamma;{\bm x}) & \Leftrightarrow & \frac{{\rm e}^{r(\bm x)+\alpha}}{1+{\rm e}^{r(\bm x)+\alpha}} = \left( \frac{{\rm e}^{r(\bm x)}}{1+{\rm e}^{r(\bm x)}}  \right)^\gamma  \\
& \Leftrightarrow & r(\bm x)+\alpha - \log \left( 1+ {\rm e}^{r(\bm x)+\alpha} \right) = \gamma \left[  r(\bm x) - \log \left( 1 + {\rm e}^{r(\bm x)}   \right) \right]  \\
& \Leftrightarrow & \gamma(\alpha) = \frac{r(\bm x) + \alpha - \log\left( 1+{\rm e}^{r(\bm x) + \alpha}  \right)}{r(\bm x) - \log \left( 1 + {\rm e}^{r(\bm x)} \right)}.
\end{eqnarray*}
Since $\alpha \sim N(0, \sigma^2)$, and equivalent BLMM would be one that maintained the form of $\pi_C(\gamma; {\bm x})$, but for which the frailty random variable had the distribution of the random variable $g_{r(\bm x)}(\alpha)$. Importantly, that distribution would need to depend on the fixed term $r(\bm x)$.
An equivalent form for $\gamma(\alpha)$ is
\begin{eqnarray}\label{relationship}
\gamma(\alpha) = \frac{\log \left( \frac{{\rm e}^{r({\bm x} + \alpha)}}{1+{\rm e}^{r({\bm x} + \alpha)}} \right)}{\log \left( \frac{{\rm e}^{r({\bm x})}}{1+{\rm e}^{r({\bm x}}} \right)} = \frac{\log \left( \pi_C( \alpha, {\bm x}) \right)}{\log \left( \pi_C(0, {\bm x}) \right)}, 
\end{eqnarray}
which highlights the connection between the conditional probabilities as defined in the GLMM model and the $\gamma$ frailty term in the BLMM model.
\newline
Given this relationship, one may ask the question of what the distribution of $\Gamma$ should be in the BLMM specification to produce a model that would be identical to the Binary GLMM with random intercept. We call such distribution the distribution of the $\Gamma^*$ random variable to distinguish it from the Gamma$(\theta, \theta)$ distribution of $\Gamma$. The transformation (\ref{relationship}) is strictly monotonically decreasing in $\alpha$, so that application of the one dimensional transformation formula yields the distribution of $\Gamma^*$ as
\begin{equation}\label{eq:exactfgamma}
  f_{\Gamma^*}(\gamma)
=
\frac{-\log p}{(1-p^\gamma)\,\sigma\sqrt{2\pi}}
\exp\!\left[
-\frac{1}{2\sigma^2}
\left(
\log\!\left(\frac{p^\gamma}{1-p^\gamma}\right)-r
\right)^2
\right],
\qquad \gamma>0.  
\end{equation}

where $p=\pi_C(0; \bm x)=\frac{{\rm e}^{r(\bm x)}}{1+{\rm e}^{r(\bm x)}}$ from the GLMM model, and where we recognize $p^\gamma$ as $\pi_C(\gamma; \bm x)$ as defined in the BLMM model. Appendix~\ref{app:glmm} derives its density.
We have performed a numerical experiment to compare, for different values of $\sigma^2$ and $\pi_C(0,\bm x)$, this distribution to the Gamma density assumed by the BLMM model.

For each combination $(\sigma^2, \pi_C(0,\bm x))$ over a grid, we have generated a sample of 10,000 random effects $\alpha$ from the N$(0,\sigma^2)$, and computed the corresponding frailty values $\gamma$ as computed from (\ref{relationship}).
To ensure comparability of the two distributions, we have estimated $\theta^*$ by the inverse of the variance of the $\gamma$ values, and compared the density (\ref{eq:exactfgamma}) to the Gamma$(\theta^*, \theta^*)$ distribution. For each combination $(\sigma^2, r(\bm x))$ we have computed: (i) the Kolmogorov-Smirnov statistic comparing the $\gamma$ values to the Gamma$(\theta^*, \theta^*)$ distribution; (ii) the empirical standard deviation of the quantities $\pi_C(\alpha;r(\bm x))$, which quantifies the effect of the heterogeneity induced by distribution of the random effect $\alpha$ on the probabilities of $Y=1$ in the GLMM model. We use of the K-S statistic to quantify the distance between the two distributions rather than to test for differences between the moment-fitted theoretical distribution and the empirical data, as the test will {\em correctly} reject in almost all cases.

\begin{figure}[!htbp]
\centering

\begin{subfigure}[t]{0.48\textwidth}
\centering
\includegraphics[width=\linewidth]{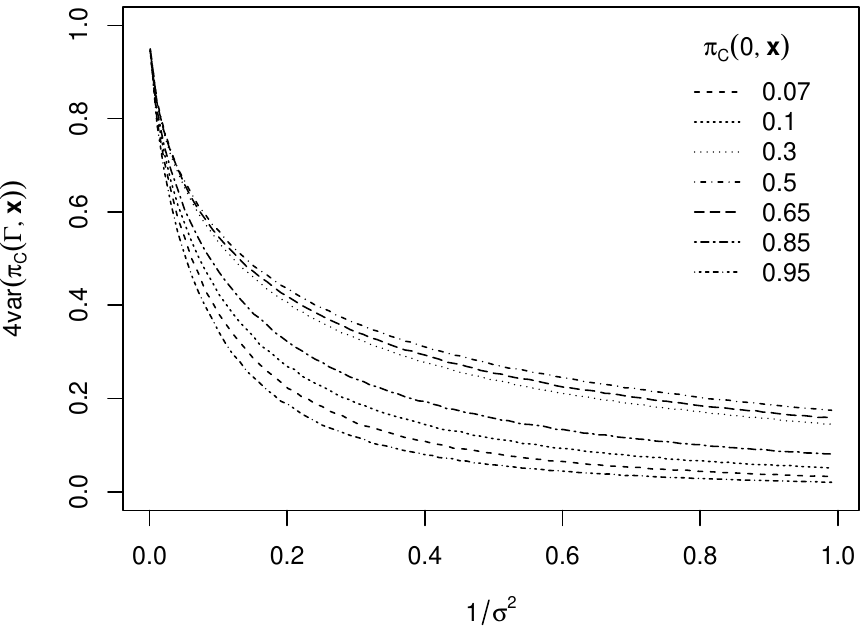}
\caption{GLMM: $4\operatorname{Var}\{\pi_C(\alpha;\bm x)\}$ versus $1/\sigma^2$.}
\label{fig:varpi-sigma2}
\end{subfigure}\hfill
\begin{subfigure}[t]{0.48\textwidth}
\centering
\includegraphics[width=\linewidth]{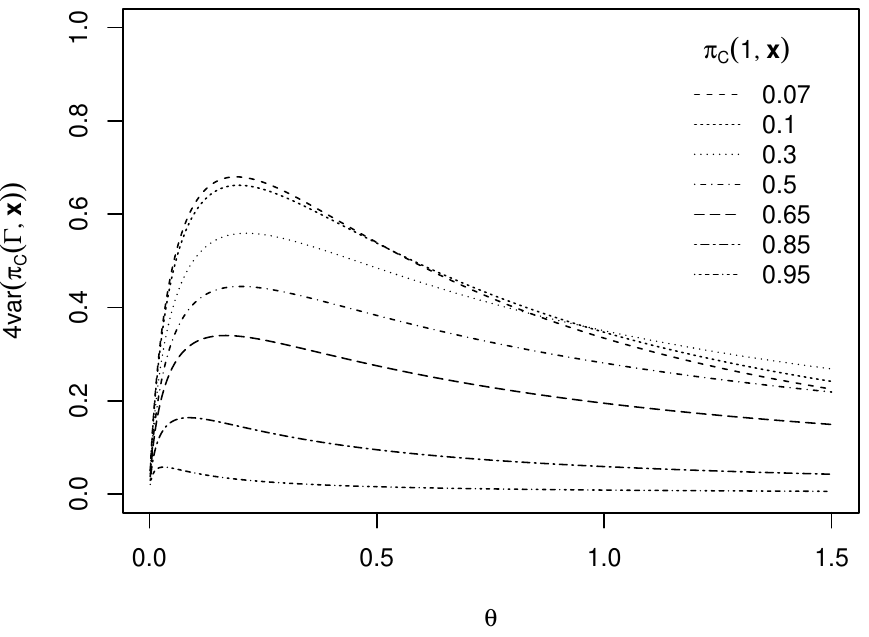}
\caption{BLMM: $4\operatorname{Var}\{\pi_C(\gamma;\bm x)\}$ versus $\theta$.}
\label{fig:varpi-theta}
\end{subfigure}

\par\medskip

\begin{subfigure}[t]{0.48\textwidth}
\centering
\includegraphics[width=\linewidth]{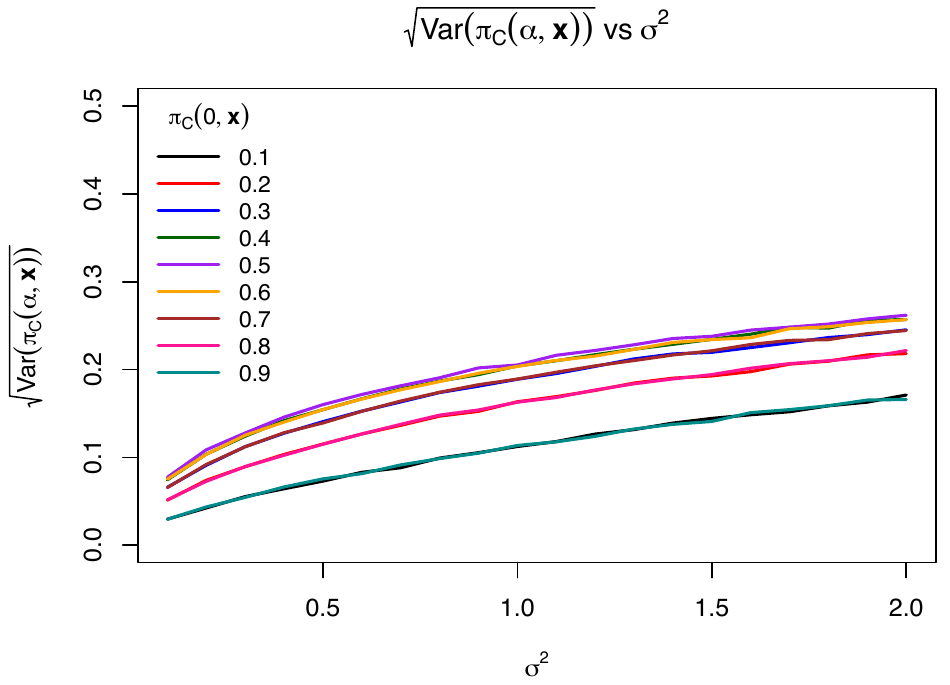}
\caption{Standard deviation of GLMM conditional probabilities.}
\label{fig:cmp-a}\label{fig:varpi-sigma-2}
\end{subfigure}\hfill
\begin{subfigure}[t]{0.48\textwidth}
\centering
\includegraphics[width=\linewidth]{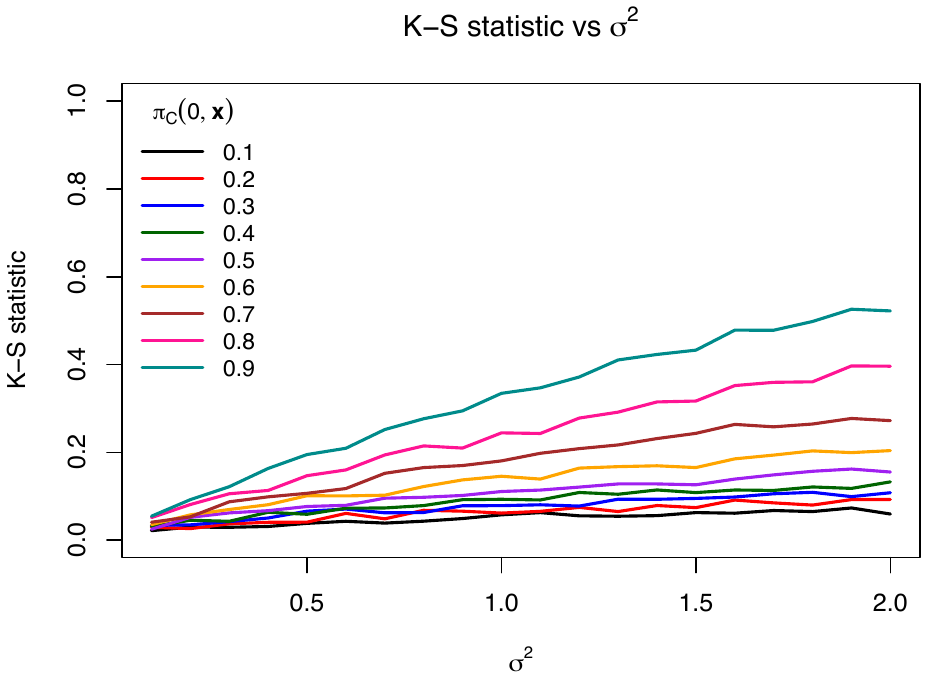}
\caption{K--S distance from the variance-matched gamma distribution.}
\label{fig:cmp-b}\label{fig:varpi-sigma-1}
\end{subfigure}

\caption{Variability of conditional probabilities and numerical comparison of the BLMM and random-intercept GLMM. In (a) and (b), the horizontal axes are inverse random-effect variances; in (c) and (d), the horizontal axis is $\sigma^2$. Curves in (a), (c), and (d) correspond to different values of $\pi_C(0;\bm x)$, whereas curves in (b) correspond to different values of $\pi_C(1;\bm x)$. The K--S distance in (d) compares the transformed GLMM random effects with a $\operatorname{Gamma}(\theta^*,\theta^*)$ distribution, where $\theta^*$ is the inverse empirical frailty variance.}
\label{fig:cmp}
\end{figure}

Table \ref{tabequiv} and Figure \ref{fig:varpi-sigma-1} and summarize the results of the experiment. We observe that the difference between the assumed Gamma$(\theta, \theta)$ distribution to the distribution of $\gamma$ that would make the BLMM model equivalent to the GLMM model increases as the heterogeneity increases. In addition, the two distributions differ the most for larger values of $pi_C(0,\bm x)$, and in particular for values above 0.5. This is confirmed by the examination of the Q-Q plots (not shown), which suggest that for the largest values of $\pi_C(0,\bm x)$ the tail behavior of the two distributions is quite different.

Figure \ref{fig:varpi-sigma-2}  shows the empirical relationship that exists between $\sigma^2$ and the standard deviation of the conditional probabilities $\pi_C(\alpha;\bm x)$. As the largest theoretical value for such standard deviation is 0.5, a value of 0.25-0.30 indicates an extremely large effect of the random effect on the probability $\pi_C(0,\bm x)$ -- especially when such probability is away from 0.5. Hence we expect a fit of the GLMM model to reasonably return an estimate of $\sigma$ of 1.0-1.5, and we have focused on the range $(0,2]$.

We now investigate and compare the effect of varying the levels of heterogeneity --as quantified by the value of the frailty/random effect variance $\sigma^2$-- on the variance var$(\pi_C(\gamma; {\bm x}))$ induced by the BLMM and the GLMM. For a fixed covariate vector $\bm x$, heterogeneity in the random effect induces variability in the conditional probability. Since $\pi_C(\gamma, \bm x)$ takes values in $[0,1]$, its variance is at most $1/4$. Hence, a {\em relative} version (taking values in $[0,1]$) of the variance of the conditional probability in both models is obtained by multiplication by 4. 

For the random-intercept GLMM, the model assumes $\alpha\sim N(0,\sigma^2)$. Letting $a={\rm e}^{{\bm \beta}'{\bm x}} >0$, we can re-write $\pi_C(\gamma; {\bm x}) = \frac{Y}{1+Y}$ with $Y \sim {\rm Lognormal}(\log a, \sigma^2)$. While the density function of this random variable can be obtained easily, its moments cannot. However, simple generation process from the model provides an approximate Montecarlo description of the relationship of interest, for different levels of the average random effect probability $\pi_C(0,{\bm x})$.  Panel~\ref{fig:varpi-sigma2} plots it against $1/\sigma^2$, allowing comparison with the BLMM.

For the BLMM, the variance is available in closed form:
\[
\operatorname{Var}\{\pi_C(\gamma;\bm x)\}
=
\left(\frac{\theta}{\theta-2\log\pi_C(1;\bm x)}\right)^\theta
-
\left(\frac{\theta}{\theta-\log\pi_C(1;\bm x)}\right)^{2\theta}.
\]
It tends to zero as $\theta\to\infty$ and as $\theta\to0$. Across different baseline probabilities shown in panel~\ref{fig:varpi-theta}, the variance decreases to zero for small values of $\theta$ (less than one), with the peak reached at the argmax, which can be found numerically. Also numerically, the argmax never exceeds $0.25$ (more precisely, a value near $0.214$) and it is often much smaller. Hence for values of $\theta>0.25$, ${\rm var} \left[ \pi_C(\gamma; \bm x) \right] \rightarrow 0$ is safely decreasing in $\theta$. Notably, the value $\theta = 0.25$ represents a situation of rather large heterogeneity: the $5th$ and $95th$ percentiles of the corresponding frailty distribution are equal to $.000017$ and $4.840465$ respectively, and they would produce very extreme modifications of the conditional probabilities $\pi_C(1;\bm x)$ into the corresponding $\pi_C(\gamma;\bm x)$: for example, $\pi_C(1;\bm x) \in \{0.1,0.5,0.9 \}$ would be transformed into $\{ 1.44E-05, 0.035, 0.600 \}$ and $\{ 1.00, 1.00, 1.00 \}$, respectively for the 0.95 and the 0.05 percentile of $\gamma$. Lastly, the value of $\theta$ for a given analysis is available after maximum likelihood estimation.

As a final remark, let us point out the modifying effect of the frailty term in the BLMM on the probability of success $\pi_C(1; {\bm x})$ of the ``baseline'' (i.e. average frailty) individual. Recall that the model assumes that $\pi_C(\gamma; {\bm x})= (\pi_C(1; {\bm x}))^\gamma$. Table \ref{tabmodif} shows, for the set of values of $\theta$ discussed in the simulation study in Section \ref{sec:simulation-study}, the modifying effect on the baseline probability of the value of $\gamma$ equal to the first, second, and third quartile of the frailty distribution.

\FloatBarrier
\section{Analysis of the adherence to the Oslo Breast Cancer Screening program}\label{sec:CaseStudy}

In this section we illustrate the use of the BLMM to longitudinal adherence (or ``attendance'') data of women in the catchment region of the Oslo breast cancer screening program in Norway (\cite{BjornsonHolenEtAl2022}). For comparison, we also fit the binary GLMM model to the same data.

All analyses were performed via remote, secure connection to the TSD system provided by the University of Oslo. Access and analysis of the data were allowed with a joint controller agreement between our University and the Norwegian Institute of Public Health (December 1, 2023), with ethical approval by both institutions (Ethical approval on September 8, 2020 by the Norwegian Regional Committee for Medical Research Ethics (REC) 150166; Bocconi Ethics Committee application FA000203.01, approved on October 24, 2022). 

Specifically, we have analyzed retrospective data collected as part of BreastScreen Norway \citep{ BreastScreen2025}. The program started in four counties in 1996 and became nationwide in 2005, after a staggered implementation. BreastScreen Norway targets about 650000 women aged 50-69, who are invited to screening every two year. The average attendance rate has been 75.9\% during the first 25 years of the programs existence \citep{ BjornsonHolenEtAl2022}. Information about all women who have attended Breast Screen Norway at least once during the study period, from January 1st, 2006 to December 31, 2019 has been initially extracted. Sociodemographic information from women participating in the program was collected through a questionnaire between August 1, 2006 and December 31, 2015. The questionnaire consisted of two parts: one questionnaire collected baseline information upon entry into the program and the other collected updates every two years \citep{ TsurudaSagstadEtAl2018}. Here we focus on the baseline information only. For additional information on the screening program we refer to \citep{ BreastScreen2025}, and specifically on the questionnaires to \citep{ EpiQuest2025}.

\subsection{The Oslo breast cancer screening adherence dataset}
The data consist of the response to the invitations sent to eligible participant to the breast cancer screening program of the Oslo region.

We have estimated the date of birth to June 15 for all subjects (only the birth year was available to us due to privacy concerns). The invitations refer to 992482 potential participants. The exact dates of the mailing of the invitation letters were not available, but only the month and year. The fixed date of the 15th of the month was provided to us for each invitation letter.
\leavevmode\newline
We have used our estimated birthdates to compute the age at the mailing of the first invitation letter. We have excluded 3551 records with missing invitation ID and missing associated date of invitation and examination.
\leavevmode\newline
\begin{table}[ht]
\centering
\caption{Selection into the final dataset}
\begin{tabular}{lcc}
\hline
Sequential selection & Subjects & Records \\
\hline
Initial & 992482 & 5771156 \\
First invitation at estimated age in $[50,51)$ & 280027 & 1578606 \\
Attended first invitation and has BL questionnaire & 50433 & 313250 \\
No post-death or post-emigration invitations & 50433 & 313249 \\
No post-diagnosis invitations & 50432 & 311245 \\
No invitations after Positive or Symptoms result & 50432 & 282737 \\
No duplicate records in subset & 50432 & 282703 \\
At least one post-baseline invitation & 46397 & 278668 \\
Born between 1955 and 1965 (included) & 46396 & 278659 \\
Remove all first invitations & 46396 & 232263 \\
\hline
\end{tabular} \label{tab:SequentialSelection}
\end{table}
We have selected the 280027 women for whom the estimated age at the mailing of the first invitation letter in the interval $[50, 51)$ years, and who have attended the first invitation {\em and} have filled out the baseline questionnaire (defined as not having in the database at least one non missing variable from the baseline questionnaire). Note that this choice is such that the results of any subsequent analysis will refer to such subjects, and any conclusions on the determinants of adherence can therefore be interpreted as being determinants of {\em continued} adherence after the first (successful) invitation.
\leavevmode\newline
To ensure that all women provided some information to the model, we removed 223 women who had only been invited (and attended) their first invitation.
\leavevmode\newline
There were a total of 313027 invitations for 50210 women.
\leavevmode\newline
Four subjects (PIDs: 169975, 514981, 629210, 988342)\footnote{Death dates were as follows. 169975: 2021-12-15; 514981: 2019-10-15; 629210: 2019-07-15; 988342: 2017-07-15.} had died with the status being shown on two different invitation records. All invitations for these subjects had mailing dates prior to the recorded death date).
\leavevmode\newline
An additional 91 subjects were recorded (once) as having died, and two as having emigrated (PIDs: 195078, 912788). Among these, only one subject (842552) was recorded as having an invitation post-death or post-emigration: the subject had died with status data 2017-11-15 but had one invitation on 2018-01-15, which we removed.
\leavevmode\newline
Then, we checked for any post-diagnosis invitations. There were 1735 recorded diagnoses across 1779 invitations, with 44 subjects having two invitations. If the two dates of diagnosis differed, we kept the earlier of the two dates. There were 62 invitations post-diagnosis recorded for 24 subjects, and we removed all such invitation records. After these adjustments, four subjects were left with only one (baseline) invitation, and they were therefore removed.
\begin{table}[tbp]
\centering
\small
\setlength{\tabcolsep}{5pt}
\renewcommand{\arraystretch}{1.15}

\begin{threeparttable}
\caption{Definitions and coding of study covariates.}
\label{tab:covariates}

\begin{tabularx}{\linewidth}{@{}l
    >{\raggedright\arraybackslash}p{0.40\linewidth}
    >{\raggedright\arraybackslash}X@{}}
\toprule
\textbf{Variable}
& \textbf{Description and source}
& \textbf{Definition / coding} \\
\midrule

\texttt{birthyearctr}
& Year of birth
  \sourcevar{fodtar}
& Mean-centered to allow for a possible birth-cohort effect. \\
\addlinespace[0.5em]

\texttt{seq\_inv}
& Invitation sequence
& Sequential invitation number; 1 = baseline. \\
\addlinespace[0.5em]

\texttt{edu}
& Education
  \sourcevar{QF2\_Utdanning}
& Low (1, 2, 3); Medium (4); High (5); NA. \\
\addlinespace[0.5em]

\texttt{men\_age}
& Age at first menstruation
  \sourcevar{QF8\_Mentruasjon\_alder}
& Early ($\leq 11$ years); Regular ($>11$ years); NA. \\
\addlinespace[0.5em]

\texttt{age\_first\_preg}
& Age at first pregnancy
  \sourcevar{QF\_25Alder\_forste\_fodsel}
& Late ($\geq 35$ years); Regular ($<35$ years); NA. \\
\addlinespace[0.5em]

\texttt{fam1}
& Family history 1: mother, sister, or daughter
  \sourcevar{QF\_30Brystkreft\_andre1}
& Yes (1); No (2); NA (missing or code 3). \\
\addlinespace[0.5em]

\texttt{fam2}
& Family history 2: grandmother, mother's sister,
  father's mother, or father's sister
  \sourcevar{QF\_32Brystkreft\_andre2}
& Yes (1); No (2); NA (missing or code 3). \\
\addlinespace[0.5em]

\texttt{ex4049}
& Exercise at ages 40--49
  \sourcevar{QF50\_Trening\_40\_49}
& None (1); Low (2, 3); High (4, 5);
  NA (missing or code 6). \\
\addlinespace[0.5em]

\texttt{alc4049}
& Alcohol use at ages 40--49
  \sourcevar{QF\_Drakk\_ikke\_40\_49}
& Yes (1); No (2); NA (missing or code 3). \\
\addlinespace[0.5em]

\texttt{smoke}
& Smoking
  \sourcevar{QF41\_Roykevaner\_periode}
& Yes (1); No (2); NA (missing or code 3). \\

\bottomrule
\end{tabularx}

\begin{tablenotes}[flushleft]
\footnotesize
\item[] \textit{Notes:} Source-variable names appear below the
descriptions. Numbers in parentheses denote original response codes
unless an age threshold is given. NA denotes the missing-data
category, including additional response codes where specified.
\end{tablenotes}
\end{threeparttable}
\end{table}
The final database thus contains data on a total of 232263 invitations for 46396 women, for an average of 6.01 invitations per woman. 
Overall, there were a total of 203,408 (87.6\%) attended invitations and 28,855 (12.4\%) invitations that were {\em not} attended. Recall that the - always attended - first invitations are {\em not} included in the analyses.

As we are interested in the determinants of adherence, we removed 3945 invitations for 853 subjects since they were sent out after a previous "P" ("Positive mammogram") or "K" ("Symptoms") result. Checking again for a minimum of one post-baseline invitation after these adjustments was such that 503 more subjects had to be removed. Table \ref{tab:SequentialSelection} summarizes the sequential selection process that we have followed to prepare the final dataset for the analyses.

The covariates that we have considered for the analyses are shown in Table \ref{tab:covariates}.

Note that the last three variables (again, measured at baseline) involve lifestyle behaviors that are likely to be related to health awareness. We did not include as a covariate having had no children, as only 38 subjects satisfied that criterion.
\leavevmode
Also, we have not included any time-varying covariates as obtained from the invitation-specific questionnaire, since those are only available if the woman attended the appointment.
Table \ref{tab:descriptive_statistics} contains a summary of the variables in the final dataset for analysis.

\begin{table}[htbp]
\centering
\caption{Descriptive statistics.}
\label{tab:descriptive_statistics}
\small
\setlength{\tabcolsep}{1.8pt}
\renewcommand{\arraystretch}{1.08}
\begin{tabular}{@{}lr@{\hspace{1.5em}}lr@{\hspace{1.5em}}lr@{}}
\toprule
\textbf{Variable} & \textbf{$n=46{,}396$}
& \textbf{Variable} & \textbf{$n=46{,}396$}
& \textbf{Variable} & \textbf{$n=46{,}396$} \\
\midrule

\multicolumn{2}{@{}l}{\textit{Invitations}}
& \multicolumn{2}{l}{\textit{Education}}
& \multicolumn{2}{l}{\textit{Family history: second degree}} \\

\quad Mean (SD) & 6.01 (1.57)
& \quad Low     & 27,211 (58.6\%)
& \quad Yes     & 7,115 (15.3\%) \\

\quad 2  & 1,471 (3.2\%)
& \quad Medium  & 11,267 (24.3\%)
& \quad No      & 29,588 (63.8\%) \\

\quad 3  & 1,308 (2.8\%)
& \quad High    & 6,914 (14.9\%)
& \quad Missing & 9,693 (20.9\%) \\

\quad 4  & 4,555 (9.8\%)
& \quad Missing & 1,004 (2.2\%)
& \multicolumn{2}{l}{\textit{Exercise, ages 40--49}} \\

\quad 5  & 10,135 (21.8\%)
& \multicolumn{2}{l}{\textit{Age at menarche}}
& \quad None    & 10,931 (23.6\%) \\

\quad 6  & 9,499 (20.5\%)
& \quad Early   & 4,805 (10.4\%)
& \quad Low     & 24,435 (52.6\%) \\

\quad 7  & 10,216 (22.0\%)
& \quad Regular & 38,968 (84.0\%)
& \quad High    & 5,616 (12.1\%) \\

\quad 8  & 8,517 (18.4\%)
& \quad Missing & 2,623 (5.6\%)
& \quad Missing & 5,414 (11.7\%) \\

\quad 9  & 694 (1.5\%)
& \multicolumn{2}{l}{\textit{Age at first pregnancy}}
& \multicolumn{2}{l}{\textit{Alcohol, ages 40--49}} \\

\quad 10 & 1 ($<0.1$\%)
& \quad Late    & 1,649 (3.6\%)
& \quad Did not drink & 34,401 (74.1\%) \\

\multicolumn{2}{@{}l}{\textit{Birth year}}
& \quad Regular & 37,868 (81.6\%)
& \quad Drank   & 6,012 (13.0\%) \\

\quad Mean (SD) & 1959.66 (2.55)
& \quad Missing & 6,879 (14.8\%)
& \quad Missing & 5,983 (12.9\%) \\

\quad Median (Q1, Q3) & 1960.00 (1958.00, 1962.00)
& \multicolumn{2}{l}{\textit{Family history: first degree}}
& \multicolumn{2}{l}{\textit{Smoking}} \\

\quad Min, Max & 1955.00, 1965.00
& \quad Yes     & 3,735 (8.0\%)
& \quad Yes     & 22,104 (47.7\%) \\

&
& \quad No      & 40,585 (87.5\%)
& \quad No      & 18,757 (40.4\%) \\

&
& \quad Missing & 2,076 (4.5\%)
& \quad Missing & 5,535 (11.9\%) \\

\bottomrule
\end{tabular}

\vspace{0.35em}
\begin{minipage}{0.95\textwidth}
\scriptsize
\end{minipage}

\end{table}

The model selection procedure proceeded with stepwise selection based on likelihood ratio tests, and it was performed on the Lehmann model due to the computational speed and stability in estimation. For comparison, we have also implemented on the final model the Binary GLMM model and the logistic regression model (with no residual heterogeneity component).

\subsection{Results}
The model selection process identified the following baseline covariates as being statistically significant to describe adherence to an invitation: \texttt{edu}; \texttt{age\_first\_preg}; \texttt{fam2}; \texttt{ex4049}; \texttt{alc4049}; \texttt{smoke}. In addition to these variables, the final model included a linear term for the
(centered) birth year (\texttt{birthyearctr}) and the sequential number of each invitation (\texttt{seq\_inv}) to capture a possible buildup of fatigue or enthusiasm.

\begin{figure}[hb!]
    \centering
    \includegraphics[width=.8\linewidth]{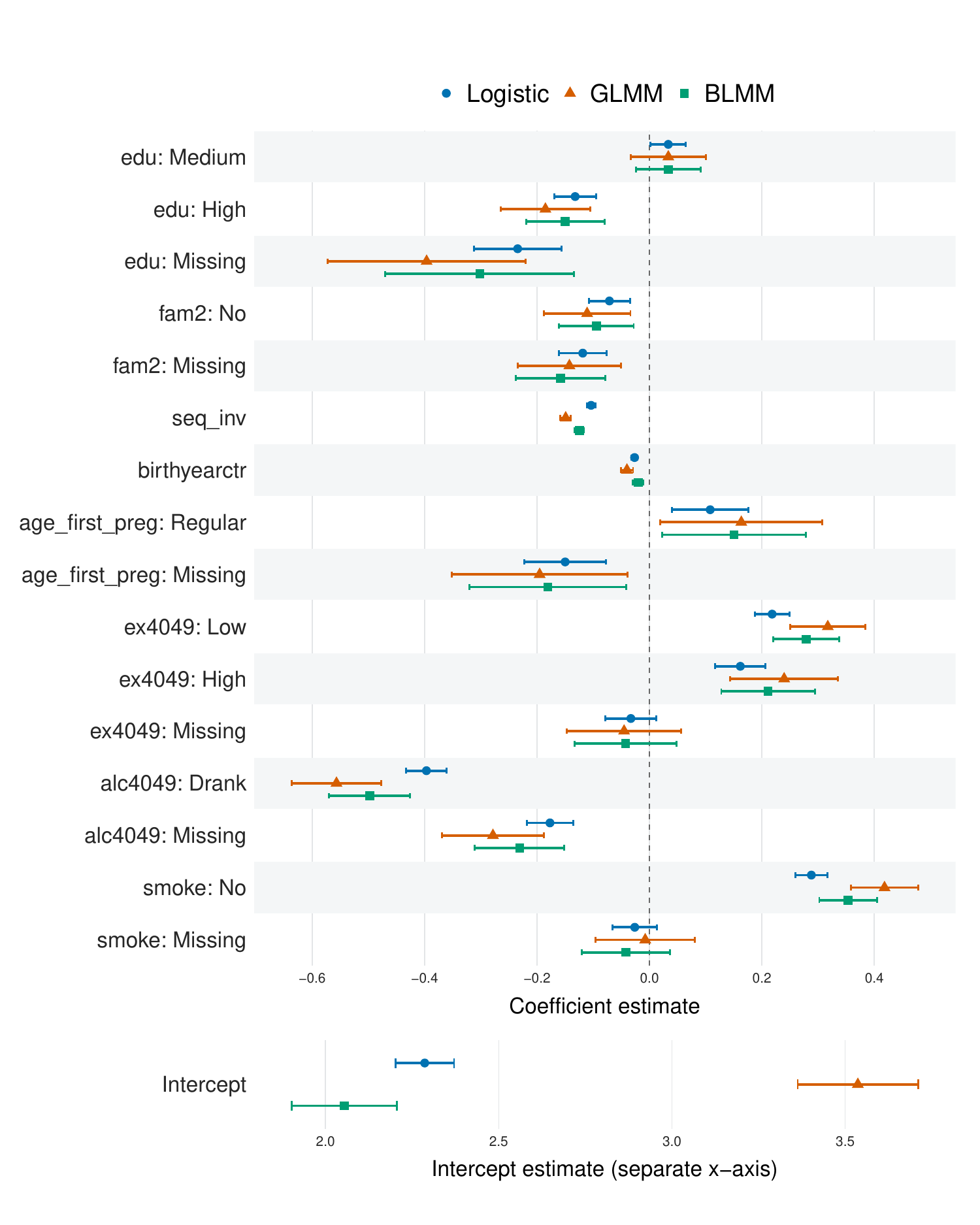}
    \caption{Point estimates and corresponding mle-based 95\% confidence intervals for the coefficients in the three models logistic regression (with no random effects), BLMM, and GLMM. Coefficients retain their model specific interpretation. Recall that for the BLMM models the exponential coefficient corresponds to the odds ratios referred to the ``average frailty'' individual, e.g an individual with $\gamma = 1$.}
    \label{fig:threemodels}
\end{figure}

The results from fitting the BLMM, GLMM, and logistic regression models are summarized in Figure \ref{fig:threemodels}. In addition to the estimates shown in the table, for the GLMM model the variance of the random effect $\sigma^2$ was estimated to be 4.052 , and for the BLMM the $\theta$ parameter was estimated to be 0.303. The complete output from the three models is reported in Section \ref{sec:alloutputs} of the Supplementary Material.

The main findings can be summarized as follows:
\begin{itemize}
\item Highly educated women are less likely to adhere to screening invitations compared to women with low education (OR = 0.86). A possible explanation is that more educated women may be screening at private clinics, where they may also receive ultrasound testing. Women with missing education value have much lower probability of adherence (OR=0.74) compared to women who declared a low educational level. This group is small and it may possibly contain many immigrants, which might explain this effect (see \cite{BhargavaTsurudaEtAl2018}, \cite{BhargavaHofvindEtAl2019}).
\item Women with ``Regular'' (non-late) first pregnancy had increased adherence (OR=1.16), and women with missing value had lower adherence (OR=0.83) when compared to women with ``late'' ($> 35$) age at first pregnancy. This may be related to education: indeed, among the 37,868 women with ``Regular'' age at first pregnancy, 22,772 (60.1\%) had low educational level, vs. 645 (39.1\%) out of the 1649 women with ``Late'' first pregnancy.
\item Women without (or missing) \texttt{fam2} family history have lower adherence (OR=0.91 and OR=85, respecively). Surprisingly, \texttt{fam2} turned out to be more relevant in the model than the \texttt{fam1} variable. This deserves further study, but possibly fam1 is measured "too early" to capture diagnoses of (mother), sisters, daughters.
\item Exercising (moderately or a lot) is associated with increased adherence probability (OR=1.32 and OR=1.23, respectively).
\item Being a ``drinker'' has a strong negative association with adherence (OR=0.61) while being a non-smoker has a strong positive association with adherence (OR=1.42).
\item The invitation number over time is associated with a statistically significant decrease in adherence probability, suggesting that fatigue occurs (OR=0.88 for subsequent invitations).
\item Lastly, more recent birth cohorts have slightly lower adherence probability. A difference of one year in the birth year reduces the odds of adherence by about 2\% (OR=.98).
\end{itemize}

The last two findings above confirm the possibility of sustaining participation through additional encouragement policies over time, and among the more recent birth cohorts. For example, since 2025 the participation rate to the screening program has been increasing after the implementation of SMS reminders.

The residual heterogeneity introduced by the random effect is large, with $\theta$ estimated at 0.3031 ($95\% \ CI: [.294,.312]$). Note that the density function of the estimated distribution of the frailty starts at $+\infty$ at zero and reaches zero in a monotonically descreasing fashion as $\gamma$ increases.

Importantly, fitting the binary GLMM with intercept random effect returned some instability warnings and did not converge. Identifiability of the model was signaled for being weak in some fitting runs, and impossible all-equal standard errors for the paramenter estimators were produced.

We implemented a cascade of two optimization algorithms (bobyqa followed by Nelder-Mead), increased the maximum number of function evaluations to 200,000, and increased from 1 to 5 to 10 the value of the parameter that governs the quadrature numerical integration (nAGQ, the number of Adaptive Gauss–Hermite Quadrature points used for scalar random effects). Increasing that parameter increases the precision of the numerical integration performed in the likelihood, and slows down the optimization. The final, most stable fit was obtained with the bobyqa algorithm and it is shown in Table \ref{fig:GLMMFit10bobyqa}.

With this fine-tuned glmm fitting algorithm, we achieved convergence and stable standard errors. Importantly, nAGQ=1 produced unreliable (all equal) standard errors. These became non trivial for nAGQ=5, and stable by comparison with the standard errors obtained from nAGQ=10. Importantly, for these three values of the parameter nAGQ the fitting time was equal to 57.8, 121.8, and 155.3 minutes, respectively. For comparison, the Lehmann model estimation algorithm converged in 3.2 minutes.

Incidentally, we note that the dataset is quite large, and that there did not seem to be issues related to sparse categories or separation.  
As can be seen from Figure \ref{fig:threemodels} and by comparing Table \ref{fig:GLMMFit10bobyqa} to Table \ref{fig:LehmannFit}, the point estimates of the model coefficients --and their standard errors -- were comparable in terms of direction and magnitude. This is reassuring, as both models essentially allow for two different kinds of perturbation of the success probabilities, which are defined for the average frailty by the same relationship with the covariates. In addition, the non-perturbed, standard logistic regression confirms the general direction and magnitude of the effects. The latter model runs very quickly -- a couple of seconds --, but it clearly does not allow for the description of a frailty distribution, or for individual risk prediction.

\begin{figure}[h!]
\centering\includegraphics[width=0.65\textwidth]{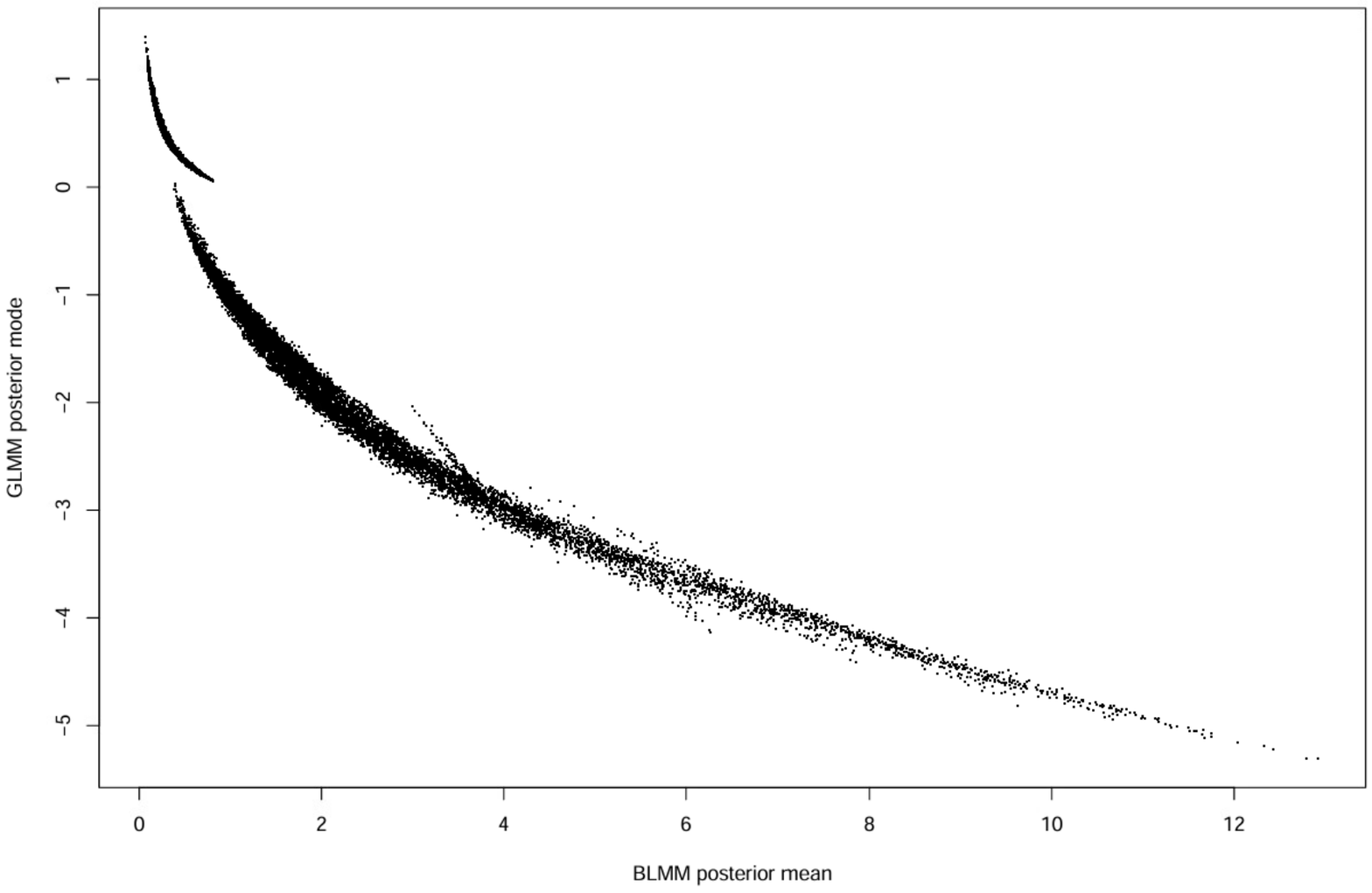}
\caption{BLMM posterior mean predictions of the subject-specific frailties and the GLMM posterior mode predictions of the subject-specific random effects.}\label{fig:comp_BLMM_GLMM}
\end{figure}

Indeed, as we have discussed our model allows for posterior prediction of the subject-specific random effect, which may be used together with the associated risk factors to identify subjects who have a large tendency to non-adherence.

Recall that for the BLMM model, a small value for the $\theta$ parameter that governs the gamma distribution of the frailties corresponds to a large variance $\sigma^2$ of the frailties in the GLMM model. The opposite case corresponds to a small heterogeneity in the two models (i.e. large $\theta$ and small $\sigma^2$). 

\begin{table}[ht!]
    \centering
    \begin{tabular}{lccc}
    \hline
      && \multicolumn{2}{c}{GLMM} \\
      \cline{3-4} 
      & & Not tail & Tail \\
 \hline
  BLMM   &  Not tail   &  46341 & 7 \\
    &  Tail   & 8  & 40 \\
  \hline  \end{tabular}
    \caption{Cross-classification table into top or bottom 0.001 tail of the predicted risk distribution for the BLMM and GLMM models}
    \label{tab:crossclass001}
\end{table}

The estimation of the subject-specific frailty is important in the screening adherence problem, but more in general for policy decision making and interventions. The frailty term can be described as a ``hesitancy'' subject-specific quantity that, when estimated from past behavior, can help target the most hesitant subjects (beyond what their covariates would suggest, that is) with an {\em individual} intervention to promote adherence. Such an intervention would be above and beyond other efforts aimed at population strata defined by the covariates.

\begin{table}[h!]
\centering
\caption{Prediction performance}
\label{tab:trainvalidBLMMGLMM}
\normalsize
\sisetup{round-mode=places,round-precision=4,table-number-alignment=center}
\begin{threeparttable}
\begin{tabular}{@{}llS[table-format=1.4]S[table-format=1.4]@{}}
\toprule
Index & Prediction summary & {BLMM} & {GLMM} \\
\midrule
\multicolumn{4}{@{}l}{\textit{Brier score}} \\
\texttt{BrierCMean} & Conditional, pre-calibration & 0.063065521 & 0.062930683 \\
\texttt{BrierCCal} & Conditional, post-calibration & 0.061103609 & 0.059496744 \\
\texttt{BrierCMed} & Conditional, posterior median & 0.063910427 & 0.062642728 \\
\texttt{BrierM} & Marginal & 0.108397663 & 0.106726875 \\
\addlinespace[0.4em]
\multicolumn{4}{@{}l}{\textit{Observed-to-expected ratio}} \\
\texttt{OE\_CMean} & Conditional, pre-calibration & 1.009311083 & 0.972133033 \\
\texttt{OE\_CCal} & Conditional, post-calibration & 0.994340305 & 0.999495123 \\
\texttt{OE\_CMed} & Conditional, posterior median & 0.971876955 & 0.975586311 \\
\texttt{OE\_M} & Marginal & 1.000940626 & 1.003133393 \\
\addlinespace[0.4em]
\multicolumn{4}{@{}l}{\textit{Integrated calibration index}} \\
\texttt{ICICMean} & Conditional, pre-calibration & 0.047452642 & 0.045914199 \\
\texttt{ICICCal} & Conditional, post-calibration & 0.023437016 & 0.019918538 \\
\texttt{ICICMed} & Conditional, posterior median & 0.038695319 & 0.049690381 \\
\texttt{ICIM} & Marginal & 0.004791623 & 0.008768513 \\
\addlinespace[0.4em]
\multicolumn{4}{@{}l}{\textit{Binary concordance}} \\
\texttt{C\_BM} & Between subjects, marginal & 0.577659029 & 0.574432632 \\
\texttt{C\_W} & Within subjects & 0.584510999 & 0.587850223 \\
\texttt{C\_BCMean} & Between subjects, pre-calibration & 0.944540063 & 0.945493686 \\
\texttt{C\_BCCAL} & Between subjects, post-calibration & 0.944675004 & 0.944865475 \\
\texttt{C\_BCMedian} & Between subjects, posterior median & 0.944497758 & 0.945545384 \\
\bottomrule
\end{tabular}
\begin{tablenotes}[flushleft]
\footnotesize
\item \textit{Notes.} Lower Brier scores and integrated calibration indices are preferable; the ideal observed-to-expected ratio is 1, and higher concordance is preferable. We denoted the posterior mean frailty and
the rank-based calibration frailty as pre-calibration and as post-calibration respectively. 
\end{tablenotes}
\end{threeparttable}
\end{table}
\FloatBarrier

Figure \ref{fig:comp_BLMM_GLMM} shows a comparison of the BLMM posterior mean predictions of the subject-specific frailties and the GLMM posterior mode predictions of the subject-specific random effects. Recall that low values of the former predictions and large values of the latter predictions correspond to subject-specific increased risk.

It is of interest to compare the quantiles of these two distribution, as the identification of the subject with increased risk beyond the measured risk factors is of interest. Firstly, Spearman's rank correlation coefficient between the posterior estimates from the two models is equal to -0.997, which shows a remarkable alignment.

In addition, one may be interested in identifying subjects that have a particularly high predicted risk. These subjects are identified by extracting the bottom (top) tail from distribution of the risks predicted by the BLMM (GLMM) model. One can quantify the agreement between such classification for different tail frequencies by computing Cohen's kappa index on the corresponding 2 by 2 table. For example, Table \ref{tab:crossclass001} shows the classification matrix when one sets a tail frequency of 0.001; Cohen's index for the table is equal to 0.842 (note that the symmetric nature of the table is due to chance). The same index computed on different tables is equal to 0.883, 0.894, 0.874, 0.908, and 0.912 for the tail frequencies 0.005, 0.01, 0.02, 0.05, 0.1, respectively. This shows a remarkable (and increasing with the tail frequency) level of agreement between the risk predictions obtained from the two models. 

We have performed the training-validation estimation of the probability-based performance/risk prediction metrics, using a training percentage of 70\% of the subjects. Table \ref{tab:trainvalidBLMMGLMM} shows the metrics computed on the validation dataset. The results show largely comparable performances by the two BLMM and GLMM models. Note that We did not report the marginal overall c-index because it was identical to the between-cluster c-index. See Appendix, Section~\ref{overall_between_cindex}. The BLMM and GLMM showed similar predictive performance. Post-calibration improved the conditional Brier score, O/E ratio, and ICI for both models, while leaving between-subject concordance nearly unchanged. The GLMM had slightly lower Brier scores, whereas the BLMM had a lower marginal ICI (0.0048 versus 0.0088); marginal discrimination was similar. Conditional between-subject concordance was much higher than marginal concordance. Conditional concordance was substantially higher than marginal concordance, reflecting the additional information from each woman’s observed response history used to estimate the frailty term. We note that this is similar to what we observed in our simulation study. This reduction is also similar in magnitude between BLMM and GLMM models. 

\section{Discussion}\label{sec:Discussion}
The modeling approach that we have discussed provides a flexible and interpretable likelihood-based approach to clustered binary data while retaining closed-form expressions for marginal probabilities, the likelihood, and several predictive quantities. Its tractability follows from the multiplicative modification of the covariate-adjusted baseline probabilities through the gamma frailties: conditional probabilities have a power form, and integration reduces to evaluating gamma Laplace transforms. This is analogous to the role of gamma integration in deriving the negative binomial distribution from a Poisson model with a gamma-distributed rate. In practice, the resulting expressions make it feasible to quickly fit and compare different models specifications even on large datasets. This includes exploring alternative covariate definitions and inclusion in the models, and the analyses of subsets of the data.

The BLMM simulation study showed generally good inferential performance, with decreasing coefficient bias as the training sample grew and confidence interval coverage generally close to its nominal level. Marginal probabilities were predicted accurately across the scenarios considered, while -- as expected -- the prediction of individual frailties improved with more repeated observations within clusters and less heterogeneity among clusters. The rank-based calibration improved the calibration measures in many scenarios but did not consistently improve frailty prediction or interval coverage. Thus, its use therefore should depends on the prediction target. 

The numerical comparison of the BLMM with the normal random-intercept binary GLMM in our motivating application showed good agreement of the risk predictions and similar substantial conclusions. In that breast cancer screening application, the BLMM identified associations between baseline characteristics, invitation number, and longitudinal adherence, while accounting for substantial between-woman heterogeneity. Its coefficient estimates were comparable in direction and magnitude to those of the fitted GLMM. For this dataset, however, BLMM estimation took substantially less computing time when compared to the GLMM fit using ten adaptive quadrature points. The fitted model may help characterize patterns of repeated participation and identify groups for whom adherence initiatives could be considered. We are planning to implement the model to a different screening dataset. Clearly, the proposed framework also applies to many other settings with correlated binary outcomes.

Several limitations of the proposed framework suggest directions for future research. Because a single shared frailty term induces only nonnegative within-cluster correlations, models with a richer frailty structure should be developed to accommodate more general patterns of dependence. Future work should assess the adequacy of the assumed frailty distribution and examine how its misspecification affects fixed-effect estimates, marginal probabilities, cluster-specific predictions, and measures of heterogeneity \citep{TchetgenCoull2006,McCullochNeuhaus2011}. A related direction is to consider other frailty distributions for which the analytical simplifications arising from the closed-form Laplace transform may be exploited, while maintaining interpretability of the resulting models.
Finally, the framework should be extended to accommodate multivariate frailties within clusters, as well as missing observations.

\subsection*{Funding}
This work has been carried out within the MUSA -- Multilayered Urban Sustainability Action -- project, funded by the European Union -- NextGenerationEU, under the National Recovery and Resilience Plan (NRRP) Mission 4 Component 2 Investment Line 1.5. It was also funded in part through a grant by the Invernizzi Foundation in Milan, Italy.

\subsection*{Conflicts of Interest}

The authors declare no conflicts of interest.


\printbibliography

\clearpage

\appendix
\section*{Appendix}
\addcontentsline{toc}{section}{Appendix}
\setcounter{subsection}{0}
\setcounter{equation}{0}
\setcounter{figure}{0}
\setcounter{table}{0}
\renewcommand{\thesubsection}{A.\arabic{subsection}}
\renewcommand{\thesubsubsection}{\thesubsection.\arabic{subsubsection}}
\numberwithin{equation}{subsection}
\numberwithin{figure}{subsection}
\numberwithin{table}{subsection}
\renewcommand{\theHsubsection}{app.\arabic{subsection}}
\renewcommand{\theHsubsubsection}{\theHsubsection.\arabic{subsubsection}}
\renewcommand{\theHequation}{app.\arabic{subsection}.\arabic{equation}}
\renewcommand{\theHfigure}{app.\arabic{subsection}.\arabic{figure}}
\renewcommand{\theHtable}{app.\arabic{subsection}.\arabic{table}}

\subsection{Covariance, correlation, and probability variation}\label{app:corr}
For distinct $j,k$, the conditional covariance given $\gamma_i$ is zero. Hence
\begin{align}
 \operatorname{Cov}(Y_{ij},Y_{ik}\mid\bm X_i)
 &=E_\Gamma(e^{-\Gamma u_{ij}}e^{-\Gamma u_{ik}})
       -E_\Gamma(e^{-\Gamma u_{ij}})E_\Gamma(e^{-\Gamma u_{ik}})\notag\\
 &=L_\theta(u_{ij}+u_{ik})-L_\theta(u_{ij})L_\theta(u_{ik}).\label{eq:acov}
\end{align}
The strict inequality
\[
 (1+u_{ij}/\theta)(1+u_{ik}/\theta)>1+(u_{ij}+u_{ik})/\theta
\]
implies positivity of~\eqref{eq:acov} after applying the negative power $-\theta$. Division by the Bernoulli marginal standard deviations proves~\eqref{eq:corr1}. At $u_{ij}=u_{ik}=u$, the covariance is exactly $\operatorname{Var}_\Gamma(e^{-\Gamma u})=L_\theta(2u)-L_\theta(u)^2$. For fixed $u>0$, this variance tends to zero both as $\theta\to\infty$ and as $\theta\downarrow0$.

\subsection{Closed-form marginal likelihood}\label{app:lik}

Let
\[
\mathcal J_{i1}=\{j:y_{ij}=1\},
\qquad
\mathcal J_{i0}=\{j:y_{ij}=0\},
\qquad
n_{i0}=|\mathcal J_{i0}|,
\]
and write
\[
\pi_{ij}(\bm\beta)
=\pi_C(1;\bm x_{ij}),
\qquad
u(\bm x_{ij}, \bm \beta)=-\log\pi_{ij}(\bm\beta).
\]
As defined in Section~\ref{sec:est}, for every
$A\subseteq\mathcal J_{i0}$,
\[
a_i(A)
=\theta+\sum_{j\in\mathcal J_{i1}\cup A}u(\bm x_{ij}, \bm \beta)
=\theta-\sum_{j\in\mathcal J_{i1}\cup A}
\log\pi_{ij}(\bm\beta).
\]

Under the stated conditional independence and shared gamma frailty
assumptions, the following derivation holds for any finite,
nonnegative function $u(\bm x_{ij}, \bm \beta)$, such as those in Table \ref{tab:links}. It uses only
$\pi_{ij}(\bm\beta)=\exp(-u(\bm x_{ij}, \bm \beta))$ and the gamma integral with
$a_i(A)=\theta+\sum_{j\in\mathcal J_{i1}\cup A}u(\bm x_{ij}, \bm \beta)>0$.

Integrating the joint density over the independent frailties,
and then separating the factors corresponding to $y_{ij}=1$
and $y_{ij}=0$, gives
\begin{equation}\label{eq:intlik}
\begin{aligned}
L(\bm\beta,\theta)
&=
\int_{(0,\infty)^n}
f(\bm y,\bm\gamma\mid\bm X;\bm\beta,\theta)
\,d\bm\gamma\\
&=
\int_{(0,\infty)^n}
\prod_{i=1}^{n}
\left[
f(\bm y_i\mid\gamma_i,\bm X_i;\bm\beta)
f_\Gamma(\gamma_i;\theta)
\right]d\bm\gamma\\
&=
\prod_{i=1}^{n}
\left[
\int_0^\infty
f(\bm y_i\mid\gamma_i,\bm X_i;\bm\beta)
f_\Gamma(\gamma_i;\theta)
\,d\gamma_i
\right]\\
&=
\prod_{i=1}^{n}
\left[
\int_0^\infty
\left\{
\prod_{j=1}^{n_i}
\bigl(\pi_{ij}(\bm\beta)^{\gamma_i}\bigr)^{y_{ij}}
\bigl(1-\pi_{ij}(\bm\beta)^{\gamma_i}\bigr)^{1-y_{ij}}
\right\}
\frac{\theta^\theta}{\Gamma(\theta)}
\gamma_i^{\theta-1}e^{-\theta\gamma_i}
\,d\gamma_i
\right]\\
&=
\prod_{i=1}^{n}
\left[
\int_0^\infty
\left\{
\prod_{j\in\mathcal J_{i1}}
\pi_{ij}(\bm\beta)^{\gamma_i}
\right\}
\left\{
\prod_{j\in\mathcal J_{i0}}
\bigl(1-\pi_{ij}(\bm\beta)^{\gamma_i}\bigr)
\right\}
\frac{\theta^\theta}{\Gamma(\theta)}
\gamma_i^{\theta-1}e^{-\theta\gamma_i}
\,d\gamma_i
\right].
\end{aligned}
\end{equation}

For each cluster, denote the integral in the last line
by $D_i(\bm\beta,\theta)$. Collecting the factors associated
with $y_{ij}=1$ gives, step by step,
\begin{equation}\label{eq:di-int}
\begin{aligned}
D_i(\bm\beta,\theta)
&=
\int_0^\infty
\left\{
\prod_{j\in\mathcal J_{i1}}
\pi_{ij}(\bm\beta)^{\gamma_i}
\right\}
\left\{
\prod_{j\in\mathcal J_{i0}}
\bigl(1-\pi_{ij}(\bm\beta)^{\gamma_i}\bigr)
\right\}
\frac{\theta^\theta}{\Gamma(\theta)}
\gamma_i^{\theta-1}e^{-\theta\gamma_i}
\,d\gamma_i\\
&=
\int_0^\infty
\left\{
\prod_{j\in\mathcal J_{i1}}
\pi_{ij}(\bm\beta)
\right\}^{\gamma_i}
\left\{
\prod_{j\in\mathcal J_{i0}}
\bigl(1-\pi_{ij}(\bm\beta)^{\gamma_i}\bigr)
\right\}
\frac{\theta^\theta}{\Gamma(\theta)}
\gamma_i^{\theta-1}e^{-\theta\gamma_i}
\,d\gamma_i\\
&=
\int_0^\infty
\exp\left\{
\gamma_i\sum_{j\in\mathcal J_{i1}}
\log\pi_{ij}(\bm\beta)
\right\}
\left\{
\prod_{j\in\mathcal J_{i0}}
\bigl(1-\pi_{ij}(\bm\beta)^{\gamma_i}\bigr)
\right\}
\frac{\theta^\theta}{\Gamma(\theta)}
\gamma_i^{\theta-1}e^{-\theta\gamma_i}
\,d\gamma_i\\
&=
\int_0^\infty
\left\{
\prod_{j\in\mathcal J_{i0}}
\bigl(1-\pi_{ij}(\bm\beta)^{\gamma_i}\bigr)
\right\}
\frac{\theta^\theta}{\Gamma(\theta)}
\gamma_i^{\theta-1}
\exp\left[
-\gamma_i
\left\{
\theta-
\sum_{j\in\mathcal J_{i1}}
\log\pi_{ij}(\bm\beta)
\right\}
\right]d\gamma_i\\
&=
\int_0^\infty
\left\{
\prod_{j\in\mathcal J_{i0}}
\bigl(1-\pi_{ij}(\bm\beta)^{\gamma_i}\bigr)
\right\}
\frac{\theta^\theta}{\Gamma(\theta)}
\gamma_i^{\theta-1}
e^{-a_i(\varnothing)\gamma_i}
\,d\gamma_i.
\end{aligned}
\end{equation}

where $\varnothing$ denotes the empty set. We now expand the product associated with $y_{ij}=0$.
Writing the terms of orders zero, one, two, and three
explicitly yields
\begin{equation}\label{eq:orders}
\begin{aligned}
\prod_{j\in\mathcal J_{i0}}
\bigl(1-\pi_{ij}(\bm\beta)^{\gamma_i}\bigr)
={}&
1
-\sum_{r\in\mathcal J_{i0}}
\pi_{ir}(\bm\beta)^{\gamma_i}
+\sum_{\substack{r,s\in\mathcal J_{i0}\\r<s}}
\{\pi_{ir}(\bm\beta)\pi_{is}(\bm\beta)\}^{\gamma_i}\\
&-\sum_{\substack{r,s,t\in\mathcal J_{i0}\\r<s<t}}
\{\pi_{ir}(\bm\beta)\pi_{is}(\bm\beta)
\pi_{it}(\bm\beta)\}^{\gamma_i}+\sum_{m=4}^{n_{i0}}(-1)^m
\sum_{\substack{A\subseteq\mathcal J_{i0}\\|A|=m}}
\left\{
\prod_{j\in A}\pi_{ij}(\bm\beta)
\right\}^{\gamma_i}.
\end{aligned}
\end{equation}
Sums over unavailable index combinations are zero. In
particular, the term of highest order, obtained by choosing
every index in $\mathcal J_{i0}$, is
\[
(-1)^{n_{i0}}
\left\{
\prod_{j\in\mathcal J_{i0}}
\pi_{ij}(\bm\beta)
\right\}^{\gamma_i}.
\]
Equivalently, the entire expansion is
\begin{equation}\label{eq:expand}
\prod_{j\in\mathcal J_{i0}}
\bigl(1-\pi_{ij}(\bm\beta)^{\gamma_i}\bigr)
=
\sum_{A\subseteq\mathcal J_{i0}}
(-1)^{|A|}
\exp\left\{
\gamma_i\sum_{j\in A}
\log\pi_{ij}(\bm\beta)
\right\}.
\end{equation}

On substituting~\eqref{eq:orders} into~\eqref{eq:di-int},
each selected subset contributes one integral. Its exponent
is determined by
\[
-a_i(A)\gamma_i =-\theta\gamma_i
+\gamma_i\sum_{j\in\mathcal J_{i1}}
\log\pi_{ij}(\bm\beta)
+\gamma_i\sum_{j\in A}
\log\pi_{ij}(\bm\beta).
\]
The first successive rates are therefore
\begin{align*}
a_i(\varnothing)
&=\theta-\sum_{j\in\mathcal J_{i1}}
\log\pi_{ij}(\bm\beta),\\
a_i(\{r\})
&=a_i(\varnothing)-\log\pi_{ir}(\bm\beta),\\
a_i(\{r,s\})
&=a_i(\varnothing)
-\log\pi_{ir}(\bm\beta)-\log\pi_{is}(\bm\beta),\\
a_i(\{r,s,t\})
&=a_i(\varnothing)
-\log\pi_{ir}(\bm\beta)
-\log\pi_{is}(\bm\beta)
-\log\pi_{it}(\bm\beta),
\end{align*}
while the highest-order term has rate
\[
a_i(\mathcal J_{i0})
=
\theta-\sum_{j=1}^{n_i}
\log\pi_{ij}(\bm\beta).
\]

Thus, retaining the separate integrals at each order,
\begin{equation}\label{eq:di-integrals}
\begin{aligned}
D_i(\bm\beta,\theta)
={}&
\int_0^\infty
\frac{\theta^\theta}{\Gamma(\theta)}
\gamma_i^{\theta-1}
e^{-a_i(\varnothing)\gamma_i}\,d\gamma_i -\sum_{r\in\mathcal J_{i0}}
\int_0^\infty
\frac{\theta^\theta}{\Gamma(\theta)}
\gamma_i^{\theta-1}
e^{-a_i(\{r\})\gamma_i}\,d\gamma_i\\
&+\sum_{\substack{r,s\in\mathcal J_{i0}\\r<s}}
\int_0^\infty
\frac{\theta^\theta}{\Gamma(\theta)}
\gamma_i^{\theta-1}
e^{-a_i(\{r,s\})\gamma_i}\,d\gamma_i -\sum_{\substack{r,s,t\in\mathcal J_{i0}\\r<s<t}}
\int_0^\infty
\frac{\theta^\theta}{\Gamma(\theta)}
\gamma_i^{\theta-1}
e^{-a_i(\{r,s,t\})\gamma_i}\,d\gamma_i\\
&+\sum_{m=4}^{n_{i0}}(-1)^m
\sum_{\substack{A\subseteq\mathcal J_{i0}\\|A|=m}}
\int_0^\infty
\frac{\theta^\theta}{\Gamma(\theta)}
\gamma_i^{\theta-1}
e^{-a_i(A)\gamma_i}\,d\gamma_i.
\end{aligned}
\end{equation}
The last term in this expansion is the integral with rate
$a_i(\mathcal J_{i0})$ and sign $(-1)^{n_{i0}}$.

Since $a_i(A)>0$, integration of each term gives
\[
\int_0^\infty
\frac{\theta^\theta}{\Gamma(\theta)}
\gamma_i^{\theta-1}e^{-a_i(A)\gamma_i}
\,d\gamma_i
=
\left\{\frac{\theta}{a_i(A)}\right\}^{\theta}.
\]
Consequently,~\eqref{eq:di-integrals} becomes
\begin{equation}\label{eq:di-orders}
\begin{aligned}
D_i(\bm\beta,\theta)
={}&
\left\{\frac{\theta}{a_i(\varnothing)}\right\}^{\theta}
-\sum_{r\in\mathcal J_{i0}}
\left\{\frac{\theta}{a_i(\{r\})}\right\}^{\theta} +\sum_{\substack{r,s\in\mathcal J_{i0}\\r<s}}
\left\{\frac{\theta}{a_i(\{r,s\})}\right\}^{\theta} \\
&-\sum_{\substack{r,s,t\in\mathcal J_{i0}\\r<s<t}}
\left\{\frac{\theta}{a_i(\{r,s,t\})}\right\}^{\theta} +\sum_{m=4}^{n_{i0}}(-1)^m
\sum_{\substack{A\subseteq\mathcal J_{i0}\\|A|=m}}
\left\{\frac{\theta}{a_i(A)}\right\}^{\theta}.
\end{aligned}
\end{equation}
The same result, with all orders collected into one sum, is
\[
D_i(\bm\beta,\theta)
=
\sum_{A\subseteq\mathcal J_{i0}}
(-1)^{|A|}
\left\{\frac{\theta}{a_i(A)}\right\}^{\theta}.
\]
Taking the product over clusters gives precisely~\eqref{eq:lik}.

The integration step can equivalently be obtained from the
moment generating function of a shape--rate gamma random
variable:
\[
M_{\alpha,\beta}(t)
=
E\{\exp(tZ)\}
=
\left(\frac{\beta}{\beta-t}\right)^\alpha,
\qquad
Z\sim\operatorname{Gamma}(\alpha,\beta),
\quad t<\beta,
\]
by setting $\alpha=\beta=\theta$ and
$t=\sum_{j\in\mathcal J_{i1}\cup A}
\log\pi_{ij}(\bm\beta)\leq0$.
\subsection{Parameter covariance and standard errors}\label{app:se}

This appendix gives the covariance transformation used in Section~\ref{sec:est}. Standard errors and confidence intervals follow from standard large-sample maximum likelihood theory. In the numerical maximization, the frailty parameter was reparameterized as
$\theta=\exp(\psi)$, so that optimization could be performed over the unconstrained parameter $\psi$ while automatically enforcing the positivity constraint on $\theta$. Let the parameter vector used in the numerical optimization be $\bm{\lambda} = (\bm{\beta}',\psi)',$
and let the parameter vector on the original scale be $\bm{\zeta} = (\bm{\beta}',\theta)'
= g(\bm{\lambda}).$ After maximization, the estimated covariance matrix of $\widehat{\bm{\lambda}}$ was obtained from the inverse of the observed information matrix, i.e. $\widehat{\mathrm{Var}}(\widehat{\bm{\lambda}})
=
\widehat{I}^{-1},$ where $\widehat{I}$ denotes the observed information matrix, computed numerically from the Hessian of the negative log-likelihood at the maximum. The covariance matrix on the original parameter scale was then obtained by the delta method:
\[
\widehat{\mathrm{Var}}(\widehat{\bm{\zeta}})
=
\widehat{\nabla}\,
\widehat{I}^{-1}
\widehat{\nabla}',
\]
where
\[
\widehat{\nabla}
=
\left.
\frac{\partial g(\bm{\lambda})}{\partial \bm{\lambda}'}
\right|_{\bm{\lambda}=\widehat{\bm{\lambda}}}.
\]
In the present case, since the fixed-effect parameters are unchanged by the transformation and only $\theta$ is obtained as $\exp(\psi)$, the Jacobian matrix is diagonal:
\[
\widehat{\nabla}
=
\mathrm{diag}\left(1,\ldots,1,\exp(\widehat{\psi})\right)
=
\mathrm{diag}\left(1,\ldots,1,\widehat{\theta}\right).
\]
The standard errors were obtained as the square roots of the diagonal elements of
$\widehat{\mathrm{Var}}(\widehat{\bm{\zeta}})$, and Wald-type confidence intervals were constructed on the original parameter scale.

\subsection{Posterior frailty distribution and summaries}\label{app:post}

This appendix derives the results in Section \ref{sec:pred}. At fixed $\bm{\beta}$ and $\theta$,
\[
f(\bm{\gamma}\mid\bm{y},\bm{X};\bm{\beta},\theta)
=\prod_{i=1}^{n}\frac{f(\bm{y}_i\mid\gamma_i,\bm{X}_i;\bm{\beta})f_{\Gamma}(\gamma_i;\theta)}{D_i(\bm{\beta},\theta)}.
\]
Thus, the cluster frailties are conditionally independent at fixed parameter values, and the normalizer for cluster $i$ is its own marginal likelihood contribution.

\paragraph{Posterior mean.}
The unnormalized first moment is
\begin{align}
K_i(\bm{\beta},\theta)
&=\int_0^\infty\gamma_i f(\bm{y}_i\mid\gamma_i,\bm{X}_i;\bm{\beta})f_{\Gamma}(\gamma_i;\theta)\,d\gamma_i\nonumber\\
&=\int_0^\infty
\gamma_i\left\{
\prod_{j=1}^{n_i}
\bigl(\pi_{ij}(\bm\beta)^{\gamma_i}\bigr)^{y_{ij}}
\bigl(1-\pi_{ij}(\bm\beta)^{\gamma_i}\bigr)^{1-y_{ij}}
\right\}
\frac{\theta^\theta}{\Gamma(\theta)}
\gamma_i^{\theta-1}e^{-\theta\gamma_i}
\,d\gamma_i\\
&=\frac{\theta^\theta}{\Gamma(\theta)}\sum_{A\subseteq\mathcal{J}_{i0}}(-1)^{|A|}
\int_0^\infty\gamma_i^{\theta}e^{-a_i(A)\gamma_i}\,d\gamma_i\nonumber\\
&=\frac{\theta^\theta}{\Gamma(\theta)}\sum_{A\subseteq\mathcal{J}_{i0}}(-1)^{|A|}
\frac{\Gamma(\theta+1)}{\{a_i(A)\}^{\theta+1}}\nonumber =\sum_{A\subseteq\mathcal{J}_{i0}}(-1)^{|A|}\left\{\frac{\theta}{a_i(A)}\right\}^{\theta+1},
\end{align}
where $\Gamma(\theta+1)=\theta\Gamma(\theta)$. In expanded form,
\begin{align*}
K_i(\bm{\beta},\theta)
={}&\left\{\frac{\theta}{a_i(\varnothing)}\right\}^{\theta+1}
-\sum_{r\in\mathcal{J}_{i0}}\left\{\frac{\theta}{a_i(\{r\})}\right\}^{\theta+1} +\sum_{\substack{r,s\in\mathcal{J}_{i0}\\r<s}}\left\{\frac{\theta}{a_i(\{r,s\})}\right\}^{\theta+1} \\
&-\sum_{\substack{r,s,t\in\mathcal{J}_{i0}\\r<s<t}}\left\{\frac{\theta}{a_i(\{r,s,t\})}\right\}^{\theta+1} +\sum_{m=4}^{n_{i0}}(-1)^m\sum_{\substack{A\subseteq\mathcal{J}_{i0}\\|A|=m}}
\left\{\frac{\theta}{a_i(A)}\right\}^{\theta+1}.
\end{align*}
Dividing by $D_i$ gives~\eqref{eq:mean}. The denominator is~\eqref{eq:di-orders}, with power $\theta$ instead of $\theta+1$.

\paragraph{Posterior CDF and density.}
Let $F_{\Gamma}(t;a,b)$ denote the CDF of a gamma random variable with shape $a$ and rate $b$. Integrating the posterior kernel up to $t\geq0$ gives
\begin{align}\label{eq:cdf-proof}
F_i(t;\bm{\beta},\theta)
&=\frac{\theta^\theta}{D_i(\bm{\beta},\theta)\Gamma(\theta)}
\sum_{A\subseteq\mathcal{J}_{i0}}(-1)^{|A|}\int_0^t\gamma_i^{\theta-1}e^{-a_i(A)\gamma_i}\,d\gamma_i\nonumber\\
&=\frac{1}{D_i(\bm{\beta},\theta)}\sum_{A\subseteq\mathcal{J}_{i0}}(-1)^{|A|}
\left\{\frac{\theta}{a_i(A)}\right\}^{\theta}F_{\Gamma}\{t;\theta,a_i(A)\}.
\end{align}
The identity used in each term is
\[
\int_0^t\gamma_i^{\theta-1}e^{-a_i(A)\gamma_i}\,d\gamma_i
=\frac{\Gamma(\theta)}{\{a_i(A)\}^{\theta}}F_{\Gamma}\{t;\theta,a_i(A)\}.
\]
The order-by-order form is
\begin{align*}
D_i(\bm{\beta},\theta)F_i(t;\bm{\beta},\theta)
={}&\left\{\frac{\theta}{a_i(\varnothing)}\right\}^{\theta}F_{\Gamma}\{t;\theta,a_i(\varnothing)\} -\sum_{r\in\mathcal{J}_{i0}}\left\{\frac{\theta}{a_i(\{r\})}\right\}^{\theta}F_{\Gamma}\{t;\theta,a_i(\{r\})\}\\
&+\sum_{\substack{r,s\in\mathcal{J}_{i0}\\r<s}}\left\{\frac{\theta}{a_i(\{r,s\})}\right\}^{\theta}F_{\Gamma}\{t;\theta,a_i(\{r,s\})\}\\
&-\sum_{\substack{r,s,q\in\mathcal{J}_{i0}\\r<s<q}}\left\{\frac{\theta}{a_i(\{r,s,q\})}\right\}^{\theta}F_{\Gamma}\{t;\theta,a_i(\{r,s,q\})\}\\
&+\sum_{m=4}^{n_{i0}}(-1)^m\sum_{\substack{A\subseteq\mathcal{J}_{i0}\\|A|=m}}
\left\{\frac{\theta}{a_i(A)}\right\}^{\theta}F_{\Gamma}\{t;\theta,a_i(A)\}.
\end{align*}
The triple index is written as $q$ to distinguish it from the CDF argument $t$. Empty sums follow the convention in Appendix~\ref{app:lik}. Replacing each gamma CDF by the corresponding density gives the posterior density. As $t\to\infty$, each gamma CDF tends to one and the numerator tends to $D_i$, so $F_i(\infty;\bm{\beta},\theta)=1$.

At the fitted parameter values, numerical inversion of $\widehat{F}_i$ gives the median and interval endpoints in~\eqref{eq:quant}. One may solve $\widehat{F}_i(t)=u$ for $u\in(0,1)$ or minimize $\{\widehat{F}_i(t)-u\}^2$ over $t\geq0$. The continuous posterior distribution implies
\[
P(L_i\leq\gamma_i\leq U_i\mid\bm{y}_i,\bm{X}_i;\widehat{\bm{\beta}},\widehat{\theta})=1-\alpha.
\]

As in Section \ref{app:lik}, under conditional independence and shared gamma frailty
assumptions, the above derivations holds for any finite,
nonnegative function $u(\bm x_{ij}, \bm \beta)$, such as those in Table \ref{tab:links}. It uses
$\pi_{ij}(\bm\beta)=\exp(-u(\bm x_{ij}, \bm \beta))$ and the gamma integral with
$a_i(A)=\theta+\sum_{j\in\mathcal J_{i1}\cup A}u(\bm x_{ij}, \bm \beta)>0$.

\subsection{Large-sample confidence interval for marginal predictive probability}
\label{app:ci}\label{app:app2}
This appendix gives the logit-scale confidence interval used in Section~\ref{sec:prob}. Plugging into expression \eqref{eq:pm} the maximum likelihood estimates of $\beta$ and $\theta$, for fixed covariate vector ${\bm x}$, one obtains the estimated predicted probability
\[
\widehat{\pi}_M({\bm x})= \left(    \frac{\widehat{\theta}}{\widehat{\theta} - \log \left( \frac{{\rm e}^{{\widehat{\bm \beta}}\,' \bm x}}{1+ {\rm e}^{{\widehat{\bm \beta}}\,' \bm  x}}  \right)}  \right)^{\widehat{\theta}} .
\]

To construct a confidence interval for $\pi_M(\bm x)$ from this quantity one needs to recall that the consistency and asymptotic normality of the resulting estimator follow from standard mle results on the vector $(\widehat{\bm \beta}', \widehat{\theta})'$ and from the delta method for transformations $\Re^{b+1} \to \Re^1$ (where $b$ is the length of $\bm \beta$ and $\bm x$).

To highlight the dependence of $\pi_C(1;\bm x)$ and of $\pi_M(\bm x)$ on the parameters $(\widehat{\bm \beta}', \widehat{\theta})'$, define $p(\bm\beta)= \pi_C(1;\bm x)= e^{\bm\beta' {\bm x}}/ \left( 1+e^{\bm\beta' {\bm x}} \right)$ and
\[
g(\theta,\bm\beta) = \pi_M(\bm x) = \left( \frac{\theta}{\theta-\log p(\bm\beta)} \right)^\theta.
\]
To implement the delta method, one needs the partial derivatives of $g(\theta, \beta)$ with respect to the elements of the parameter vector $\beta$ and with respect to $\theta$.

For the $h-th$ element $\beta_h$ of $\bm\beta$,
\[
\frac{\partial \log(p(\bm\beta))}{\partial \beta_h} = \frac{\partial}{\partial{\beta_h}} \left[ \bm\beta' {\bm x} - \log \left( 1 + {\rm e}^{\bm\beta' {\bm x}} \right)  \right] = x_h - \frac{{\rm e}^{\bm\beta' {\bm x}} \, x_h}{1+{\rm e}^{\bm\beta' {\bm x}}} = (1-p(\bm\beta)) x_h,
\]
so that $\nabla_{\bm\beta} \log p(\bm\beta) = \left(1-p(\bm\beta)\right){\bm x}$.
Then
\begin{eqnarray*}
\nabla_{\bm\beta} \, g(\theta, \bm\beta)  &=& \theta \left( \frac{\theta}{\theta - \log p(\bm\beta)}  \right)^{\theta-1} \theta \left(  \theta - \log p(\bm\beta) \right)^{-2} \nabla_{\bm\beta} \log p(\bm\beta) = g(\theta, \bm\beta) \frac{\theta (1-p(\bm\beta))}{\theta - \log p(\bm\beta)} \bm x.
\end{eqnarray*}
Let us now turn to $\theta$. One has
\begin{align*}
 \quad \frac{\partial g(\theta, \bm\beta)}{\partial \theta} &= g(\theta, \bm\beta) \, \frac{\partial}{\partial \theta} \log \left( g(\theta, \bm\beta) \right) = g(\theta, \bm\beta) \, \frac{\partial}{\partial \theta} \left[ \theta \log \left( \frac{\theta}{\theta - \log p(\bm \beta)} \right) \right] \\
&\hspace{-0.5cm} = g(\theta, \bm\beta) \, \left[ \log \left( \frac{\theta}{\theta - \log p(\bm\beta)} \right) + \theta  \left( \frac{\theta - \log p(\bm \beta)}{\theta} \right) \left( \left( \theta - \log p(\bm \beta) \right)^{-1} - \theta \left( \theta - \log p(\bm \beta) \right)^{-2} \right) \right] \\
&\hspace{-0.5cm} = g(\theta, \bm \beta) \left[ \log \left( \frac{\theta}{\theta - \log p(\bm\beta)} \right)  - \frac{\log p(\bm \beta)}{\theta - \log p(\bm \beta)} \right].
\end{align*}
Now, let
\begin{align*}
M &= \left[  
\begin{array}{c} \nabla_{\bm\beta} \, g(\theta, \bm\beta)  \\ \ \\ \frac{\partial g(\theta, \bm\beta)}{\partial \theta}  \end{array} \right] = g(\theta, \bm\beta) \left[ \begin{array}{c}  \frac{\theta (1-p(\bm\beta))}{\theta - \log p(\bm\beta)} \bm x \\ \ \\  \log \left( \frac{\theta}{\theta - \log p(\bm\beta)} \right)  - \frac{\log p(\bm \beta)}{\theta - \log p(\bm \beta)} \end{array} \right] \\ \ \\
&= \pi_M(\bm x) \left[ \begin{array}{c}  \frac{\theta (1-p(\bm\beta))}{\theta - \log p(\bm\beta)} \bm x \\ \ \\  \log \left( \frac{\theta}{\theta - \log p(\bm\beta)} \right)  - \frac{\log p(\bm \beta)}{\theta - \log p(\bm \beta)} \end{array} \right],
\end{align*}
and $\widehat{M}$ the vector $M$ with plugged-in estimates $(\widehat{\theta}, \widehat{\bm \beta}')'$. For a general function $u(\bm x_{ij}, \bm \beta)$, the vector $M$ is
\begin{align*}
 M=&\pi_M(\bm x)
\left[
\begin{array}{c} 
-\dfrac{\theta}{\theta+u(\bm x_{ij}, \bm \beta)}\bm x\\ \ \\
 \log\left(\dfrac{\theta}{\theta+u(\bm x_{ij}, \bm \beta)}\right)+\dfrac{u(\bm x_{ij}, \bm \beta)}{\theta+u(\bm x_{ij}, \bm \beta)}
 \end{array}
\right].
\end{align*}

If we denote by $\widehat{\Sigma}$ the
$(b+1)\times(b+1)$-dimensional variance-covariance matrix of $(\widehat{\bm{\beta}}',\widehat{\theta})'$, then the consistently estimated variance of $\widehat{\pi}_M(\bm x) = g( \widehat{\bm \beta}, \widehat{\theta})$ is

\begin{align}
& \hspace{-.3cm} \widehat{\rm Var}[\widehat{\pi}_M(\bm x)] 
= \widehat{M}' \, \widehat{\Sigma} \, \widehat{M} \nonumber \\
& = 
 (\widehat{\pi}_M(\bm x))^2  
\left[ \! \! \!  \begin{array}{cc}  \frac{\widehat{\theta} (1-p(\widehat{\bm\beta}))}{\widehat{\theta} - \log p(\widehat{\bm\beta})} \bm x' ,\, \log \left( \frac{\widehat{\theta}}{\widehat{\theta} - \log p(\widehat{\bm\beta})} \right)  - \frac{\log p(\widehat{\bm \beta})}{\widehat{\theta} - \log p(\widehat{\bm \beta})} \end{array} \! \! \!  \right] 
\widehat{\Sigma} 
\left[  \! \! \!  \begin{array}{c}  \frac{\widehat{\theta} (1-p(\widehat{\bm\beta}))}{\widehat{\theta} - \log p(\widehat{\bm\beta})} \bm x \\ \ \\  \log \left( \frac{\widehat{\theta}}{\widehat{\theta} - \log p(\widehat{\bm\beta})} \right)  - \frac{\log p(\widehat{\bm \beta})}{\widehat{\theta} - \log p(\widehat{\bm \beta})} \end{array} \! \! \! \right].    \label{varpiM}   
\end{align}

One may use this to construct confidence intervals for $\pi_M(\bm x)$. To avoid having the extremes of the interval fall outside the $[0,1]$ range, the delta method can be applied to the logit transformation
\[
\xi = \operatorname{logit}({\pi}_M(\bm x))=\log\left\{\frac{{\pi}_M(\bm x)}{1-{\pi}_M(\bm x)}\right\}.
\]
Given an estimate $\widehat{\pi}_M(\bm x)$ with asymptotic variance $\widehat{\rm Var}[\widehat{\pi}_M(\bm x)]$ as in (\ref{varpiM}), the delta method gives
\[
\widehat{\rm Var}[\widehat{\xi}]
\approx
\left\{\frac{1}{\widehat{\pi}_M(\bm x)(1-\widehat{\pi}_M(\bm x))}\right\}^{2}
\widehat{\rm Var}[\widehat{\pi}_M(\bm x)]
\]
as well as the asymptotic normality of $\sqrt{n} \left( \widehat{\xi} - \xi \right)$, which justifies the construction of the approximate level $1-\alpha$ for $\xi$ as
\[
\xi_L =
\widehat{\xi}
-
z_{1-\alpha/2}
\sqrt{\widehat{\rm Var}[\widehat{\xi}]},
\qquad
\xi_U =
\widehat{\xi}
+
z_{1-\alpha/2}
\sqrt{\widehat{\rm Var}[\widehat{\xi}]}.
\]
Finally, the confidence interval on the probability scale can then be obtained by back-transformation as
\[
\left[
\frac{\exp(\xi_L)}{1+\exp(\xi_L)},
\frac{\exp(\xi_U)}{1+\exp(\xi_U)}
\right],
\]
which will be contained in the $(0,1)$ interval.

\subsection{Evaluation of $c_{BM}$ and $c_{BC}$ under the Lehmann binary model}
\label{app:cb}\label{app:app3}

\paragraph{Evaluation of $c_{BM}$.} We establish the between-cluster comparisons introduced in Section~\ref{sec:simulation-study}. Consider a randomly selected inter-cluster pair with observation $j$ from cluster $i$ 
and observation $q$ from cluster $p \neq i$, for which $y_{ij}=1$ and $y_{pq}=0$. The $c_{BM}$ index is defined as (Lesaffre, 2012)
\begin{align}
c_{BM}
=E_{Y_{ij},Y_{pq},\bm X_{ij},\bm X_{pq},\gamma_j,\gamma_p}
\left[
I\left\{
\pi_M(\bm X_{ij}) > \pi_M(\bm X_{pq}) \right\}
\,\middle|\,
Y_{ij}=1 \text{ and } Y_{pq}=0
\right]
\nonumber
\end{align}
where $I(\cdot)$ is the usual indicator function. 
The evaluation of $c_{BM}$ for this pair involves the evaluation of whether
$\pi_M(\bm x_{ij}) > \pi_M(\bm x_{pq})$. For the binary mixed model, this can also be written as
\[
\pi_M(\bm x_{ij}) > \pi_M(\bm x_{pq})
\Longleftrightarrow
E_\gamma\left\{
h^{-1}\left(\gamma_i+\bm\beta'\bm x_{ij}\right)
\right\}
>
E_\gamma\left\{
h^{-1}\left(\gamma_p+\bm\beta'\bm x_{pq}\right)
\right\}
\Longleftrightarrow
\bm\beta'\bm x_{ij} > \bm\beta'\bm x_{pq},
\]
since $h^{-1}(\cdot)$ is a positive monotone function and $\gamma_i$ and $\gamma_p$ are the {\em i.i.d.} $N(0,\sigma^2)$ random intercepts (Lesaffre, 2012). Let us recall here that while in this model a large value for $\gamma$ corresponds to an increased probability that $Y=1$, in the Lehmann binary model a large $\gamma$ (which we have therefore called frailty) corresponds to a decreased probability that $Y=1$.

{\bf Remark.} For the Lehmann binary mixed model, one has that for fixed $\theta >0$, 
\begin{equation*}
    \pi_M({\bm x}) = \left(    \frac{\theta}{\theta - \log \left( \frac{{\rm e}^{\bm \beta' \bm x}}{1+ {\rm e}^{\bm \beta' \bm  x}}  \right)}  \right)^\theta
\end{equation*} 
is a positive monotone function of the argument $\eta = \beta' {\bm x}$. 

Indeed, letting $ p(\eta)=\exp(\eta) / \left( 1+\exp(\eta) \right)$ one can re-write $\pi_M(\eta) = \left(
\frac{\theta}{\theta-\log (p(\eta))}
\right)^\theta$, one has
\begin{eqnarray}
\frac{d\pi_M(\eta)}{d\eta}
&=& \frac{\partial \pi_M(\eta)}{\partial p(\eta)}
\frac{dp(\eta)}{d\eta} =
\frac{\theta^{\theta+1}}
{
p(\eta)\left[\theta-\log\{p(\eta)\}\right]^{\theta+1}
} p(\eta)\{1-p(\eta)\}  \nonumber \\
&=& \frac{\theta^{\theta+1}\{1-p(\eta)\}}
{
\left[\theta-\log\{p(\eta)\}\right]^{\theta+1}
}  = \frac{
\theta^{\theta+1}
}
{
\{1+\exp(\eta)\}
\left[
\theta-\log\left\{
\frac{\exp(\eta)}{1+\exp(\eta)}
\right\}
\right]^{\theta+1}} > 0  \label{monotone}
\end{eqnarray}
since $\theta>0$ and 
$ \log\left\{
\frac{\exp(\eta)}{1+\exp(\eta)}
\right\}<0$.

Thus, also for the Lehmann binary model, $c_{BM}$ can equivalently be defined as
\begin{align*}
c_{BM} =
E_{Y_{ij},Y_{pq},\bm X_{ij},\bm X_{pq},\gamma_j,\gamma_p}
\left[
I\left\{
\bm\beta'\bm X_{ij} > \bm\beta'\bm X_{pq}
\right\}
\,\middle|\,
Y_{ij}=1 \text{ and } Y_{pq}=0
\right].
\end{align*}
\paragraph{Evaluation of $c_{BC}$.}In order to calculate $c_{BC}$ for a pair of observations from two different clusters (say, $i$ and $p$), 
$\pi_C(\gamma_i,\bm x_{ij}) > \pi_C(\gamma_p,\bm x_{pq})$ needs to be evaluated. 
In the traditional binary mixed model, this can also be reduced to a comparison of the linear predictors plus the frailty term :
\begin{align*}
c_{BC}
&=
E_{Y_{ij},Y_{pq},\bm X_{ij},\bm X_{pq},\gamma_j,\gamma_p}
\left[
I\left\{
\pi_C(\gamma_i,\bm X_{ij}) > \pi_C(\gamma_p,\bm X_{pq})
\right\}
\,\middle|\,
Y_{ij}=1 \text{ and } Y_{pq}=0
\right]
\nonumber\\
&=
E_{Y_{ij},Y_{pq},\bm X_{ij},\bm X_{pq},\gamma_j,\gamma_p}
\left[
I\left\{
\bm\beta'\bm X_{ij} + \gamma_j
>
\bm\beta'\bm X_{qp} + \gamma_q
\right\}
\,\middle|\,
Y_{ji}=1 \text{ and } Y_{qp}=0
\right].
\end{align*}
Note that for the Lehmann binary model this comparison cannot be reduced to the comparison of the linear predictors plus the frailty terms, but one should instead keep the comparison in terms of the probabilities:
\[
\pi_C(\gamma_i,\bm x_{ij}) > \pi_C(\gamma_p,\bm x_{pq})
\Longleftrightarrow
\left( \frac{{\rm e}^{\bm\beta'\bm x_{ij}}}{1+ {\rm e}^{\bm \beta'\bm x_{ij}}}  \right)^{\gamma_i}
>
\left( \frac{{\rm e}^{\bm\beta'\bm x_{pq}}}{1+ {\rm e}^{\bm \beta'\bm x_{pq}}}  \right)^{\gamma_p} \Longleftrightarrow \frac{\gamma_i}{\gamma_p} < \frac{\log \left( \pi_C(1;{\bm x}_{pq}) \right)}{\log \left( \pi_C(1;{\bm x}_{ij}) \right)},
\]
where $\pi_C(1;{\bm x}_{ij}) = \frac{{\rm e}^{\beta' {\bm x}_{ij}}}{1+{\rm e}^{\beta' {\bm x}_{ij}}}$ represents the probability of $Y=1$ for a cluster with average frailty. Hence for the Lehmann binary model one has
\[
c_{BC}
=
E_{Y_{ij},Y_{pq},\bm X_{ij},\bm X_{pq},\gamma_j,\gamma_p}
\left[
I \left\{
\frac{\gamma_i}{\gamma_p} < \frac{\log \left( \pi_C(1;{\bm x}_{pq}) \right)}{\log \left( \pi_C(1;{\bm x}_{ij}) \right)}
\right\}
\,\middle|\,
Y_{ij}=1 \text{ and } Y_{pq}=0
\right].
\nonumber
\]

Importantly, it is easy to check that for both models, the same monotonicity of $\pi_C(1;{\bm x}_{ij})$ and of the random effect (frailty) on the concordance holds. Indeed, when the frailty or random effect favors cluster $i$ vs. cluster $p$ {\em and} the same happens for $\pi_C(1;{\bm x}_{ij})$ vs. $\pi_C(1;{\bm x}_{pq})$, then concordance always occurs. Similarly, when the disadvantage of $i$ vs. $p$ occurs both on the frailty and on the $\pi_C(1;{\bm x}_{ij})$ level, then concordance is never true. For the other cases, the inequality must be assessed based on the frailty and $\pi_C(1;{\bm x}_{ij})$ values.

\subsection{Equality of $c_{WC}$ and $c_{WM}$ under the Lehmann binary model}
\label{app:cw}\label{app:app4}
\paragraph{Evaluation of $c_{WC}$.} This appendix proves the equality stated in Section~\ref{sec:simulation-study}. Consider a randomly selected intra-cluster pair, observation $j$ and observation $p$ belonging to same cluster $i$, for which $y_{ij}=1$ and $y_{ip}=0$.
In order to calculate $c_{WC}$ for such a pair, 
$\pi_C(\gamma_i,\bm x_{ij}) > \pi_C(\gamma_i,\bm x_{ip})$ needs to be evaluated. 
In the binary mixed model, this can be reduced to a comparison of the linear predictors only (Lesaffre, 2012):
\[
\pi_C(\gamma_i,\bm x_{ij}) > \pi_C(\gamma_i,\bm x_{ip})
\Longleftrightarrow
h^{-1}\left(\gamma_i+\bm\beta'\bm x_{ij}\right)
>
h^{-1}\left(\gamma_i+\bm\beta'\bm x_{ip}\right)
\Longleftrightarrow
\bm\beta'\bm x_{ij} > \bm\beta'\bm x_{ip},
\]
since $h^{-1}(\cdot)$ is a positive monotone function and since observation $j$ 
and observation $p$ belong to the same cluster. We note that the same result holds also for the Lehmann binary model. Indeed
\[
\pi_C(\gamma_i,\bm x_{ij}) > \pi_C(\gamma_i,\bm x_{ip})
\Longleftrightarrow
\left( \frac{{\rm e}^{\bm\beta'\bm x_{ij}}}{1+ {\rm e}^{\bm \beta'\bm x_{ij}}}  \right)^{\gamma_i}
>
\left( \frac{{\rm e}^{\bm\beta'\bm x_{ip}}}{1+ {\rm e}^{\bm \beta'\bm x_{ip}}}  \right)^{\gamma_i}
\Longleftrightarrow
\bm\beta'\bm x_{ij} > \bm\beta'\bm x_{ip}.
\]

\paragraph{Evaluation of $c_{WM}$.} The evaluation of $c_{WM}$ for  observation $j$ and $p$ follows the same logic of the evaluation of $c_{BM}$ since it involves marginal probabilities. In the binary mixed model, the comparison between a pair of observation within the same cluster can be reduced to a comparison of the linear predictors only (Lesaffre, 2012)
\[
\pi_M(\bm x_{ij}) > \pi_M(\bm x_{ip})
\Longleftrightarrow
E_\gamma\left\{
h^{-1}\left(\gamma_i+\bm\beta'\bm x_{ij}\right)
\right\}
>
E_\gamma\left\{
h^{-1}\left(\gamma_p+\bm\beta'\bm x_{ip}\right)
\right\}
\Longleftrightarrow
\bm\beta'\bm x_{ij} > \bm\beta'\bm x_{ip}.
\]
By the monotonicity remark (\ref{monotone}), the same important result holds for the Lehmann binary model. 
Thus
\[
c_{WM} = c_{WC} = E_{Y_{ij},Y_{ip},\bm X_{ij},\bm X_{ip},\gamma}
\left[
I\left\{
\bm\beta'\bm X_{ij}
>
\bm\beta'\bm X_{ip}
\right\}
\,\middle|\,
Y_{ij}=1 \text{ and } Y_{ip}=0
\right]
\nonumber,
\]
is also true for the Lehmann binary model.

\subsection{Analytical relationship with the normal random-intercept GLMM}\label{app:glmm}

This appendix derives the transformation in~\eqref{eq:exactfgamma} and the density used in Supplementary Section~\ref{sup:comp}. As in Section~\ref{sec:comp}, $\alpha\sim N(0,\sigma^2)$ denotes the GLMM random intercept and $\gamma$ denotes the BLMM frailty. Fix $\bm{x}$ and let $r(\bm{x})=\bm{\beta}'\bm{x}$. Equating the two conditional probabilities gives
\begin{align*}
\frac{e^{r(\bm{x})+\alpha}}{1+e^{r(\bm{x})+\alpha}}
&=\left\{\frac{e^{r(\bm{x})}}{1+e^{r(\bm{x})}}\right\}^{\gamma} \Longleftrightarrow\quad\\
r(\bm{x})+\alpha-\log\{1+e^{r(\bm{x})+\alpha}\}
&=\gamma\left[r(\bm{x})-\log\{1+e^{r(\bm{x})}\}\right] \Longleftrightarrow\quad\\
\gamma(\alpha)
&=\frac{r(\bm{x})+\alpha-\log\{1+e^{r(\bm{x})+\alpha}\}}
{r(\bm{x})-\log\{1+e^{r(\bm{x})}\}}.
\end{align*}
This is equivalent to~\eqref{relationship}. In the GLMM notation used locally for this comparison, it is also
\[
\gamma(\alpha)=\frac{\log\{\pi_C(\alpha;\bm{x})\}}{\log\{\pi_C(0;\bm{x})\}}.
\]
The baseline probability in this display is evaluated at a zero GLMM intercept; it equals the BLMM probability at frailty one for the same $\bm{\beta}$ and $\bm{x}$.

Write $r=r(\bm{x})$ and $p=\pi_C(0;\bm{x})=e^r/(1+e^r)$ in the GLMM. The induced positive random variable is denoted by $\Gamma^*$. Since $\log p<0$ and the GLMM success probability is increasing in $\alpha$, the transformation $\gamma(\alpha)$ is strictly decreasing. Its inverse and derivative are
\[
\alpha(\gamma)=\log\left(\frac{p^\gamma}{1-p^\gamma}\right)-r,
\qquad
\alpha'(\gamma)=\frac{\log p}{1-p^\gamma},
\qquad\gamma>0.
\]
The normal density is
\[
f_\alpha(a)=\frac{1}{\sigma\sqrt{2\pi}}\exp\left(-\frac{a^2}{2\sigma^2}\right).
\]
Applying the change-of-variable formula yields
\begin{equation}\label{eq:gstar}
\begin{aligned}
f_{\Gamma^*}(\gamma)
&=f_\alpha\{\alpha(\gamma)\}|\alpha'(\gamma)|\\
&=\frac{-\log p}{(1-p^\gamma)\sigma\sqrt{2\pi}}
\exp\left[-\frac{1}{2\sigma^2}
\left\{\log\left(\frac{p^\gamma}{1-p^\gamma}\right)-r\right\}^{2}\right],
\qquad\gamma>0.
\end{aligned}
\end{equation}
The density depends on $r(\bm{x})$ through both $r$ and $p$. It describes the frailty transformation needed to reproduce the GLMM probability at a fixed covariate vector. The numerical comparison with the gamma frailty family is reported in Supplementary Section~\ref{sup:comp}, Figure~\ref{fig:cmp}, and Table~\ref{tabequiv}.

\clearpage

\section*{Supplementary material}
\addcontentsline{toc}{section}{Supplementary materials}
\setcounter{subsection}{0}
\setcounter{subsubsection}{0}
\setcounter{equation}{0}
\setcounter{figure}{0}
\setcounter{table}{0}
\renewcommand{\thesubsection}{S.\arabic{subsection}}
\renewcommand{\theHsubsection}{supp.\arabic{subsection}}
\renewcommand{\theHsubsubsection}{\theHsubsection.\arabic{subsubsection}}
\renewcommand{\theHequation}{supp.\arabic{subsection}.\arabic{equation}}
\renewcommand{\theHfigure}{supp.\arabic{subsection}.\arabic{figure}}
\renewcommand{\theHtable}{supp.\arabic{subsection}.\arabic{table}}

\subsection{Numerical comparison of the induced and gamma frailty distributions}\label{sup:comp}

Table~\ref{tabequiv} reports the numerical results. The distance between the induced frailty distribution and the variance-matched gamma distribution generally increases with heterogeneity. Differences are greater for larger baseline probabilities, particularly above $0.5$. The Q--Q plots (not shown) also indicated more pronounced tail differences at larger baseline probabilities. These findings provide the numerical context for the distinction between the model specifications discussed in Section~\ref{sec:comp}.

\input{table_portrait_all_results.tex}
\FloatBarrier

\subsection{Effect of gamma frailty quantiles on success probabilities}\label{sup:quart}

For the BLMM in~\eqref{eq:pc}, a frailty value $\gamma$ changes the mean-frailty probability $\pi_C(1;\bm{x})$ into $\{\pi_C(1;\bm{x})\}^{\gamma}$. Table~\ref{tab:quart} illustrates this transformation using the first, second, and third quartiles of $\operatorname{Gamma}(\theta,\theta)$ for the values of $\theta$ considered in Section~\ref{sec:simulation-study}. The columns labelled $Q_3$, $Q_2$, and $Q_1$ report the transformed success probabilities at the corresponding frailty quartiles. Their ordering reflects the decrease in success probability as the frailty increases. This table supplements the interpretation of the BLMM frailty in Section~\ref{sec:blmm}.

\begin{table}  
\centering
\caption{Modified probabilities $\pi_C(\gamma; {\bm x})$ for different baseline probabilities $\pi_C(1; {\bm x})$ and quartiles of the Gamma$(\theta, \theta)$ frailty distribution, for different values of the heterogeneity parameter $\theta$.}
\label{tab:quart}\label{tab:baseline-prob-theta-split}
\normalsize
\setlength{\tabcolsep}{4pt}
\renewcommand{\arraystretch}{1.15}
\label{tabmodif}
\begin{adjustbox}{max width=\textwidth}
\begin{tabular}{c *{4}{ccc@{\hspace{1.2em}}} ccc}
\toprule
& \multicolumn{3}{c}{$\pi_C(1; {\bm x})=0.1$}
& \multicolumn{3}{c}{$\pi_C(1; {\bm x})=0.2$}
& \multicolumn{3}{c}{$\pi_C(1; {\bm x})=0.3$}
& \multicolumn{3}{c}{$\pi_C(1; {\bm x})=0.4$}
& \multicolumn{3}{c}{$\pi_C(1; {\bm x})=0.5$} \\
\cmidrule(lr){2-4}
\cmidrule(lr){5-7}
\cmidrule(lr){8-10}
\cmidrule(lr){11-13}
\cmidrule(lr){14-16}
$\theta$
& $Q_3$ & $Q_2$ & $Q_1$
& $Q_3$ & $Q_2$ & $Q_1$
& $Q_3$ & $Q_2$ & $Q_1$
& $Q_3$ & $Q_2$ & $Q_1$
& $Q_3$ & $Q_2$ & $Q_1$  \\
\midrule
0.5
& 0.05 & 0.35 & 0.79
& 0.12 & 0.48 & 0.85
& 0.20 & 0.58 & 0.88
& 0.30 & 0.66 & 0.91
& 0.40 & 0.73 & 0.93 \\
1
& 0.04 & 0.20 & 0.52
& 0.11 & 0.33 & 0.63
& 0.19 & 0.43 & 0.71
& 0.28 & 0.53 & 0.77
& 0.38 & 0.62 & 0.82 \\
2
& 0.05 & 0.14 & 0.33
& 0.11 & 0.26 & 0.46
& 0.20 & 0.36 & 0.56
& 0.29 & 0.46 & 0.64
& 0.39 & 0.56 & 0.72 \\
5
& 0.06 & 0.12 & 0.21
& 0.13 & 0.22 & 0.34
& 0.22 & 0.32 & 0.44
& 0.32 & 0.42 & 0.54
& 0.42 & 0.52 & 0.63 \\
\bottomrule
\end{tabular}
\end{adjustbox}

\vspace{0.8em}

\begin{adjustbox}{max width=\textwidth}
\begin{tabular}{c *{3}{ccc@{\hspace{1.2em}}} ccc}
\toprule
& \multicolumn{3}{c}{$\pi_C(1; {\bm x})=0.6$}
& \multicolumn{3}{c}{$\pi_C(1; {\bm x})=0.7$}
& \multicolumn{3}{c}{$\pi_C(1; {\bm x})=0.8$}
& \multicolumn{3}{c}{$\pi_C(1; {\bm x})=0.9$} \\
\cmidrule(lr){2-4}
\cmidrule(lr){5-7}
\cmidrule(lr){8-10}
\cmidrule(lr){11-13}
$\theta$
& $Q_3$ & $Q_2$ & $Q_1$
& $Q_3$ & $Q_2$ & $Q_1$
& $Q_3$ & $Q_2$ & $Q_1$
& $Q_3$ & $Q_2$ & $Q_1$ \\
\midrule
0.5
& 0.51 & 0.79 & 0.95
& 0.62 & 0.85 & 0.96
& 0.74 & 0.90 & 0.98
& 0.87 & 0.95 & 0.99 \\
1
& 0.49 & 0.70 & 0.86
& 0.61 & 0.78 & 0.90
& 0.73 & 0.86 & 0.94
& 0.86 & 0.93 & 0.97 \\
2
& 0.50 & 0.65 & 0.78
& 0.62 & 0.74 & 0.84
& 0.74 & 0.83 & 0.90
& 0.87 & 0.92 & 0.95 \\
5
& 0.53 & 0.62 & 0.71
& 0.64 & 0.72 & 0.79
& 0.76 & 0.81 & 0.86
& 0.88 & 0.91 & 0.93 \\
\bottomrule
\end{tabular}
\end{adjustbox}

\end{table}
\FloatBarrier

\subsection{Additional simulation results }\label{app:simuresults}

In this section we report on some additional results from our simulation study.

\begin{table}[htbp]
\centering
\caption{Mean outcome proportion across 100 simulated datasets
for each simulation scenario. Column headings give the training
sample size, $n_{\mathrm{train}}$.}
\label{tab:mean-outcome}

\begin{tabular}{@{}rr rrrr@{}}
\toprule
$J$ & $\theta$ & $500$ & $1000$ & $5000$ & $10000$ \\
\midrule
3 & 0.5 & 0.6794 & 0.6788 & 0.6787 & 0.6789 \\
3 & 1.0 & 0.6290 & 0.6285 & 0.6285 & 0.6285 \\
3 & 2.0 & 0.5941 & 0.5930 & 0.5951 & 0.5945 \\
3 & 5.0 & 0.5688 & 0.5714 & 0.5691 & 0.5692 \\
\addlinespace
6 & 0.5 & 0.6900 & 0.6901 & 0.6899 & 0.6905 \\
6 & 1.0 & 0.6437 & 0.6455 & 0.6433 & 0.6437 \\
6 & 2.0 & 0.6124 & 0.6118 & 0.6119 & 0.6115 \\
6 & 5.0 & 0.5874 & 0.5859 & 0.5874 & 0.5869 \\
\addlinespace
10 & 0.5 & 0.7060 & 0.7055 & 0.7067 & 0.7057 \\
10 & 1.0 & 0.6646 & 0.6629 & 0.6621 & 0.6629 \\
10 & 2.0 & 0.6341 & 0.6321 & 0.6332 & 0.6323 \\
10 & 5.0 & 0.6122 & 0.6110 & 0.6113 & 0.6114 \\
\bottomrule
\end{tabular}
\end{table}

\FloatBarrier

\begin{figure}[!htbp]
    \centering

    \begin{subfigure}{\textwidth}
        \centering
        \makebox[\textwidth][c]{%
            \includegraphics[width=0.65\textwidth]{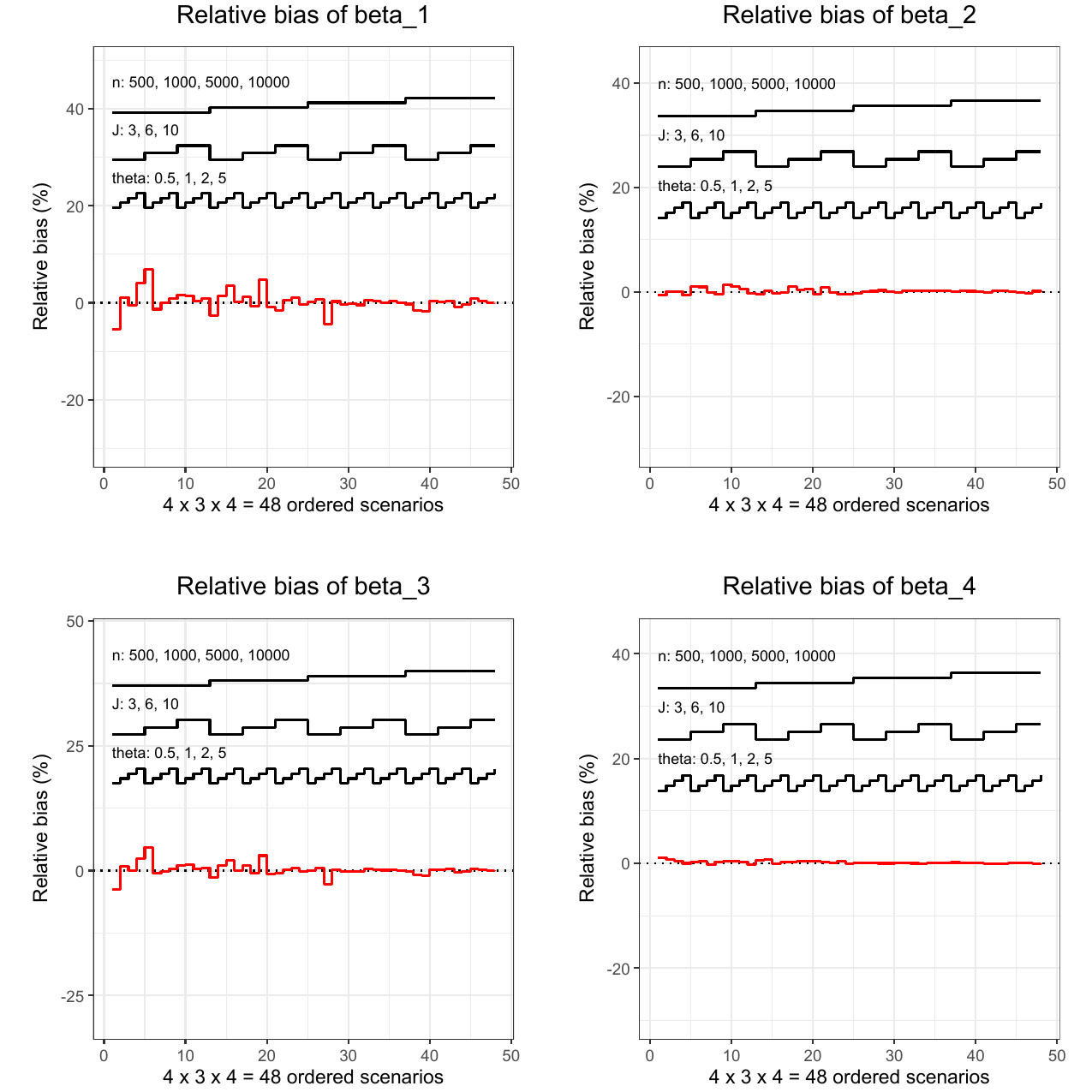}
        }
        \caption{}
        \label{fig:beta_relative_bias}
    \end{subfigure}

    \vspace{0.5cm}

    \begin{subfigure}{\textwidth}
        \centering
        \makebox[\textwidth][c]{%
            \includegraphics[width=0.7\textwidth]{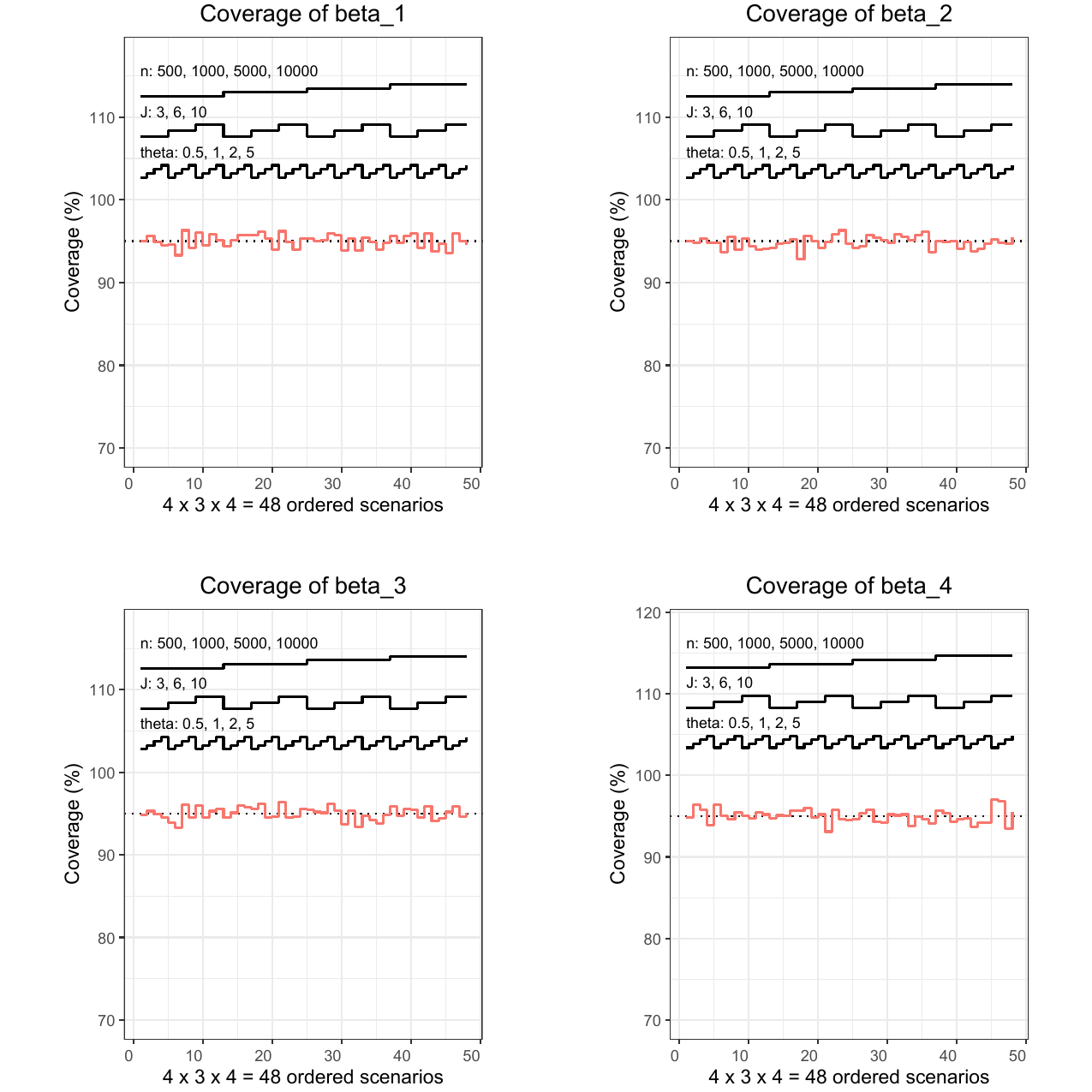}
        }
        \caption{}
        \label{fig:beta_coverage}
    \end{subfigure}

    \caption{Inferential performance of the regression coefficient estimates across the 48 simulation scenarios. Panel (a) shows the relative bias, while panel (b) shows the empirical coverage of the 95\% Wald confidence intervals. Note that $n$ corresponds to the number of cluster of the training dataset, i.e. the number of subjects.}
    \label{fig:beta_inferential_performance}
\end{figure}

\FloatBarrier

\begin{figure}[!htbp]
    \centering
    \includegraphics[width=\textwidth]{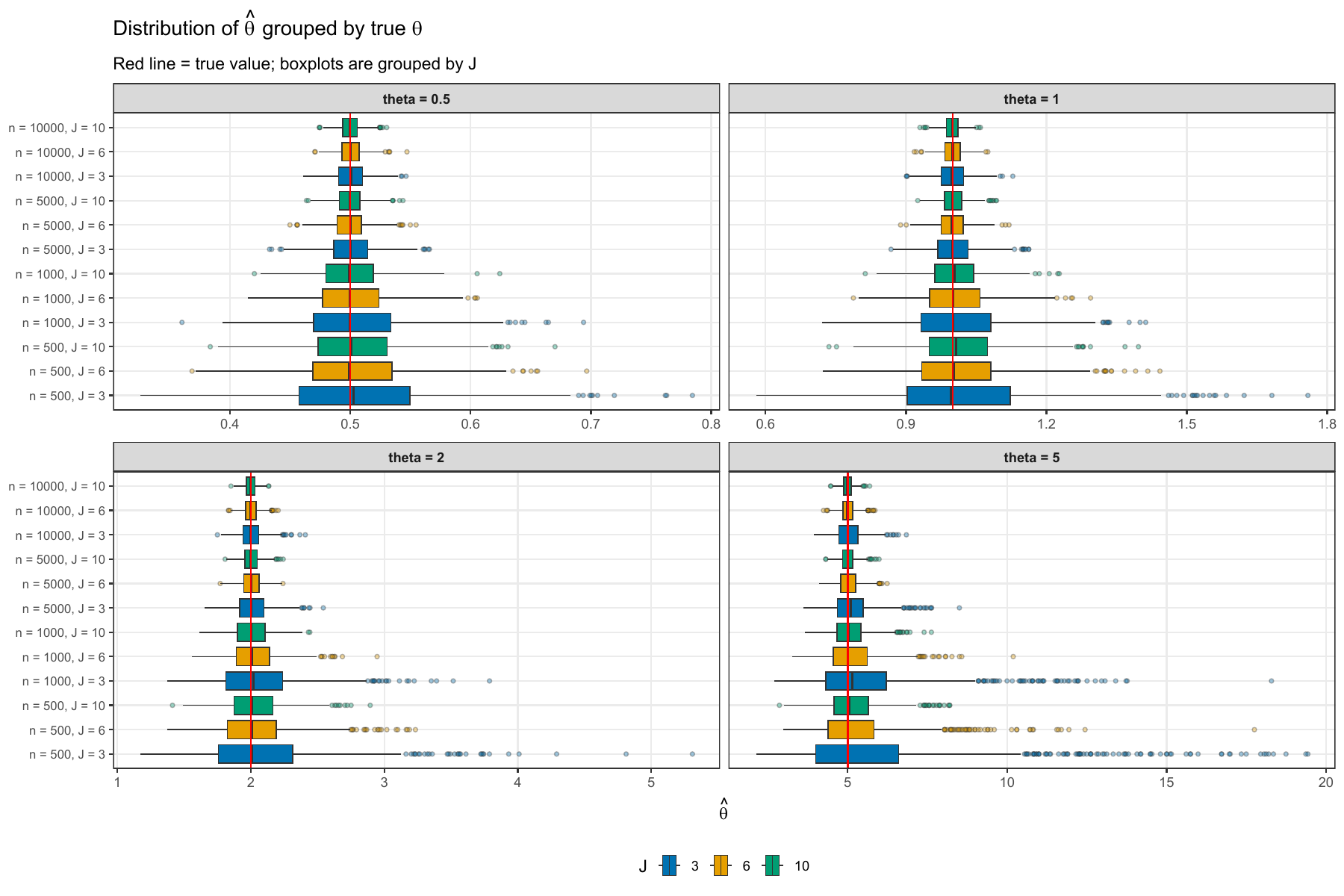}
    \caption{Distribution of $\widehat{\theta}$ by simulation scenario and true value of $\theta$. The red vertical line denotes the true value of $\theta$. Note that $n$ corresponds to the number of cluster of the training dataset, i.e. the number of subjects.}
    \label{fig:theta_boxplot}
\end{figure}

\begin{figure}[!htbp]
    \centering
    \includegraphics[width=0.8\textwidth]{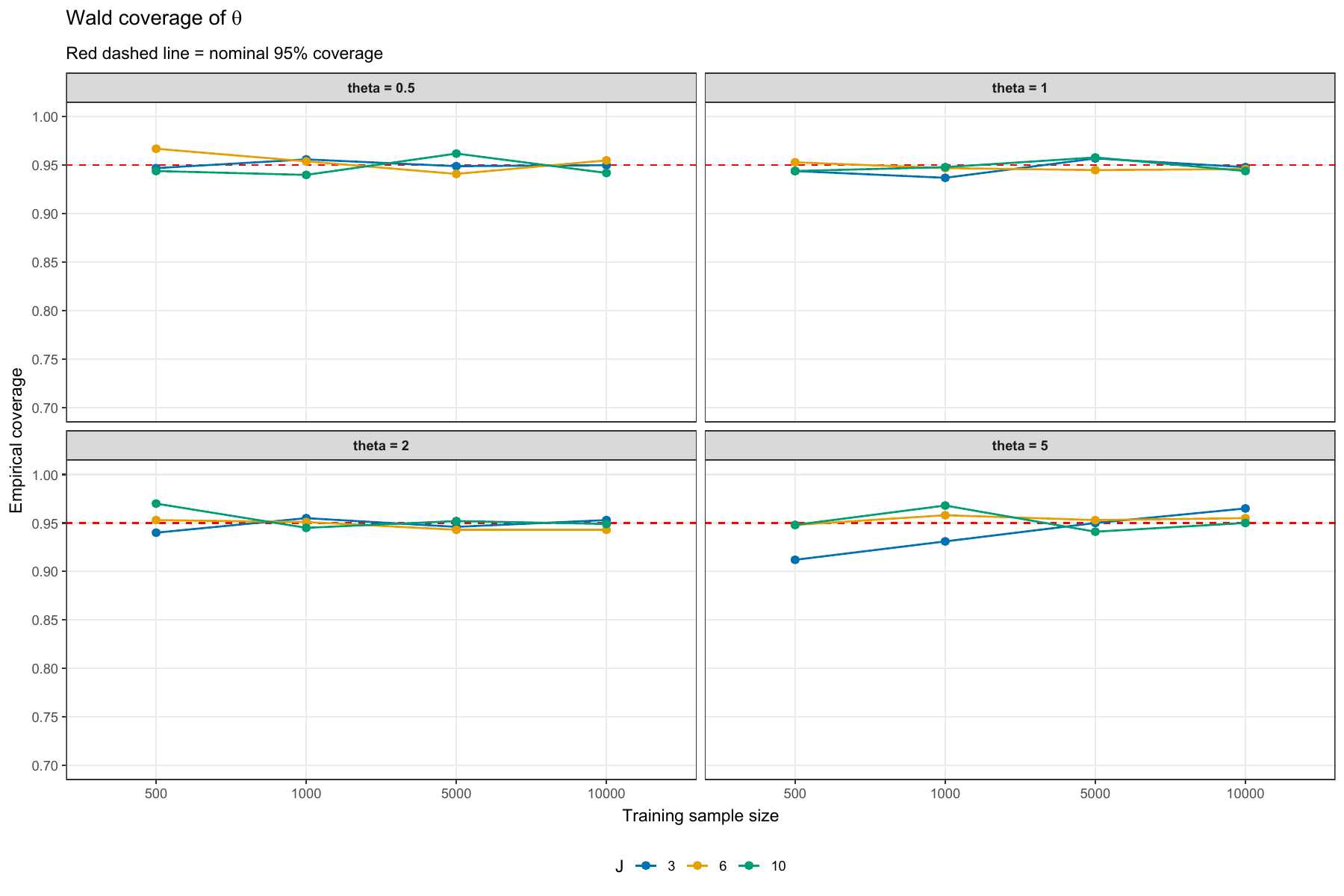}
    \caption{Empirical coverage of the 95\% Wald confidence intervals for $\theta$ across simulation scenarios. The red dashed line denotes the nominal 95\% coverage level.}
    \label{fig:theta_coverage}
\end{figure}

\begin{figure}[h]
    \centering
    \makebox[\textwidth][c]{%
    \includegraphics[width=.9\textwidth]{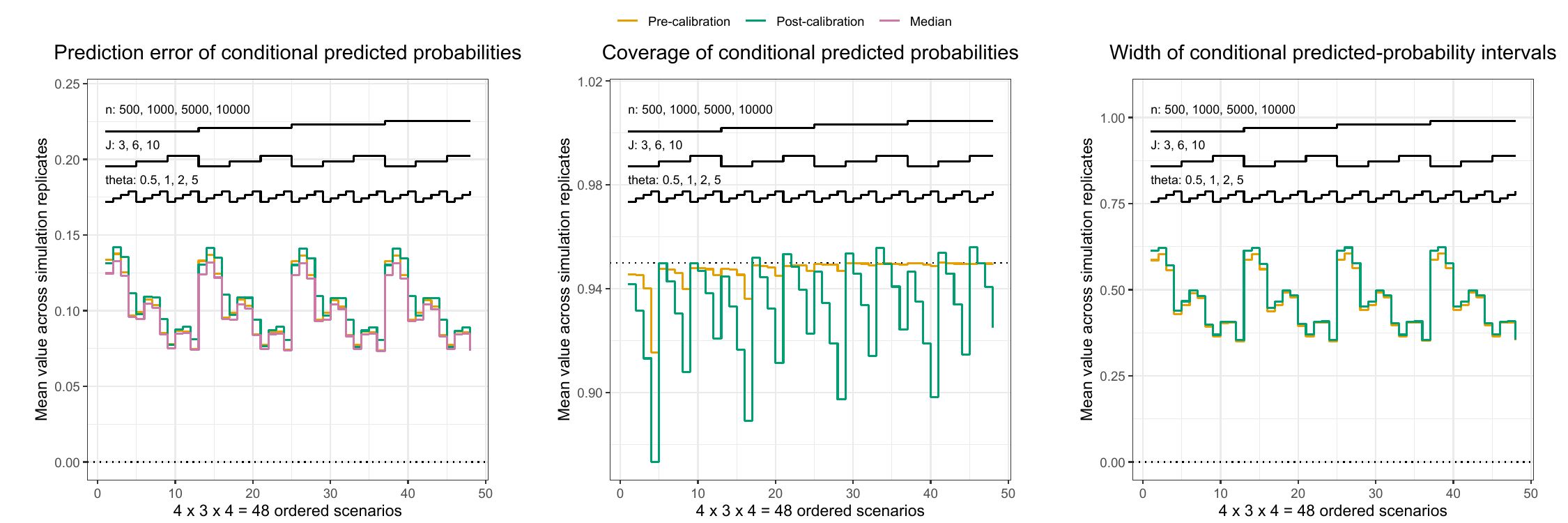}
    }
    \caption{Prediction error, empirical coverage, and interval width for the conditional predicted probabilities. We denoted the posterior mean frailty and the rank-based calibration frailty as pre-calibration and  as post-calibration respectively. Note that $n$ corresponds to the number of cluster of the training dataset, i.e. the number of subjects.}
\label{fig:conditional_probability_prediction}
\end{figure}

\begin{figure}[h]
    \centering
    \makebox[\textwidth][c]{%
    \includegraphics[width=.9\textwidth]{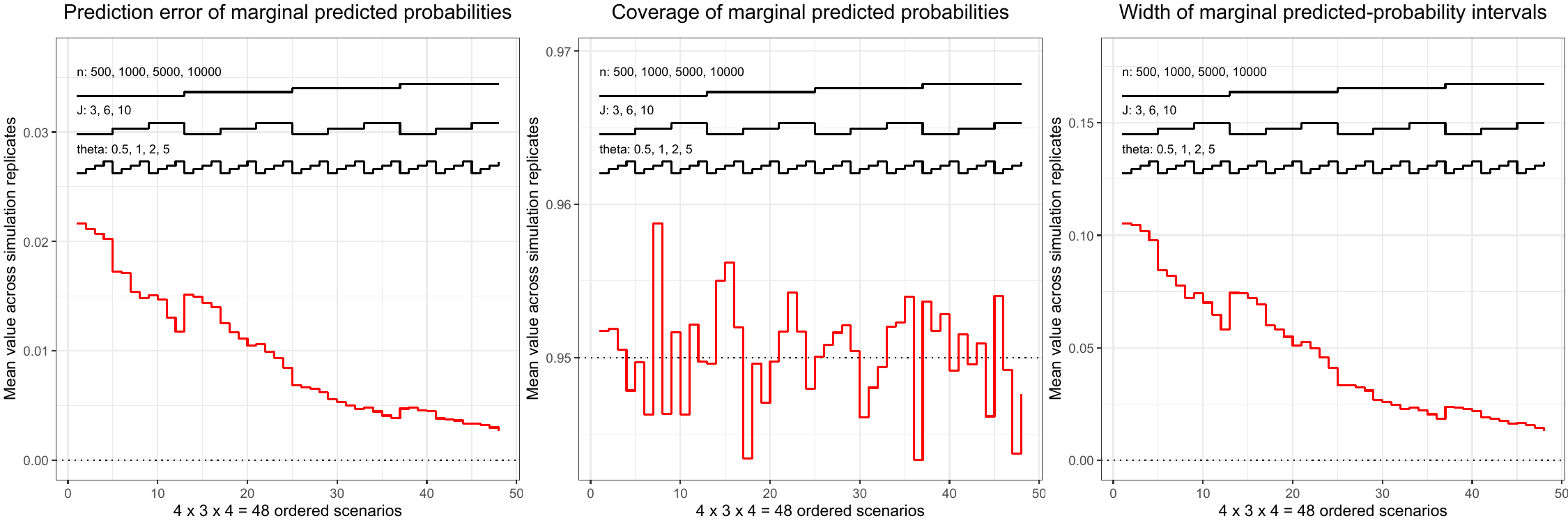}
    }
    \caption{Prediction error, empirical coverage, and interval width for the marginal predicted probabilities. Note that $n$ corresponds to the number of cluster of the training dataset, i.e. the number of subjects.}
    \label{fig:marginal_probability_prediction}
\end{figure}

\begin{figure}[!htbp]
    \centering
    \makebox[\textwidth][c]{%
    \includegraphics[width=.9\textwidth]{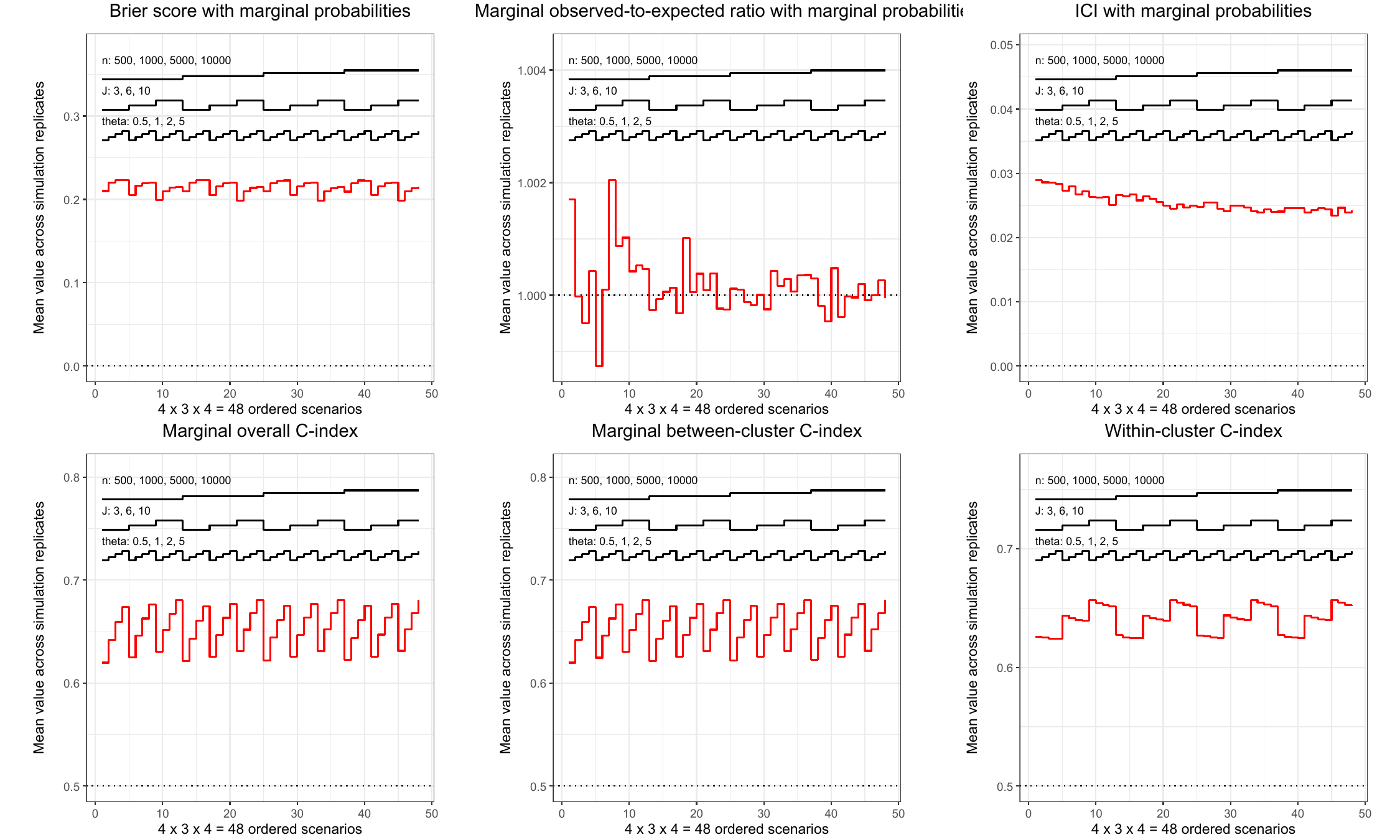}
    }
    \caption{Calibration and discrimination measures for the marginal predicted probabilities: marginal Brier score, marginal observed-to-expected ratio, marginal ICI, marginal overall C-index, marginal between-cluster C-index, and within-cluster C-index. Note that $n$ corresponds to the number of cluster of the training dataset, i.e. the number of subjects.}
    \label{fig:marginal_calibration_discrimination}
\end{figure}
\FloatBarrier

\subsubsection{Numerical equality between overall c-index and between-cluster c-index}\label{overall_between_cindex}
The overall c-index can be interpreted as a weighted combination of discrimination between observations belonging to different clusters and discrimination between observations belonging to the same cluster. In the present simulation setting, each dataset contains $n$ clusters (the individuals), each with $j$ observations (the repeated individual measures). If two observations are sampled at random from the total set of $n\times j$ observations, the probability that they belong to the same cluster is
\[
\frac{n\binom{j}{2}}{\binom{nj}{2}}
=
\frac{nj(j-1)/2}{nj(nj-1)/2}
=
\frac{j-1}{nj-1}.
\]
Therefore, the probability that the two observations belong to different clusters is
\[
1-\frac{j-1}{nj-1}.
\]

\begin{figure}[!htbp]
    \centering
    \makebox[\textwidth][c]{%
    \includegraphics[width=.7\textwidth]{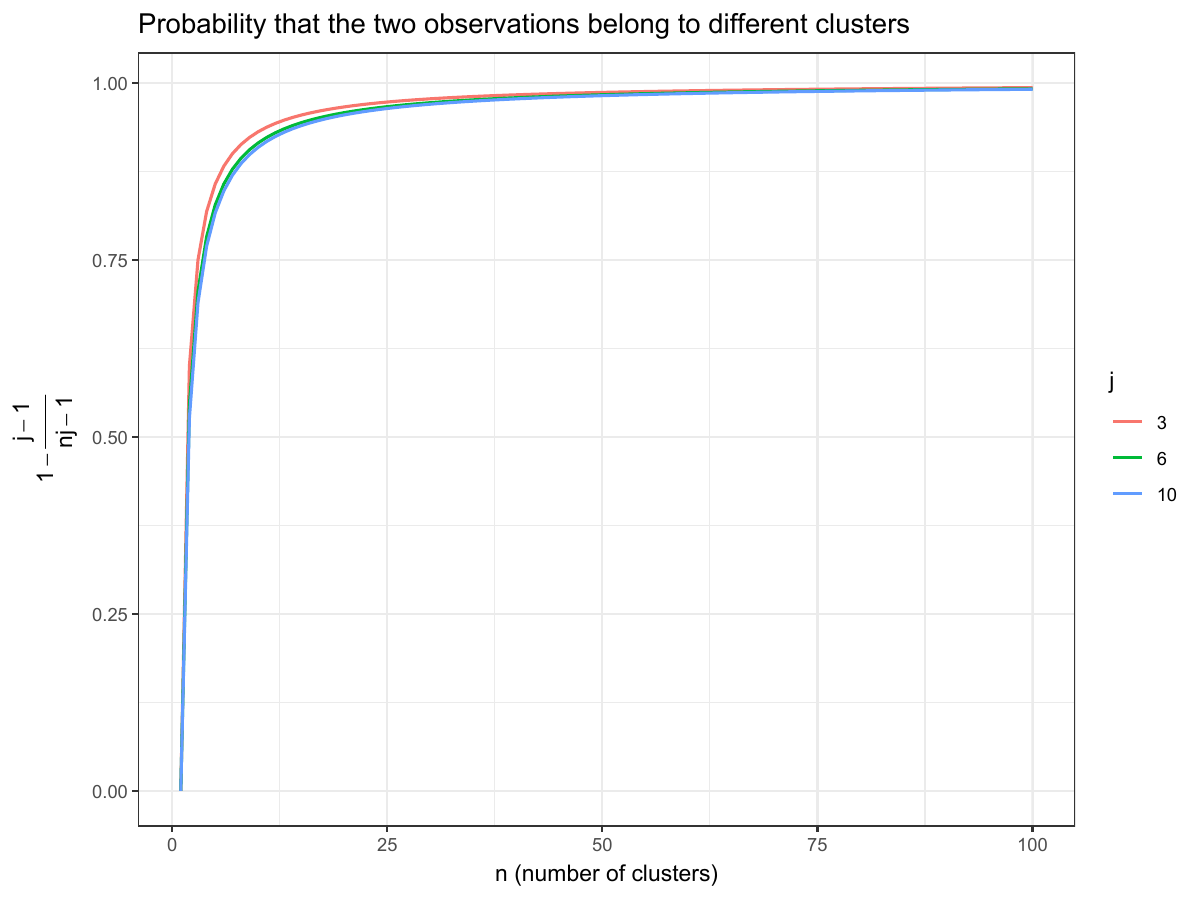}
    }
    \caption{}
    \label{fig:probbelongsclusters}
\end{figure}

For fixed $j$ and large $n$,
\[
\frac{j-1}{nj-1}
\approx
\frac{j-1}{nj}
=
\frac{1}{n}\left(1-\frac{1}{j}\right),
\]
which is of order $1/n$. Thus, even when $j$ is small, the probability of sampling two observations from the same cluster becomes negligible as $n$ increases. We plotted this probability in Figure \ref{fig:probbelongsclusters}.

When $j$ is large, this probability further simplifies to
\[
\frac{j-1}{nj-1}
\approx
\frac{1}{n}.
\]

Consequently, in the simulation scenarios considered here, where the number of clusters (subjects) is large relative to the number of repeated observations per cluster, almost all comparable pairs of observations are between-cluster pairs. This explains why the overall c-index and the between-cluster c-index are nearly identical in the simulation results: the within-cluster component contributes only a negligible fraction of all possible observation pairs.

\newpage \subsection{Output from fitting the three models Logistic Regression, BLMM, and GLMM to the Oslo Breast Cancer Screening adherence data} \label{sec:alloutputs}
In this section we report the details of the result of fitting the three models logistic regression, BLMM, and GLMM to the Oslo breast cancer screening adherence dataset.

In particular, the parameter estimates and associated standard errors are behind the confidence intervals that are shown in Figure \ref{fig:threemodels} of the main manuscript. 
\bigskip \bigskip

\begin{figure}[h!]
    \centering
    \includegraphics[width=.75\linewidth]{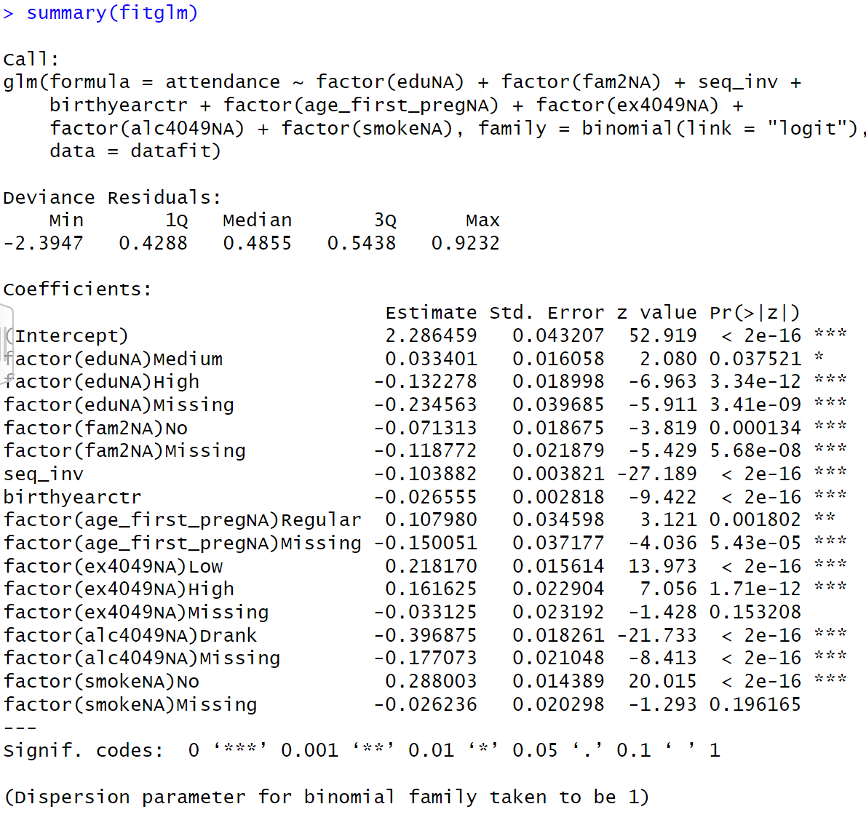}
    \caption{Fit of the logistic regression model (no random effect for risk prediction).}
    \label{fig:LogisticFit}
\end{figure}

\begin{figure}[h!]
    \centering
\includegraphics[width=.75\linewidth]{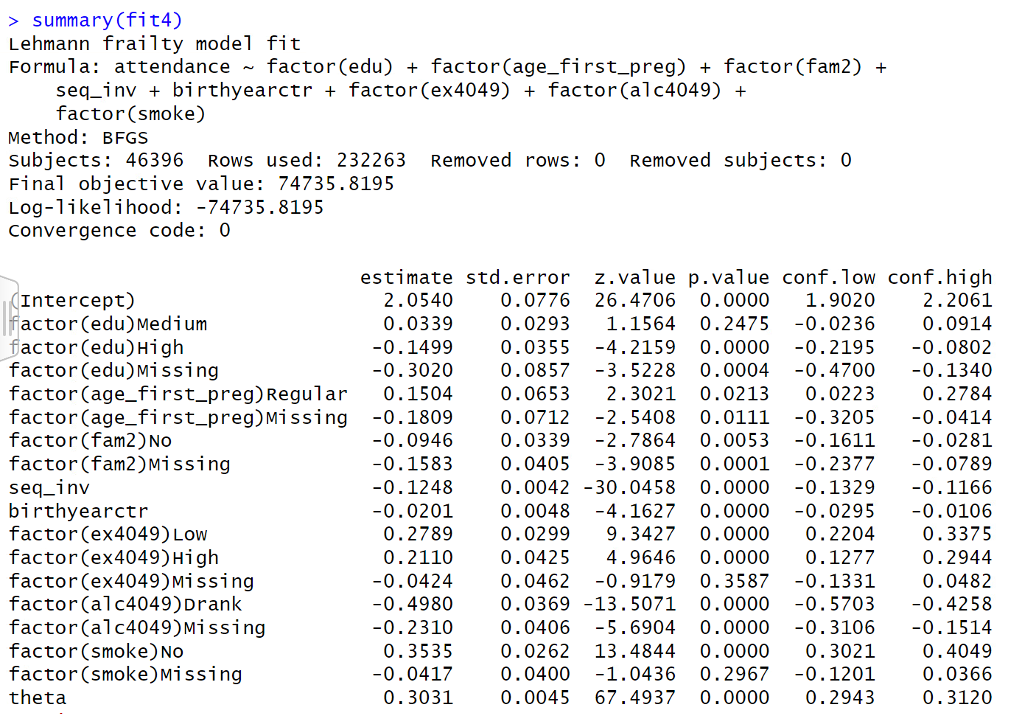}
    \caption{Fit of the BLMM model.}
    \label{fig:LehmannFit}
\bigskip \bigskip
\end{figure}

\begin{figure}[h!]
    \centering
\includegraphics[width=.75\linewidth]{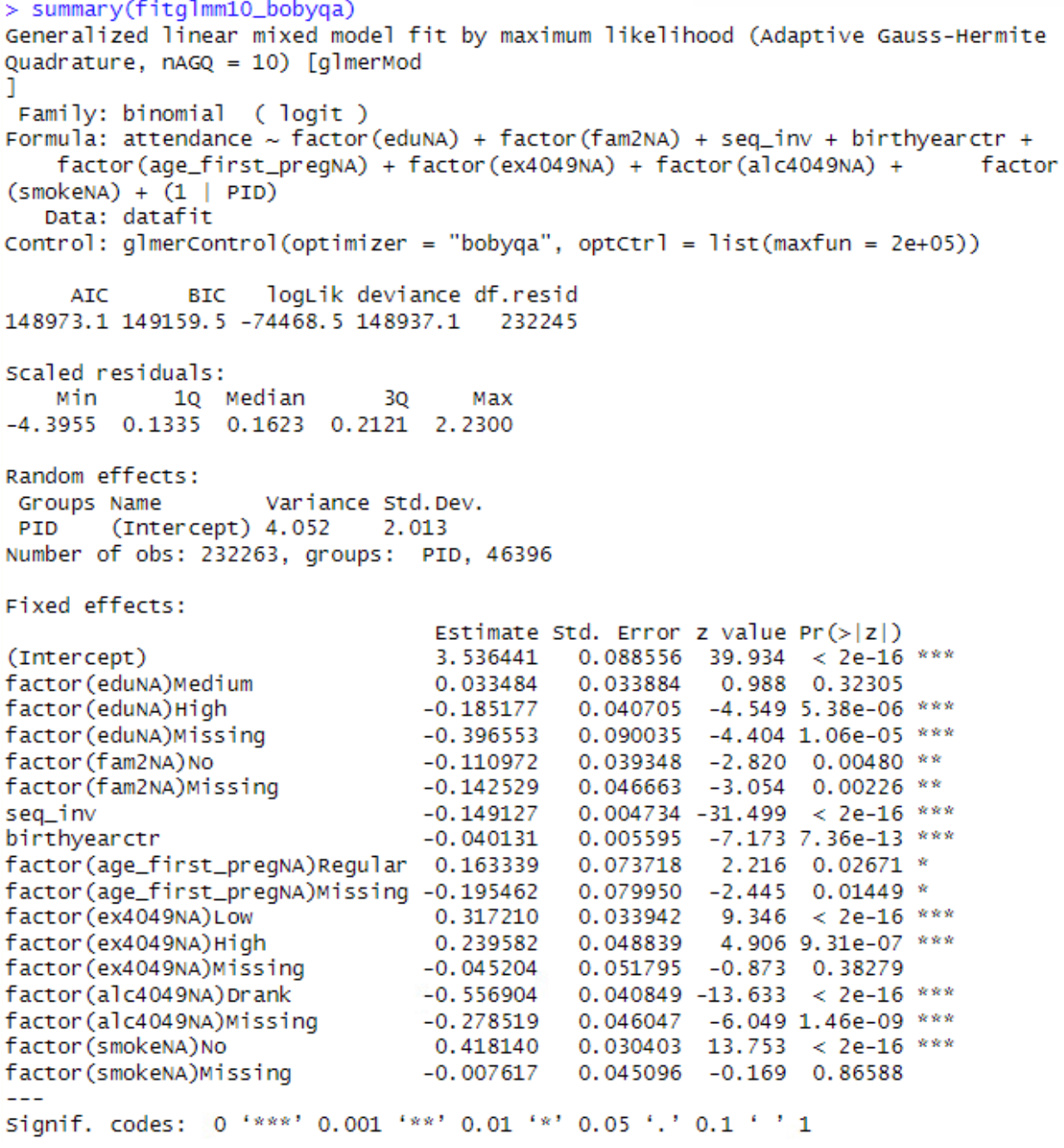}
    \caption{Fit of the binary GLMM model (nAGQ=10, bobyqa algorithm).}
    \label{fig:GLMMFit10bobyqa}
\end{figure}


\end{document}

%% file: table_portrait_all_results.tex
\begin{table}[!h]
\centering
\caption{KS statistics and standard deviations of $\pi_C(\alpha; \bm{x})$ by $\sigma^2$ and $\pi_C(0; \bm{x})$\label{tabequiv}}
\centering
\resizebox{\ifdim\width>\linewidth\linewidth\else\width\fi}{!}{
\fontsize{14}{16}\selectfont
\begin{tabular}[t]{>{}rcccccccccccccccccc}
\toprule
\multicolumn{1}{c}{ } & \multicolumn{18}{c}{$\pi_C(0; \bm{x})$} \\
\cmidrule(l{3pt}r{3pt}){2-19}
\multicolumn{1}{c}{ } & \multicolumn{2}{c}{0.1} & \multicolumn{2}{c}{0.2} & \multicolumn{2}{c}{0.3} & \multicolumn{2}{c}{0.4} & \multicolumn{2}{c}{0.5} & \multicolumn{2}{c}{0.6} & \multicolumn{2}{c}{0.7} & \multicolumn{2}{c}{0.8} & \multicolumn{2}{c}{0.9} \\
\cmidrule(l{3pt}r{3pt}){2-3} \cmidrule(l{3pt}r{3pt}){4-5} \cmidrule(l{3pt}r{3pt}){6-7} \cmidrule(l{3pt}r{3pt}){8-9} \cmidrule(l{3pt}r{3pt}){10-11} \cmidrule(l{3pt}r{3pt}){12-13} \cmidrule(l{3pt}r{3pt}){14-15} \cmidrule(l{3pt}r{3pt}){16-17} \cmidrule(l{3pt}r{3pt}){18-19}
$\sigma^2$ & KS & $\mathrm{SD}$ & KS & $\mathrm{SD}$ & KS & $\mathrm{SD}$ & KS & $\mathrm{SD}$ & KS & $\mathrm{SD}$ & KS & $\mathrm{SD}$ & KS & $\mathrm{SD}$ & KS & $\mathrm{SD}$ & KS & $\mathrm{SD}$\\
\midrule
\textbf{0.1} & 0.02 & 0.03 & 0.03 & 0.05 & 0.03 & 0.07 & 0.03 & 0.07 & 0.02 & 0.08 & 0.04 & 0.08 & 0.04 & 0.07 & 0.05 & 0.05 & 0.05 & 0.03\\
\textbf{0.2} & 0.03 & 0.04 & 0.03 & 0.07 & 0.03 & 0.09 & 0.05 & 0.10 & 0.05 & 0.11 & 0.06 & 0.10 & 0.05 & 0.09 & 0.08 & 0.07 & 0.09 & 0.04\\
\textbf{0.3} & 0.03 & 0.06 & 0.04 & 0.09 & 0.04 & 0.11 & 0.04 & 0.12 & 0.06 & 0.13 & 0.07 & 0.13 & 0.09 & 0.11 & 0.11 & 0.09 & 0.12 & 0.05\\
\textbf{0.4} & 0.03 & 0.06 & 0.04 & 0.10 & 0.05 & 0.13 & 0.06 & 0.14 & 0.07 & 0.15 & 0.08 & 0.14 & 0.10 & 0.13 & 0.11 & 0.10 & 0.16 & 0.07\\
\textbf{0.5} & 0.04 & 0.07 & 0.04 & 0.12 & 0.07 & 0.14 & 0.06 & 0.15 & 0.08 & 0.16 & 0.10 & 0.15 & 0.11 & 0.14 & 0.15 & 0.11 & 0.19 & 0.08\\
\addlinespace
\textbf{0.6} & 0.04 & 0.08 & 0.06 & 0.13 & 0.07 & 0.15 & 0.07 & 0.17 & 0.08 & 0.17 & 0.10 & 0.17 & 0.12 & 0.15 & 0.16 & 0.13 & 0.21 & 0.08\\
\textbf{0.7} & 0.04 & 0.09 & 0.05 & 0.14 & 0.06 & 0.16 & 0.07 & 0.18 & 0.10 & 0.18 & 0.10 & 0.18 & 0.15 & 0.16 & 0.19 & 0.14 & 0.25 & 0.09\\
\textbf{0.8} & 0.04 & 0.10 & 0.07 & 0.15 & 0.06 & 0.17 & 0.08 & 0.19 & 0.10 & 0.19 & 0.12 & 0.19 & 0.17 & 0.17 & 0.21 & 0.15 & 0.28 & 0.10\\
\textbf{0.9} & 0.05 & 0.11 & 0.07 & 0.15 & 0.08 & 0.18 & 0.09 & 0.19 & 0.10 & 0.20 & 0.14 & 0.20 & 0.17 & 0.18 & 0.21 & 0.15 & 0.29 & 0.10\\
\textbf{1.0} & 0.06 & 0.11 & 0.06 & 0.16 & 0.08 & 0.19 & 0.09 & 0.20 & 0.11 & 0.21 & 0.15 & 0.20 & 0.18 & 0.19 & 0.24 & 0.16 & 0.33 & 0.11\\
\addlinespace
\textbf{1.1} & 0.06 & 0.12 & 0.07 & 0.17 & 0.08 & 0.20 & 0.09 & 0.21 & 0.11 & 0.22 & 0.14 & 0.21 & 0.20 & 0.20 & 0.24 & 0.17 & 0.35 & 0.12\\
\textbf{1.2} & 0.06 & 0.13 & 0.07 & 0.18 & 0.08 & 0.20 & 0.11 & 0.22 & 0.12 & 0.22 & 0.16 & 0.22 & 0.21 & 0.20 & 0.28 & 0.18 & 0.37 & 0.12\\
\textbf{1.3} & 0.05 & 0.13 & 0.06 & 0.18 & 0.09 & 0.21 & 0.10 & 0.22 & 0.13 & 0.23 & 0.17 & 0.22 & 0.22 & 0.21 & 0.29 & 0.18 & 0.41 & 0.13\\
\textbf{1.4} & 0.06 & 0.14 & 0.08 & 0.19 & 0.09 & 0.22 & 0.11 & 0.23 & 0.13 & 0.24 & 0.17 & 0.23 & 0.23 & 0.22 & 0.31 & 0.19 & 0.42 & 0.14\\
\textbf{1.5} & 0.06 & 0.14 & 0.07 & 0.19 & 0.09 & 0.22 & 0.11 & 0.23 & 0.13 & 0.24 & 0.17 & 0.23 & 0.24 & 0.22 & 0.32 & 0.19 & 0.43 & 0.14\\
\addlinespace
\textbf{1.6} & 0.06 & 0.15 & 0.09 & 0.20 & 0.10 & 0.23 & 0.11 & 0.24 & 0.14 & 0.24 & 0.19 & 0.24 & 0.26 & 0.23 & 0.35 & 0.20 & 0.48 & 0.15\\
\textbf{1.7} & 0.07 & 0.15 & 0.08 & 0.21 & 0.11 & 0.23 & 0.11 & 0.25 & 0.15 & 0.25 & 0.19 & 0.25 & 0.26 & 0.23 & 0.36 & 0.21 & 0.48 & 0.15\\
\textbf{1.8} & 0.06 & 0.16 & 0.08 & 0.21 & 0.11 & 0.24 & 0.12 & 0.25 & 0.16 & 0.25 & 0.20 & 0.25 & 0.26 & 0.23 & 0.36 & 0.21 & 0.50 & 0.16\\
\textbf{1.9} & 0.07 & 0.16 & 0.09 & 0.22 & 0.10 & 0.24 & 0.12 & 0.26 & 0.16 & 0.26 & 0.20 & 0.25 & 0.28 & 0.24 & 0.40 & 0.21 & 0.53 & 0.17\\
\textbf{2.0} & 0.06 & 0.17 & 0.09 & 0.22 & 0.11 & 0.25 & 0.13 & 0.26 & 0.16 & 0.26 & 0.20 & 0.26 & 0.27 & 0.24 & 0.40 & 0.22 & 0.52 & 0.17\\
\bottomrule
\end{tabular}}
\end{table}

%% file: references.bib
@article{AustinSteyerberg2019,
  author = {Austin, Peter C. and Steyerberg, Ewout W.},
  title = {The Integrated Calibration Index ({ICI}) and Related Metrics for Quantifying the Calibration of Logistic Regression Models},
  journaltitle = {Statistics in Medicine},
  year = {2019},
  volume = {38},
  number = {21},
  pages = {4051--4065},
  doi = {10.1002/sim.8281}
}

@article{BalanPutter2020,
  author       = {Balan, Theodor A. and Putter, Hein},
  title        = {A tutorial on frailty models},
  journaltitle = {Statistical Methods in Medical Research},
  date         = {2020},
  volume       = {29},
  number       = {11},
  pages        = {3424--3454},
  doi          = {10.1177/0962280220921889}
}

@article{BarrenadaEtAl2026,
  author       = {Barre{\~n}ada, Lasai and De Cock Campo, Bavo and Wynants, Laure and Van Calster, Ben},
  title        = {Clustered flexible calibration plots for binary outcomes using random effects modeling},
  journaltitle = {Research Synthesis Methods},
  date         = {2026},
  volume       = {17},
  pages        = {567--588},
  doi          = {10.1017/rsm.2025.10046}
}

@article{BhargavaHofvindEtAl2019,
  title = {Gender, Letters, Relatives, and {{God}}: Mediating Actors in Mammographic Screening among {{Pakistani}} Women in {{Norway}}},
  shorttitle = {Gender, Letters, Relatives, and {{God}}},
  author = {Bhargava, Sameer and Hofvind, Solveig and Moen, K{\aa}re},
  year = 2019,
  month = sep,
  journal = {Acta Radiologica Open},
  volume = {8},
  number = {9},
  pages = {2058460119875015},
  issn = {2058-4601},
  doi = {10.1177/2058460119875015},
  langid = {english},
  pmcid = {PMC6743200},
  pmid = {31548913},
}

@article{BhargavaTsurudaEtAl2018,
  title = {Lower Attendance Rates in Immigrant versus Non-Immigrant Women in the {{Norwegian Breast Cancer Screening Programme}}},
  author = {Bhargava, Sameer and Tsuruda, Kaitlyn and Moen, K{\aa}re and Bukholm, Ida and Hofvind, Solveig},
  year = 2018,
  month = sep,
  journal = {Journal of Medical Screening},
  volume = {25},
  number = {3},
  pages = {155--161},
  issn = {1475-5793},
  doi = {10.1177/0969141317733771},
  langid = {english},
  pmid = {29059006},
}

@techreport{BjornsonHolenEtAl2022,
  title = {{{BreastScreen Norway}}: 25 Years of Organized Screening},
  author = {Bj{\o}rnson, Elin W{\o}lner and Holen, {\AA}sne S{\o}rlien and Sagstad, Silje and Larsen, Marthe and Thy, Jonas and Mangerud, Gunhild and Ertzaas, Anne Kathrin and Hofvind, Solveig},
  editor = {Bj{\o}rnson, Elin W{\o}lner and Holen, {\AA}sne S{\o}rlien and Hofvind, Solveig},
  year = 2022,
  number = {978-82-93804-03-1},
  address = {Oslo},
  institution = {Cancer Registry of Norway},
}

@article{BoothHobert1999,
  author       = {Booth, James G. and Hobert, James P.},
  title        = {Maximizing generalized linear mixed model likelihoods with an automated {Monte Carlo EM} algorithm},
  journaltitle = {Journal of the Royal Statistical Society. Series B (Statistical Methodology)},
  date         = {1999},
  volume       = {61},
  number       = {1},
  pages        = {265--285},
  doi          = {10.1111/1467-9868.00176}
}

@article{BouwmeesterEtAl2013,
  author       = {Bouwmeester, Walter and Twisk, Jos W. R. and Kappen, Teus H. and van Klei, Wilton A. and Moons, Karel G. M. and Vergouwe, Yvonne},
  title        = {Prediction models for clustered data: Comparison of a random intercept and standard regression model},
  journaltitle = {BMC Medical Research Methodology},
  date         = {2013},
  volume       = {13},
  eid          = {19},
  doi          = {10.1186/1471-2288-13-19}
}

@misc{BreastScreen2025,
  author = {{Norwegian Institute of Public Health}},
  year = 2025,
  month = jun,
  journal = {Norwegian Institute of Public Health},
  urldate = {2026-07-07},
  howpublished = {www.fhi.no/en/cancer/screening/breastscreen/},
  langid = {english},
}

@article{BreslowClayton1993,
  author       = {Breslow, N. E. and Clayton, D. G.},
  title        = {Approximate inference in generalized linear mixed models},
  journaltitle = {Journal of the American Statistical Association},
  date         = {1993},
  volume       = {88},
  number       = {421},
  pages        = {9--25},
  doi          = {10.1080/01621459.1993.10594284}
}

@misc{Codex2026,
  title = {Codex},
  author = {OpenAI},
  year = 2026
}

@article{Conaway1990,
  author       = {Conaway, Mark R.},
  title        = {A Random Effects Model for Binary Data},
  journal      = {Biometrics},
  year         = {1990},
  volume       = {46},
  number       = {2},
  pages        = {317--328},
  doi          = {10.2307/2531437},
}

@article{CoullHousemanBetensky2006,
  author       = {Coull, Brent A. and Houseman, E. Andres and Betensky, Rebecca A.},
  title        = {A Computationally Tractable Multivariate Random Effects Model for Clustered Binary Data},
  journal      = {Biometrika},
  year         = {2006},
  volume       = {93},
  number       = {3},
  pages        = {587--599},
  doi          = {10.1093/biomet/93.3.587},
}

@book{Demidenko2013,
  author    = {Demidenko, Eugene},
  title     = {Mixed Models},
  subtitle  = {Theory and Applications with {R}},
  edition   = {2},
  date      = {2013},
  series    = {Wiley Series in Probability and Statistics},
  publisher = {John Wiley \& Sons},
  location  = {Hoboken, NJ},
  doi       = {10.1002/9781118651537}
}

@article{EddelbuettelFrancois2011,
  title = {Rcpp : {{Seamless R}} and {{C}}++ {{Integration}}},
  shorttitle = {{\textbf{Rcpp}}},
  author = {Eddelbuettel, Dirk and Fran{\c c}ois, Romain},
  year = 2011,
  journal = {Journal of Statistical Software},
  volume = {40},
  number = {8},
  issn = {1548-7660},
  doi = {10.18637/jss.v040.i08},
  urldate = {2026-06-24},
  langid = {english},
}

@misc{EpiQuest2025,
  author = {{Norwegian Institute of Public Health}},
  year = 2025,
  month = may,
  journal = {Norwegian Institute of Public Health},
  urldate = {2026-07-07},
  howpublished = {www.fhi.no/en/cancer/screening/breastscreen/epidemilogical-questionnaire-q2006/},
  langid = {english},
}

@article{Heagerty1999,
  author       = {Heagerty, Patrick J.},
  title        = {Marginally specified logistic-normal models for longitudinal binary data},
  journaltitle = {Biometrics},
  date         = {1999},
  volume       = {55},
  number       = {3},
  pages        = {688--698},
  doi          = {10.1111/j.0006-341X.1999.00688.x}
}

@book{JiangNguyen2021,
  author    = {Jiang, Jiming and Nguyen, Thuan},
  title     = {Linear and Generalized Linear Mixed Models and Their Applications},
  edition   = {2},
  date      = {2021},
  series    = {Springer Series in Statistics},
  publisher = {Springer},
  location  = {New York},
  doi       = {10.1007/978-1-0716-1282-8}
}

@article{KazemiHassanzadeh2021,
  author       = {Kazemi, Iraj and Hassanzadeh, Fatemeh},
  title        = {Marginalized random-effects models for clustered binomial data through innovative link functions},
  journaltitle = {AStA Advances in Statistical Analysis},
  date         = {2021},
  volume       = {105},
  number       = {2},
  pages        = {197--228},
  doi          = {10.1007/s10182-021-00400-0}
}

@article{Lehmann1953,
  author = {Lehmann, E. L.},
  title = {The Power of Rank Tests},
  journaltitle = {The Annals of Mathematical Statistics},
  year = {1953},
  volume = {24},
  number = {1},
  pages = {23--43},
  doi = {10.1214/aoms/1177729080}
}

@article{LiangZeger1986,
  author       = {Liang, Kung-Yee and Zeger, Scott L.},
  title        = {Longitudinal data analysis using generalized linear models},
  journaltitle = {Biometrika},
  date         = {1986},
  volume       = {73},
  number       = {1},
  pages        = {13--22},
  doi          = {10.1093/biomet/73.1.13}
}

@article{LiuPierce1994,
  author       = {Liu, Qing and Pierce, Donald A.},
  title        = {A note on {Gauss--Hermite} quadrature},
  journaltitle = {Biometrika},
  date         = {1994},
  volume       = {81},
  number       = {3},
  pages        = {624--629},
  doi          = {10.1093/biomet/81.3.624}
}

@article{McCulloch1997,
  author       = {McCulloch, Charles E.},
  title        = {Maximum likelihood algorithms for generalized linear mixed models},
  journaltitle = {Journal of the American Statistical Association},
  date         = {1997},
  volume       = {92},
  number       = {437},
  pages        = {162--170},
  doi          = {10.1080/01621459.1997.10473613}
}

@book{MolenberghsVerbeke2005,
  author    = {Molenberghs, Geert and Verbeke, Geert},
  title     = {Models for Discrete Longitudinal Data},
  date      = {2005},
  series    = {Springer Series in Statistics},
  publisher = {Springer},
  location  = {New York},
  doi       = {10.1007/0-387-28980-1}
}

@online{MunozEtAl2025,
  author       = {Munoz, Johanna and Moons, Karel G. M. and Debray, Thomas P. A. and de Jong, Valentijn M. T.},
  title        = {Deriving calibration plots in large clustered datasets},
  date         = {2025-06-05},
  organization = {Research Square},
  version      = {1},
  note         = {Preprint},
  doi          = {10.21203/rs.3.rs-6740452/v1}
}

@article{PinheiroBates1995,
  author       = {Pinheiro, Jos{\'e} C. and Bates, Douglas M.},
  title        = {Approximations to the log-likelihood function in the nonlinear mixed-effects model},
  journaltitle = {Journal of Computational and Graphical Statistics},
  date         = {1995},
  volume       = {4},
  number       = {1},
  pages        = {12--35},
  doi          = {10.1080/10618600.1995.10474663}
}

@misc{R2026,
  title = {R: {{A Language}} and {{Environment}} for {{Statistical Computing}}},
  author = {R Core Team},
  year = 2026,
  publisher = {R Foundation for Statistical Computing},
  address = {VIenna, Austria}
}

@article{RodriguezGoldman1995,
  author       = {Rodr{\'i}guez, Germ{\'a}n and Goldman, Noreen},
  title        = {An assessment of estimation procedures for multilevel models with binary responses},
  journaltitle = {Journal of the Royal Statistical Society. Series A (Statistics in Society)},
  date         = {1995},
  volume       = {158},
  number       = {1},
  pages        = {73--89},
  doi          = {10.2307/2983404}
}

@article{SkrondalRabeHesketh2009,
  author       = {Skrondal, Anders and Rabe-Hesketh, Sophia},
  title        = {Prediction in multilevel generalized linear models},
  journaltitle = {Journal of the Royal Statistical Society. Series A (Statistics in Society)},
  date         = {2009},
  volume       = {172},
  number       = {3},
  pages        = {659--687},
  doi          = {10.1111/j.1467-985X.2009.00587.x}
}

@article{SunEtAl2018,
  author       = {Sun, Hanmei and Nguyen, Thuan and Luan, Yihui and Jiang, Jiming},
  title        = {Classified mixed logistic model prediction},
  journaltitle = {Journal of Multivariate Analysis},
  date         = {2018},
  volume       = {168},
  pages        = {63--74},
  doi          = {10.1016/j.jmva.2018.06.004}
}

@article{TamuraGiampaoli2010,
  author       = {Tamura, Karin Ayumi and Giampaoli, Viviana},
  title        = {Prediction in multilevel logistic regression},
  journaltitle = {Communications in Statistics---Simulation and Computation},
  date         = {2010},
  volume       = {39},
  number       = {6},
  pages        = {1083--1096},
  doi          = {10.1080/03610911003790106}
}

@article{TamuraGiampaoli2013,
  author       = {Tamura, Karin Ayumi and Giampaoli, Viviana},
  title        = {New prediction method for the mixed logistic model applied in a marketing problem},
  journaltitle = {Computational Statistics \& Data Analysis},
  date         = {2013},
  volume       = {66},
  pages        = {202--216},
  doi          = {10.1016/j.csda.2013.04.006}
}

@article{TierneyKadane1986,
  author       = {Tierney, Luke and Kadane, Joseph B.},
  title        = {Accurate approximations for posterior moments and marginal densities},
  journaltitle = {Journal of the American Statistical Association},
  date         = {1986},
  volume       = {81},
  number       = {393},
  pages        = {82--86},
  doi          = {10.1080/01621459.1986.10478240}
}

@article{TsurudaSagstadEtAl2018,
  title = {Validity and Reliability of Self-Reported Health Indicators among Women Attending Organized Mammographic Screening},
  author = {Tsuruda, Kaitlyn M. and Sagstad, Silje and Sebu{\o}deg{\aa}rd, Sofie and Hofvind, Solveig},
  year = 2018,
  month = nov,
  journal = {Scandinavian Journal of Public Health},
  volume = {46},
  number = {7},
  pages = {744--751},
  publisher = {SAGE Publications Ltd STM},
  issn = {1403-4948},
  doi = {10.1177/1403494817749393},
  urldate = {2026-07-07},
  langid = {english}
}

@article{VanOirbeekLesaffre2012,
  author       = {Van Oirbeek, Robin and Lesaffre, Emmanuel},
  title        = {Assessing the predictive ability of a multilevel binary regression model},
  journaltitle = {Computational Statistics \& Data Analysis},
  date         = {2012},
  volume       = {56},
  number       = {6},
  pages        = {1966--1980},
  doi          = {10.1016/j.csda.2011.11.023}
}

@article{VaupelEtAl1979,
  author       = {Vaupel, J. W. and Manton, K. G. and Stallard, E.},
  title        = {The impact of heterogeneity in individual frailty on the dynamics of mortality},
  journaltitle = {Demography},
  date         = {1979},
  volume       = {16},
  number       = {3},
  pages        = {439--454},
  doi          = {10.2307/2061224}
}

@article{WangLouis2003,
  author       = {Wang, Zengri and Louis, Thomas A.},
  title        = {Matching conditional and marginal shapes in binary random intercept models using a bridge distribution function},
  journaltitle = {Biometrika},
  date         = {2003},
  volume       = {90},
  number       = {4},
  pages        = {765--775},
  doi          = {10.1093/biomet/90.4.765}
}

@article{WynantsEtAl2015,
  author       = {Wynants, Laure and Bouwmeester, Walter and Moons, Karel G. M. and Moerbeek, Mirjam and Timmerman, Dirk and Van Huffel, Sabine and Van Calster, Ben and Vergouwe, Yvonne},
  title        = {A simulation study of sample size demonstrated the importance of the number of events per variable to develop prediction models in clustered data},
  journaltitle = {Journal of Clinical Epidemiology},
  date         = {2015},
  volume       = {68},
  number       = {12},
  pages        = {1406--1414},
  doi          = {10.1016/j.jclinepi.2015.02.002}
}

@article{WynantsEtAl2018,
  author       = {Wynants, Laure and Vergouwe, Yvonne and Van Huffel, Sabine and Timmerman, Dirk and Van Calster, Ben},
  title        = {Does ignoring clustering in multicenter data influence the performance of prediction models? A simulation study},
  journaltitle = {Statistical Methods in Medical Research},
  date         = {2018},
  volume       = {27},
  number       = {6},
  pages        = {1723--1736},
  doi          = {10.1177/0962280216668555}
}

@article{morris2019simulation,
  author = {Morris, Tim P. and White, Ian R. and Crowther, Michael J.},
  title = {Using Simulation Studies to Evaluate Statistical Methods},
  journaltitle = {Statistics in Medicine},
  year = {2019},
  volume = {38},
  number = {11},
  pages = {2074--2102},
  doi = {10.1002/sim.8086}
}

@article{JiangLahiri2001,
  author        = {Jiang, Jiming and Lahiri, P.},
  title         = {{Empirical Best Prediction for Small Area Inference with Binary Data}},
  year          = {2001},
  journal       = {Annals of the Institute of Statistical Mathematics},
  volume        = {53},
  number        = {2},
  pages         = {217--243},
  doi           = {10.1023/a:1012410420337}
}

@article{Jiang2003,
  author        = {Jiang, Jiming},
  title         = {{Empirical best prediction for small-area inference based on generalized linear mixed models}},
  year          = {2003},
  journal       = {Journal of Statistical Planning and Inference},
  volume        = {111},
  number        = {1--2},
  pages         = {117--127},
  doi           = {10.1016/S0378-3758(02)00293-8}
}

@book{RaoMolina2015,
  author        = {Rao, J. N. K. and Molina, Isabel},
  title         = {{Small Area Estimation}},
  year          = {2015},
  publisher     = {John Wiley \& Sons},
  doi           = {10.1002/9781118735855},
  edition       = {2nd}
}

@article{NeuhausMcCullochBoylan2025,
  author        = {Neuhaus, John and McCulloch, Charles and Boylan, Ross},
  title         = {{Improved prediction and flagging of extreme random effects for non-Gaussian outcomes using weighted methods}},
  year          = {2025},
  journal       = {Biometrics},
  volume        = {81},
  number        = {3},
  pages         = {ujaf094},
  doi           = {10.1093/biomtc/ujaf094}
}

@article{Robinson1991,
  author        = {Robinson, G. K.},
  title         = {{That BLUP Is a Good Thing: The Estimation of Random Effects}},
  year          = {1991},
  journal       = {Statistical Science},
  volume        = {6},
  number        = {1},
  doi           = {10.1214/ss/1177011926},
  pages         = {15--32}
}

@book{SkrondalRabeHesketh2004,
  author        = {Skrondal, Anders and Rabe-Hesketh, Sophia},
  title         = {{Generalized Latent Variable Modeling: Multilevel, Longitudinal, and Structural Equation Models}},
  year          = {2004},
  publisher     = {Chapman \& Hall/CRC},
  doi           = {10.1201/9780203489437}
}

@misc{ZhangChan2014,
  author        = {Zhang, Rui and Chan, Kwun Chuen Gary},
  title         = {{Marginalizable conditional model for clustered ordinal data}},
  year          = {2014},
  howpublished  = {arXiv preprint arXiv:1403.7812},
  doi           = {10.48550/arXiv.1403.7812},
  url           = {https://arxiv.org/abs/1403.7812},
  eprint        = {1403.7812},
  archivePrefix = {arXiv},
  primaryClass  = {stat.ME}
}

@article{RiceJohnsonStrawderman2019,
  author        = {Rice, John D and Johnson, Brent A and Strawderman, Robert L},
  title         = {{Modeling the rate of HIV testing from repeated binary data amidst potential never-testers}},
  year          = {2019},
  journal       = {Biostatistics},
  volume        = {20},
  number        = {2},
  pages         = {183--198},
  doi           = {10.1093/biostatistics/kxx071}
}

@article{TchetgenCoull2006,
  author  = {Tchetgen, Eric J. and Coull, Brent A.},
  title   = {A diagnostic test for the mixing distribution in a generalised linear mixed model},
  journal = {Biometrika},
  year    = {2006},
  volume  = {93},
  number  = {4},
  pages   = {1003--1010},
  doi     = {10.1093/biomet/93.4.1003}
}

@article{McCullochNeuhaus2011,
  author  = {McCulloch, Charles E. and Neuhaus, John M.},
  title   = {Misspecifying the Shape of a Random Effects Distribution: Why Getting It Wrong May Not Matter},
  journal = {Statistical Science},
  year    = {2011},
  volume  = {26},
  number  = {3},
  pages   = {388--402},
  doi     = {10.1214/11-STS361}
}
